\documentclass{aa}

\usepackage[colorlinks=true,linkcolor=blue,citecolor=blue,filecolor=blue,urlcolor=blue]{hyperref}
\usepackage{txfonts}
\usepackage{natbib}
\usepackage{inputenc}

\newcommand{\m}[1]{\boldsymbol{#1}}

\newcommand{\gcc}{~g~cm$^{-3}$}
\newcommand{\Su}{~g~cm$^{-2}$}
\newcommand{\Teq}{T_{\rm eq}}
\newcommand{\Mj}{M_{\rm j}}
\newcommand{\Rj}{R_{\rm j}}
\newcommand{\Me}{M_\oplus}

\begin{document}

\title{Inflated hot Jupiters: Inferring average atmospheric velocity via Ohmic models coupled with internal dynamo evolution}

\author{Daniele Vigan\`{o}\inst{1,2,3}\thanks{E-mail: vigano@ice.csic.es}, Soumya Sengupta\inst{1}, Cl\`{a}udia Soriano-Guerrero\inst{1,2}, Rosalba Perna\inst{4}, Albert Elias-L\'{o}pez\inst{1,2}, Sandeep Kumar\inst{5}, Taner Akg\"{u}n\inst{1}}
\authorrunning{Vigan\`o et al.}
\titlerunning{Ohmic dissipation in HJs}

\institute{Institut de Ci\`encies de I'Espai (ICE-CSIC), Campus UAB, Carrer de Can Magrans s/n, 08193 Cerdanyola del Vallès, Barcelona, Catalonia, Spain
\and
Institut d’Estudis Espacials de Catalunya (IEEC), 08860 Castelldefels, Barcelona, Catalonia, Spain
\and Institute of Applied Computing \& Community Code (IAC3), University of the Balearic Islands, Palma, 07122, Spain
\and Department of Physics and Astronomy, Stony Brook University, Stony Brook, NY 11794-3800, USA
\and Department of Physics, Indian Institute of Technology Jammu, Jammu 181221, India}

\date{Received 19 April 2025 / Accepted 17 July 2025}
\date{}

\abstract
{}
{The inflated radii observed in hundreds of hot Jupiters (HJ) represent a long-standing open issue. In this study, we quantitatively investigate this phenomenon within the framework of Ohmic dissipation arising from magnetic induction in the atmosphere, one of the most promising mechanisms for explaining the radius anomaly.}
{Using {\tt MESA}, we simulated the evolution of irradiated giant planets spanning the observed range of masses and equilibrium temperatures, incorporating an internal source of Ohmic dissipation that extends to deep layers of the envelope. We considered the heat-flux-dependent evolution of the deep-seated dynamo field on which the induced field depends. We adopted a state-of-the-art  electrical conductivity, considering the thermal ionisation of alkali metals in the outer layers and the pressure-ionisation in the interior and the corresponding solutions for the induced currents across the planet.}
{We inferred that, in order to reproduce the range of observed radii, the atmospheric wind intensities averaged in the region $p\lesssim 10$ bar have to be in the range 0.01-1 km/s and to decrease roughly linearly with planetary mass and much more steeply with equilibrium temperature. This is consistent with the expected effects of magnetic drag from the induced field, which is higher for more intense irradiation, via conductivity, and for larger masses, which have higher dynamo fields. Due to the evolution of the dynamo field and the proportionality of the induced currents on it, the Ohmic efficiency typically decreases by at least one order of magnitude from 0.1 to 10 Gyr, which is in contrast with the common assumption of a constant-in-time value. Notably, the extent of the main convective region and the associated heat flux supporting the dynamo is reduced in the presence of strong Ohmic dissipation, which in turn depends on the dynamo field strength, generating a non-trivial coupling of the latter with the atmospheric induction and potentially leading to the oscillatory behaviour of the field strength. These findings remain generally valid even when accounting for a long-term increase in the main-sequence host star luminosity, although this case can more readily lead to HJ re-inflation, consistent with previous studies.}
{}


\maketitle

\section{Introduction}

The intriguing exoplanet class of hot Jupiters (HJs) currently consists of several hundred objects, discovered and characterised in terms of mass and radius. These gas giant planets have orbital periods of a few days, corresponding to separations within $\sim$0.1~AU from their host stars (for a review, see \citealt{heng15} and \citealt{fortney21}). Such proximity implies tidal locking, with strong stellar irradiation of their day sides, which results in typical equilibrium temperatures of up to $3000$~K, with a few that are even hotter. The intense irradiation flux coupled with the strong thermal gradients due to the tidal locking gives rise to powerful atmospheric flows, with equatorial super-rotation and strong day-night flows that try to redistribute the heat across the planetary surface (e.g. \citealt{cho08,dobbs08,showman09,heng11,rauscher12,perna12,rauscher13,perez13,parmentier13,rogers14a,showman15,kataria15,koll18,beltz22,komacek22,sengupta23}).

Observations of HJs have revealed that a substantial fraction of them have radii that are significantly larger than those predicted by standard cooling evolutionary models (e.g. \citealt{showman02} and \citealt{wang15}), even when accounting for the effect of irradiation, which can inflate the radius by up to $\sim$20$\%$ \citep{guillot96,arras06,fortney07}. Several explanations have been proposed \citep{spiegel13}, and they can be broadly classified into two categories (see \citealt{thorngren24} for a review). The first one invokes planetary evolution with a delayed cooling, which has been proposed to be due to large atmospheric opacities \citep{burrows07} or due to the inhibition of large-scale convection \citep{chabrier07}. As the opacity increases, cooling becomes inefficient and planets can retain more internal heat. The second category relies on the presence of additional heat sources in the convective regions of the planet (e.g. considered in \citealt{ginzburg15,lopez16,komacek17,komacek20}). Various mechanisms have been proposed for this heating enhancement: hydrodynamic dissipation (involving the conversion of the stellar flux into the kinetic energy of the global atmospheric flow due to the large day-night temperature gradient (e.g. \citealt{showman02} and \citealt{showman09}), tidal dissipation \citep{bodenheimer01}, fluid instabilities \citep{li10}, forced turbulence in the radiative layer \citep{youdin10}, advection of potential temperature \citep{tremblin17}, and Ohmic dissipation. 
It is worth noting that some processes, particularly tidal and Ohmic dissipation, primarily require additional heat deposition in the convective region, whereas others exert their main influence in the radiative zone.

Among these proposed mechanisms, Ohmic dissipation has received considerable attention  \citep{batygin10,batygin11,perna10a,perna10b,huang12,wu13,rauscher13,rogers14a,rogers14b,ginzburg16,rogers17,hardy22,beltz22,benavides22,knierim22}.
The deep-seated magnetic field, generated in the metallic hydrogen region \citep{french12}, is a fundamental ingredient since the atmospheric flow of thermally ionised particles (mostly free electrons from alkali metals) advects its field lines, twisting them considerably in case of high magnetic Reynolds numbers, as it arguably happens in a large fraction of HJs \citep{dietrich22,soriano25}. Although the shear and the corresponding magnetic field winding happen in the shallow atmospheric layers (at pressures $p\lesssim 10$ bar), the twisting of the lines propagates into deeper convective regions, as modelled in previous works for cold giant planets \citep{liu08} and HJs \citep{perna10a,perna10b,batygin10,batygin11,wu13,ginzburg16}, with different assumptions.
This results in an additional heat source for the planet that slows down the cooling and shrinking, leading to a larger radius than that expected from standard cooling. While the extent of the magnetic drag \citep{perna10a} in reducing flow speeds (and hence the magnitude of Ohmic dissipation) is still debated \citep{menou12,knierim22}, this heating source appears consistent with observations \citep{laughlin11,thorngren18}. There is also the possibility that more than one mechanism may operate simultaneously \citep{sarkis21}.

The goal of this paper is to study the long-term evolution (on gigayear timescales) of HJs. Using a specifically modified version of the public code {\tt MESA}\footnote{\url{https://github.com/MESAHub/mesa}} \citep{paxton11, paxton13, paxton15, paxton18, paxton19, jermyn23}, we ran a grid of simulations for a range of planetary masses, equilibrium temperatures, and intensity of the currents, varying also the composition and core masses. Similarly to \cite{batygin11,wu13}, for example, we used a temperature- and density-dependent conductivity, which makes the total Ohmic dissipation rate vary in time.

We introduce three main novelties. The first is the proportionality of the induced currents with the deep-seated dynamo magnetic field, which evolves in time via widely used scaling laws that account for the internal heat flux at a given age \citep{christensen09}. The second is the parametrisation of the complex, intrinsically 3D wind pattern in the uppermost layer, with the average wind velocity in the atmospheric region. Our aim is to constrain its value.
This approach is similar but also complementary to what has been done by \cite{knierim22}, who nevertheless used a constant dynamo magnetic field (like previous studies) and focused more on the exploration of the detailed wind profile in the atmospheric region.

The third novelty is the inclusion of the most updated values of temperature- and density-dependent electrical conductivity, taking into account both the thermal ionisation of alkali metals, dominant at $\rho \lesssim 10^{-2}$ \gcc \, \citep{kumar21}, and the latest calculation of electron conductivity arising from the pressure-ionisation of H and He \citep{bonitz24}, which are important at higher densities. Finally, we propose the use of semi-analytical piecewise power-law solutions for the current density profiles, applicable to arbitrary combinations of magnetic and wind velocities. 

The paper is organised as follows: Sect.~\ref{sec:obs} briefly presents the observational data. In Sect.~\ref{sec:theory} we describe the standard evolutionary models for HJs as implemented in {\tt MESA}. Sect.~\ref{sec:ohmic} presents our Ohmic heating model and its major ingredients. Sect.~\ref{sec:results} shows representative results of the simulations, and we discusses some implications. We provide a discussion and the main conclusions in Sect.~\ref{sec:summary}.

\begin{figure*}
\centering{\includegraphics[width=.59\textwidth]{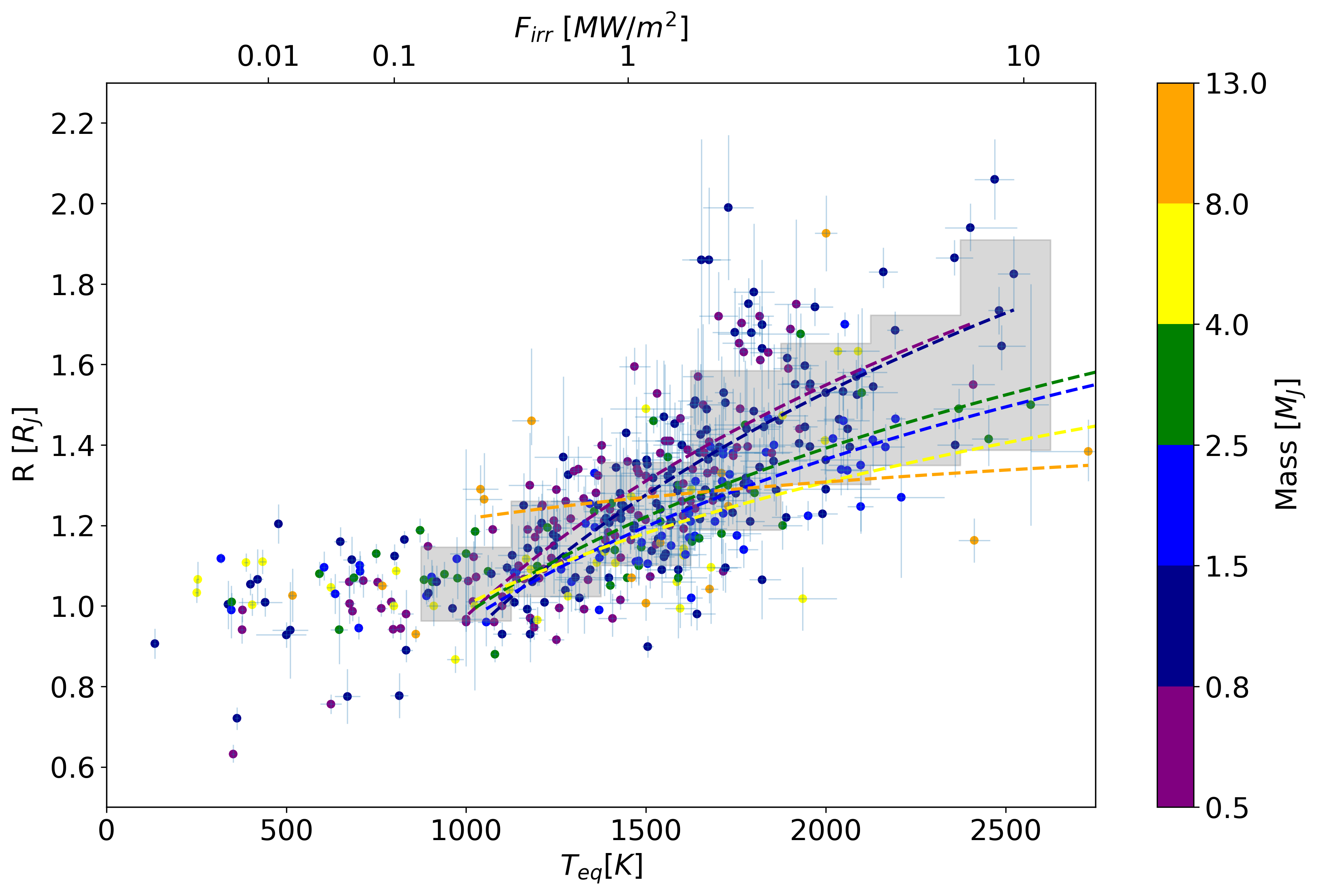}
\centering{\includegraphics[width=.8\textwidth]{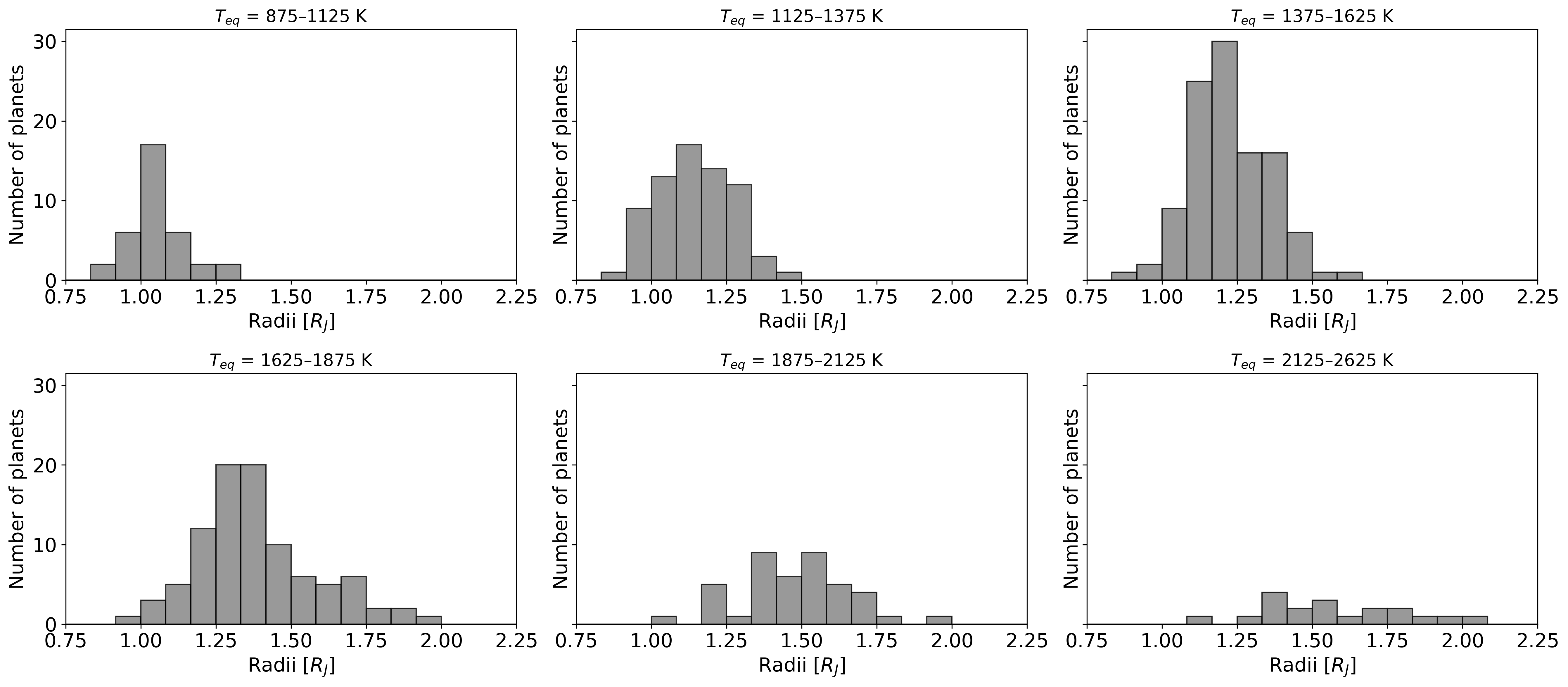}}
}
\caption{Hot Jupiter inflated radii observational trends for the 424 Jupiter-like exoplanets in the database with available estimated mass $M\geq 0.5~\Mj$; age $t \geq 100$ Myr; radius, $R$; and equilibrium temperature, $\Teq$. We consider only the planets with relative errors in a radius of less than $25\%$ and available uncertainties for the mass. We also excluded seven planets younger than 100 Myr. {\em Top:} Planetary radius, $R$, versus, $\Teq$, or, equivalently, irradiated flux $F_{\rm irr}=4\sigma_{\rm sb}\Teq^4$. The colour of the data points identify the mass by ranges as follows: purple [0.5,0.8) $\Mj$, dark blue [0.8,1.5) $\Mj$, blue [1.5,2.5) $\Mj$, green [2.5,4) $\Mj$, yellow [4,8) $\Mj$ and orange [8,13] $\Mj$. With the same colour coding, the lines represent the best-fit reported in Table~\ref{tab:hj_trends}, for the planets within each mass range, and considering only $\Teq > 1000$ K. For the sake of better visibility of the bulk of planets, we left out three ultra-HJ with $\Teq>3300$ K from the plot, but they are considered in the statistical trends. The grey background boxes indicate the mean and standard deviation of the observed radius $\bar R \pm \delta_R$ for all HJs within a 250 K-wide range of $\Teq$. {\em Bottom:} Histograms of the observed radii for the same $\Teq$ ranges per each $\Teq$ range of the grey boxes of the top panel. Due to the small size of the hottest HJ sample, we grouped two $\Teq$ bins in the last histogram.}
\label{fig:obs_hj}
\end{figure*}

\section{Observational sample of Hot Jupiters}\label{sec:obs}

Here we present a short overview of the observational data on HJs obtained from the NASA Exoplanet Archive. As of June 1, 2025, there were a total of 1899 confirmed gas giants with known masses in the range considered here, $M \in [0.5,13]~\Mj$ (we denote the Jovian mass and radius by $\Mj$ and $\Rj$, respectively). The upper limit of the range is simply the Deuterium-burning boundary with brown dwarfs. The lower limit is conservatively in agreement with other studies, such as \cite{sestovic18}, who excluded the lightest planets whose masses and radii might be affected by mass loss.

\begin{table}
\caption{Observed HJ trends as a function of mass range.}
\centering
\begin{tabular}{cccc}
\hline 
Mass  [$\Mj$]  & Number & $A$ & $C$ \\
\hline
0.5-0.8 & 118 & $0.48\pm0.04$ & $0.97\pm 0.03$ \\
0.8-1.5 &150  & $0.51\pm 0.04$ & $0.92 \pm 0.04$ \\
1.5-2.5 &65  & $0.34\pm 0.04$ & $0.96\pm 0.04$ \\
2.5-4   &  41& $0.34\pm 0.05$ & $0.98\pm0.05$ \\
4-8     & 33 & $0.25\pm 0.08$ & $1.00\pm0.07$ \\
8-13    &  15& $0.08\pm 0.15$ & $1.22\pm0.14$ \\
\hline
\end{tabular} 
\tablefoot{Columns indicate: mass ranges, number of objects with $\Teq\geq 1000$ K and best-fit parameters of the functional form $R/\Rj = C + A\log(F/F_{s})$, where $F_s$ corresponds to the irradiation flux of $\Teq=1000$ K.}
\label{tab:hj_trends}
\end{table}

Among that sample, only 470 planets have available information on the quantities we are interested in: mass $M$, radius $R$, age $t$, and equilibrium temperature $\Teq$, defined as 
    \begin{equation}
    T_{\rm eq} = \left( \frac{R_{\star}}{2a} \right)^{1/2} T_\star \, ,
    \end{equation}
where $R_\star$ and $T_\star$ are the observed radius and effective temperature of the host star, $a$ is the star-planet separation, and the definition assumes negligible planetary albedo and a perfectly homogeneous heat redistribution over the entire planetary surface.

Since the radius is our main observable to compare with, we excluded data with high errors in radius and only selected those planets with errors of less than $25\%$. Moreover, we have discarded those planets with only upper limits on the mass and without associated mass uncertainties. We also exclude seven planets younger than $100$ Myr, since at those ages the planet shrinking has some memory of the initial entropy and the very early ages, which is unknown, therefore introducing additional free parameters in the problem.

These selection criteria yield a data set of 424 gas giants. Fig.~\ref{fig:obs_hj} shows the measured values of $R$ versus $\Teq$ (top panel) and the distribution of radii for different planets grouped by different ranges of temperature with 250 K-wide bins centred around values in the range $\Teq = [1000,2500]$ K (six small panels in the bottom), with the associated mean and standard deviation values, $\bar{R}$ and $\sigma_R$ represented by grey boxes in the top panel. As is well known (e.g. \citealt{menou12}), the radii of the planets increase as the equilibrium temperature rises, for planets with $T_{\rm eq}\gtrsim 1000$~K. The average radius increases up to $\bar R = 1.48 \pm 0.18 ~\Rj$ for the $\Teq \in [1875,2125]$ K sample, and up to $\bar R = 1.53 \pm 0.19~\Rj$ for the $\Teq \in [2125,2375]$ K sample (which is, however, sparsely populated). The coloured lines in the figure show the best-fitting trends of the form $R/\Rj = C + A\log (F_{\rm irr}/F_{s})$, where we fix $F_{s}=0.227$ MW m$^{-2}$, i.e. the irradiation flux corresponding to $T_\mathrm{eq} = 1000$~K. The fit is performed for planets with $T_\mathrm{eq} \gtrsim 1000$~K, across different mass ranges, as reported in  Table~\ref{tab:hj_trends}. Such fits are comparable to the values provided by \cite{sestovic18}, who fitted the same functional form for planets with $F\geq F_s$, but they let $F_s$ to vary. Other older studies \citep{laughlin11,enoch12} use different functional forms, all non-linear with $\Teq$ as well, with a sample a few times smaller. Considering the differences in samples and functional forms, our results can be considered overall consistent.

These trends show that lighter HJs ($M\sim 0.5-1.5~\Mj$) have a statistically significant steeper dependence with irradiation ($A\sim 0.5$), compared to the heavier ones. The extremely massive HJs, $M\gtrsim 8~\Mj$ (close to the border with brown dwarfs) barely show signs of inflation. This trend with mass is well known, (e.g. \citealt{sestovic18,thorngren24} and references within) and can initially be understood by the gravity counteracting inflation for a given rate of deposited heat (see also e.g. \citealt{lopez16} for a brief discussion).

Below we assess this dispersion in terms of the most important factors intervening in the planetary long-term cooling, considering the dependence of the modelled radius evolution on observables ($\Teq,M$) and on the deposition of heat. Factors like composition, mass core can also contribute to the dispersion of the sample, but in this study we do not explore them. We refer the reader to App.~\ref{app:benchmark} for a brief exploration of composition, core mass and density, with references to the extended literature on this aspect.

\section{Overview of the evolutionary model}\label{sec:theory}

\subsection{Structure equations}

We modelled the evolution of HJs using the publicly available code {\tt MESA} \citep{paxton11,paxton13,paxton19} and solving the one-dimensional time-dependent equations of internal structure applied to gas giants. The first equation is the conservation of mass,
    \begin{equation}
    \frac{dm}{dr} = 4 \pi r^2 \rho \, ,
    \end{equation}
where $m$ is the mass enclosed within the radius, $r$, and $\rho$ is the density. The second is the hydrostatic equilibrium,
    \begin{equation}
    \frac{dp}{dm} = - \frac{G m}{4\pi r^4} \, ,
    \end{equation}
where $p$ is the pressure and $G$ is the gravitational constant. The energy transport equation is given by
    \begin{equation}
    \frac{dT}{dm} = - \frac{G m T}{4\pi r^4 p} \nabla \, ,
    \end{equation}
where $T$ is the temperature and $\nabla = d \ln T / d \ln p$ is the logarithmic temperature gradient, set to the smallest value between the adiabatic and the radiative gradients.

The equations above define the internal structure at a given time; its evolution is driven by energy conservation, which in terms of the internal luminosity, $L$, reads as
    \begin{equation}
    \frac{dL}{dm} = \epsilon_{\rm grav} + \epsilon_{\rm irr} + \epsilon_{\rm j} \, ,
    \label{eq_luminosity}
    \end{equation}
where $\epsilon_{\rm grav} = - T dS/dt$ is the heating term due to gravitational contraction or inflation, with $S$ as the specific entropy; $\epsilon_{\rm irr}$ is the heating of the outermost layers due to irradiation (see Sect.~\ref{sec:irradiation}); and $\epsilon_{\rm j}$ represents any additional heat deposition rate, which in our case comes from Ohmic dissipation of currents and varies in depth and time, as specified in Sect.~\ref{sec:ohmic}. The planetary radius $R$ varies in time, according to these structure equations. {\tt MESA} typically solves the equations from the centre to the outermost layers (located at $p\sim 0.1$ bar).

The equations above are closed using the {\tt MESA} equation of state \citep{paxton19}, which consists of the interpolation of OPAL and Saumon--Chabrier--van Horn \citep{saumon95} for the temperatures and densities relevant for gas giants. If not stated otherwise, we employ a solar-like composition ($Y = 0.24,Z = 0.02$), homogeneous throughout the envelope. In this respect, we neglect important metal enrichment in the envelope \citep{thorngren16}. We also neglect the pressure-induced immiscibility of hydrogen and helium \citep{stevenson77a,stevenson77b,stevenson80,lorenzen09,morales13}, arguably relevant for Saturn and Jupiter, for which the adiabat can intersect the phase curve separation at $p\sim$ Mbar, $T\sim 5000$ K \citep{brygoo21}. For typical HJs, the adiabats are typically hotter and above such phase separation, but we cannot exclude that an adiabat intersection may happen in mildly irradiated, weakly heated, old planets. Since there are anyway large uncertainties, and our main focus is rather on the hottest sample of giant planets, we thus neglect the H-He immiscibility. Another ingredient we neglect is the double-diffusive convection, expected to happen under the presence of a sufficiently strong compositional stratification \citep{chabrier07,leconte13}.

An important free parameter in solving the structure is the total planetary mass $M=m(R)$, which is assumed to be constant throughout the evolution. We do not consider any mass loss due to, for example, wind stripping of the atmosphere (see e.g. \citealt{lazovik23}). Such processes could explain at least part of the HJs with $R\lesssim 1~\Rj$ and have been inferred in some short-orbit exoplanets, via Lyman-$\alpha$ and helium triplet observations (e.g. \citealt{salz15,lavie17,tyler24}). For HJs, observations indicate a relatively small mass loss ($\lesssim 1\%$) accumulated over their lifetime. However, this might be an observational bias, and HJ evaporation could affect the observed mass distribution \citep{kurokawa14}, similar to the atmospheric loss observed in hot super-Earths, which can leave behind smaller, gas-depleted rocky remnants (e.g. \citealt{owen16}). These effects should be more important for relatively light HJs and hot Saturns \citep{thorngren23}. Therefore, this study mostly focuses on the upper range of the radius values.

\subsection{Core mass}

{\tt MESA} allows for the presence of an inert core at the centre, with a constant, non-evolving density. This assumption does not significantly affect the evolving radius as long as the core represents only a small fraction of the total mass. An inert core is likely a simplification of a more complex, multi-layer structure, including a diluted core and a metal-enriched envelope \citep{thorngren16}, favoured by the Jovian models fitting Juno gravity data (e.g. \citealt{wahl17,debras19}).
In these models, the Jovian core has a range between 7 and 25 $\Me$ ($\Me$ being the mass of the Earth), depending on the equation of state and on whether it is diluted or not. \cite{wahl17} compared evolutionary models with different internal structure and, for Jupiter, they obtained compositions close to the solar ones if the core is assumed as compact, or higher values of metallicity if the core is diluted in the envelope. \cite{yildiz24} recently performed a {\tt MESA} study of the interior of solar system giants, using similar tools to this study (same free parameters $M_c$ and $\rho_c$). They find best-fitting values for Jupiter and Saturn of $M_c=40$ and 25 $\Me$, with $\rho_c=10$~\gcc, and they explore ranges of up to $\rho_c=60$~\gcc.
See also e.g. \cite{burrows07} for a general discussion about the role of core mass in cooling models and inferred best-fitting values.

In this paper, we fix the core mass to $M_c = 10\sqrt{M/\Mj}~\Me$,  compatible with the trend inferred from the sample of cold Jupiters by \cite{thorngren16}.
In order to fix the core density $\rho_c$, we consider what is the central density obtained at 10 Gyr with a model having the same total mass $M$, but in absence of both core and irradiation (i.e. the largest value for central density, since it increases with age and decreases with irradiation and age). In our models, such a central density goes approximately as $\sim 2.7 (M/\Mj)$~\gcc \, between 0.3 and 5 $\Mj$ (becomes slightly steeper for higher masses). Therefore, we conservatively fix a core density $\rho_c = 10 (M/\Mj)$~\gcc, which recovers values often fixed in previous works using {\tt MESA} for a $1~\Mj$ planet, e.g. \cite{komacek17}. 
This practical prescription ensures that the central density, $\rho_c$, never falls below the density of the deepest envelope layers at any time ($t \leq 10$ Gyr) or for any model parameters ($M$ and $\Teq$), thereby avoiding unphysical values.
Moreover, these values are well below what can be extrapolated by the iron equation of state (the most dense case, e.g. \citealt{smith18}), meaning that the core here considered is compatible with representing material less dense than iron. In any case, for the core masses here considered, the role of $\rho_c$ is minor, since it is only used to fix the innermost values of enclosed mass, $m_i=M_c$, and radius, $r_i=[3M_c/(4\pi\rho_c)]^{1/3}$.

In App.~\ref{app:benchmark} we briefly validate standard cooling results and show the sensitivity of results (without internal heating) for planets of different core mass, core density, and composition. Since these are well-known results and are not the objective of the paper, hereafter we vary only the observables, $M$ and $\Teq$, fixing the composition to Solar values, and the values of core mass and density as described above.

\subsection{Irradiation}\label{sec:irradiation}

{\tt MESA} provides a few different ways to implement an irradiated atmosphere. As in \cite{komacek17}, we use a specific energy absorption rate, $\epsilon_{\rm irr} = F_{\rm irr} / (4 \Sigma_\star)$, applied uniformly through the outermost mass column $m(r) \leqslant \Sigma_\star$, where $F_{\rm irr}=4\sigma_{\rm sb}T_{\rm eq}^4$ is the irradiation flux, $\sigma_{\rm sb}$ is the Stefan-Boltzmann constant, and $\Sigma_\star$ is the heated column mass, a value which parametrises the atmospheric absorption. We consider a constant value of stellar flux $F_{\rm irr}$, thus neglecting its increase during the stellar evolution. This effect has been considered instead in e.g. \cite{komacek20} and \cite{lopez16}, who find that planetary reinflation is possible (assuming a constant Ohmic efficiency), especially for $\lesssim 1~\Mj$ planets around stars at the end of the main sequence phase. Observational evidence in this sense might have been found \citep{hartman16}.

A more detailed atmospheric modelling, accounting quantitatively for chemical composition and opacity, would give a different, non-uniform and non-grey heat deposition profile $\epsilon_{\rm irr}$. We also note that since the model is 1D, we cannot model the differences of irradiation across the surface. Hence, the local values of the temperature in the outer layer must be interpreted as averages. While these caveats would be important if one wants to fit, for example, the atmospheric properties of a given object, this paper is focused on the deep deposition of heat and aims at an exploration of the main parameters to explain the trends in the entire sample. Therefore, for our purposes, this simple implementation is sufficient.

We fixed the absorbing mass column to $\Sigma_\star = 250$~\Su, which approximately corresponds to a visible grey opacity $\kappa_{\rm vis}=4\times 10^{-3}$~cm$^2$~g$^{-1}$ (as in previous similar works, e.g. \citealt{fortney08,guillot10,owen16,komacek17}). This value, for an $M=1 \Mj$ planet at gigayear ages, corresponds to absorbing stellar energy homogeneously (per unit mass) down to a depth of $p \sim 1$ bar. We show in App.~\ref{app:benchmark} that the results depend only slightly on the values of $\Sigma_\star$ within a reasonable range, with planetary radius variations of a few percent at most, much less than the main parameters of the model ($M$, $\Teq$, and the Ohmic parameters introduced below). 

Practically, we considered the previously built planetary structure (fixing the mass, core mass and density, composition) and found a relaxed solution considering irradiation. Such a profile is the starting point for the evolution. This extra step helps the convergence at early times, especially for high values of irradiation and low masses. We started the simulations with an initial radius of $\sim$4 $\Rj$, a value that is close to what was chosen in other works \citep{paxton13} and much larger than the evolved values at the relevant ages so that the results at gigayears are insensitive to the specific choice.

\section{Ohmic heating model}\label{sec:ohmic}

We include the specific heating term due to Ohmic dissipation as a source in Eq.~(\ref{eq_luminosity}):
\begin{equation}
\epsilon_{\rm j} = \frac{Q_{\rm j}}{\rho} =  \frac{J^2}{\sigma~\rho}  \, ,
\label{eq:eps_joule}
\end{equation}
where $Q_{\rm j}=J^2/\sigma$ is the volumetric Ohmic heating rate, $J$ is the electrical current density, and $\sigma$ is the electrical conductivity. Obviously, the radial profiles of these quantities represent the averages over each sphere, $r$. The Ohmic model we describe hereafter is implemented as a modification of the {\tt {\tt MESA} make\_planet} suite \citep{paxton13}, in which we define the extra heat source, $\epsilon_{\rm j}$, with all the ingredients and calculations explained in the following subsections. The module is available\footnote{\url{https://github.com/danielevigano/mesa_ohmic_hj}} and has been developed for {\tt {\tt MESA} v24.08.1}, the version we use in this work.

\subsection{Electrical conductivity}\label{sec:sigma}

Within shallow layers of HJs ($\rho \lesssim 10^{-2}$\gcc, i.e. $p\lesssim$ kbar for Gyr ages), the electrical conductivity is dominated by the thermal ionisation of the alkali metals (e.g. \citealt{kumar21}). As the density increases ($\rho\gtrsim 10^{-2}$ \gcc), the electrons start becoming degenerate and ions becoming strongly coupled (e.g. \citealt{redmer97,ramakrishna24}), so that the pressure-ionisation from hydrogen and helium (the most abundant species), negligible at low densities, produces a sudden increase of conductivity with density, which is more drastic at lower temperatures. Pressure ionisation is a many-particle effect that arises due to the lowering of ionisation energy in a high-density plasma. When degeneracy and strong correlations plays a role, more detailed calculations, usually based on density functional theory and molecular dynamics (DFT-MD), are needed. The recent review by \cite{bonitz24} (see also the extensive list of references within) gives an overview on the transport properties calculations for such warm dense matter at high densities, $\rho\gtrsim 10^{-2}$ \gcc.

Above $\rho\gtrsim 1$ \gcc, the conductivity has a reduced dependence on the temperature and follows the trend shown for the Jovian adiabat by \cite{french12}, with precise theoretical values typically differing by a factor of a few from one model to another. On the other hand, at intermediate depths $\rho \sim 10^{-2}-1$ \gcc, the pressure ionisation contribution to the conductivity is highly sensitive on temperature, and the available points in the $\rho$-$T$ plane are few and especially sparse at $T\lesssim 10^4$ K, which is the regime of our interest. In this range, the conductivity changes abruptly with temperature (at a given $\rho$) and becomes very steep, see e.g.~\citealt{holst11}.

To incorporate all these results and ensure a correct transition from thermally dominated to pressure-dominated ionisation, we linearly interpolated\footnote{
We interpolate in the logarithm of the variables, as they span several orders of magnitude and are strictly positive. Due to the scarcity and uncertainty of available data across broad regions of the $\rho$–$T$ parameter space of interest higher-order interpolations are not warranted.}
the corresponding tables for $\sigma$ in the $\log(\rho)$-$\log(T)$ plane: the conductivity values from \cite{kumar21} (for $\rho \leq 10^{-2}$ \gcc) and Fig. 42 of \cite{bonitz24} (for $\rho \geq 10^{-1}$ \gcc). In the intermediate region ($\rho \sim 10^{-2}$-$10^{-1}$ \gcc), we interpolated between the closest extremes of the two tables, again linearly with logarithmic quantities. We note that the latest DFT-MD results differs at deep layers from the values of $\sigma$ used in the HJ inflation study by e.g. \cite{huang12} and \cite{wu13}, who considered the thermal and pressure ionisation via the electron fraction from the \cite{saumon95} equation of state but employed the electron–neutral collision cross-section from \cite{draine83} and an analytical expression for electron-proton scattering. The expression used for electron-proton scattering in their conductivity calculations may lack accuracy over the temperature and density ranges relevant to Ohmic dissipation studies. Furthermore, their approach neglects the influence of strong coupling effects on electrical conductivity. 

\begin{figure}
\centerline{\includegraphics[width=.5\textwidth]{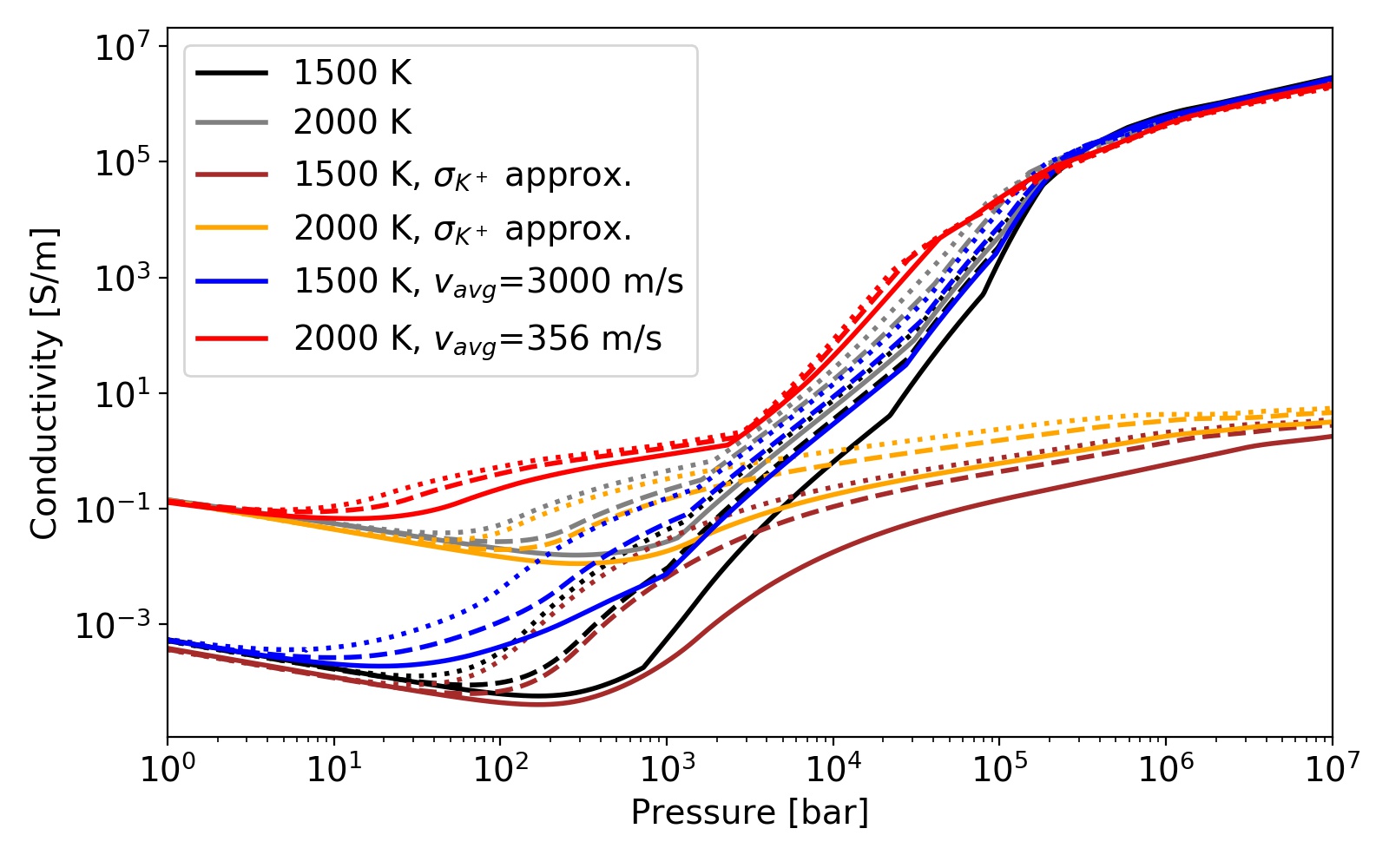}}
\caption{Electrical conductivity, $\sigma$, as a function of pressure for different planetary models. We show ages of $t\simeq 0.4$ (dots), 1.1 (dashes), and 5 Gyr (solid lines) for a planet with $M=1~\Mj$, showing two cases, $T_{\rm eq}=1500,2000$ K, without (black, grey) and with Ohmic heating, parameter $v_{\rm avg}=3000$ and 356 m/s, respectively (blue, red). We also show (brown and gold lines), for comparison, the simplified $K^+$-only thermal ionisation formula, which neglects pressure ionisation, for the cases without Ohmic heating.} 
\label{fig:cond}
\end{figure}

Figure~\ref{fig:cond} shows the conductivity profiles obtained at three ages ($t\sim 0.4,1.1,5$ Gyr) for four different models: $\Teq=1500,2000$ K, and with (blue and red lines) or without (black and grey lines) the inclusion of Ohmic heating (see below). While in the external part the values are highly sensitive to the local temperature, which is higher for highly irradiated and/or internally heated cases, at $p\gtrsim 10^5$ bar the curves tend to similar, very high values of $\sigma$, as in \cite{french12}. In the same plot, we also show for comparison (brown, gold lines), the widely used \citep{perna10a,perna10b,soriano23} analytical approximation for the thermal ionisation of potassium (the dominant one in the outermost regions of the envelope) neglecting degeneracy and strong coupling effects, and valid for not too high temperatures, $T\lesssim 2000$ K (see discussion in \citealt{soriano23}). It can be seen how substantial differences are encountered at $p\gtrsim 1000$ bar. 

Finally, we note that the dissociation of molecular into atomic hydrogen is taken into account in the conductivity calculations \citep{kumar21} here used, but not as a source of heat, which might become relevant in the uppermost layers for $\Teq \gtrsim 2000$ K \citep{bell18, gandhi20}. At those shallow layers, in any case, there a number of non-ideal MHD effects can take place \citep{koskinen14}, but the heat dissipation does not affect the radius inflation, which is much more sensitive on deep deposition.

\subsection{Atmospherically induced current}

Our aim is to estimate the spherically averaged radial profile of the current density, $J(r,t):=<|\vec{J}|>_{\theta,\phi}(r,t)$, where $\vec{J}=(\vec{\nabla}\times\vec{B})/\mu_0$ and $\mu_0$ is the magnetic permeability, in order to evaluate the radial profile of the local Joule dissipation rate, $Q_{\rm j}(r,t)=J(r,t)^2/\sigma(r,t)$, in the 1D code. As in the first works on Ohmic dissipation in HJs \citep{batygin10,batygin11}, we quantified the electrical currents by imposing a stationary solution to the induction equation given by the balance between the advection (itself given in terms of the thermal wind $\vec{v}$) and the resistive term (neglecting the Hall and ambipolar non-ideal terms, see \citealt{perna10a,koskinen14,soriano25}):
\begin{equation}
\frac{d\vec{B}_{\rm ind}}{dt} = \vec{\nabla}\times\left(\vec{v}\times \vec{B} - \frac{\vec{J}}{\sigma} \right) = 0~.\label{eq:steady_induction}
\end{equation}
Here, the magnetic field, $\vec{B}$, includes the dynamo-generated background field, $\vec{B}_{\rm dyn}$, and the atmospherically induced field itself, $\vec{B}_{\rm ind}$. Following e.g. \cite{batygin10}, at equilibrium, the solution of the induced current density could then be defined by considering the electrical potential, $\Phi$,
\begin{equation}\label{eq:current}
    \vec{J} = \sigma(\vec{v}\times\vec{B} - \vec{\nabla}\Phi)~,
\end{equation}
and imposing the continuity equation, $\vec{\nabla}\cdot\vec{J}=0$, so that
\begin{equation}\label{eq:continuity_j}
    \sigma\nabla^2\Phi + \vec{\nabla}\sigma\cdot\vec{\nabla}\Phi = \vec{\nabla}\cdot[\sigma(\vec{v}\times\vec{B})]~.
\end{equation}
We then expanded $\Phi$ in spherical harmonics, $\Phi=\Sigma_{lm} \Phi_{lm}(r) Y_{l}^m(\theta,\phi)$, and solved for the radial dependence of each multipole, $\Phi_{lm}(r)$, for a given prescription of $\vec{v}$, $\vec{B}$, and $\sigma$ --- which in turn depends on $T$ and density, i.e. on $p(T)$. \cite{batygin10}, \cite{batygin11}, and related works solve such an elliptic equation, considering first the uppermost regions where $\vec{v}\neq 0$ and assuming axial symmetry (i.e. $\Phi_{lm}=0$ for any $m \neq 0$) and simplified geometries for $\vec{v}(r,\theta)$, $\vec{B}(r,\theta)$ for a given profile $\sigma(r)$. The simplest configuration is a purely zonal wind $v_\phi\sim \sin\theta$ ($\theta\in[0,\pi]$ being the co-latitude) and a dipolar, aligned field, which means that the only possible induced current multipole is the quadrupolar one, $\Phi_{20}\neq 0$.
Here we propose new families of semi-analytical solutions for the deep layers where $\vec{v}\sim 0$, treating the complex atmospheric layers in an effective, parametrised way and allowing for any generic profile $\sigma(r)$.

\begin{figure}
\centerline{\includegraphics[width=.5\textwidth]{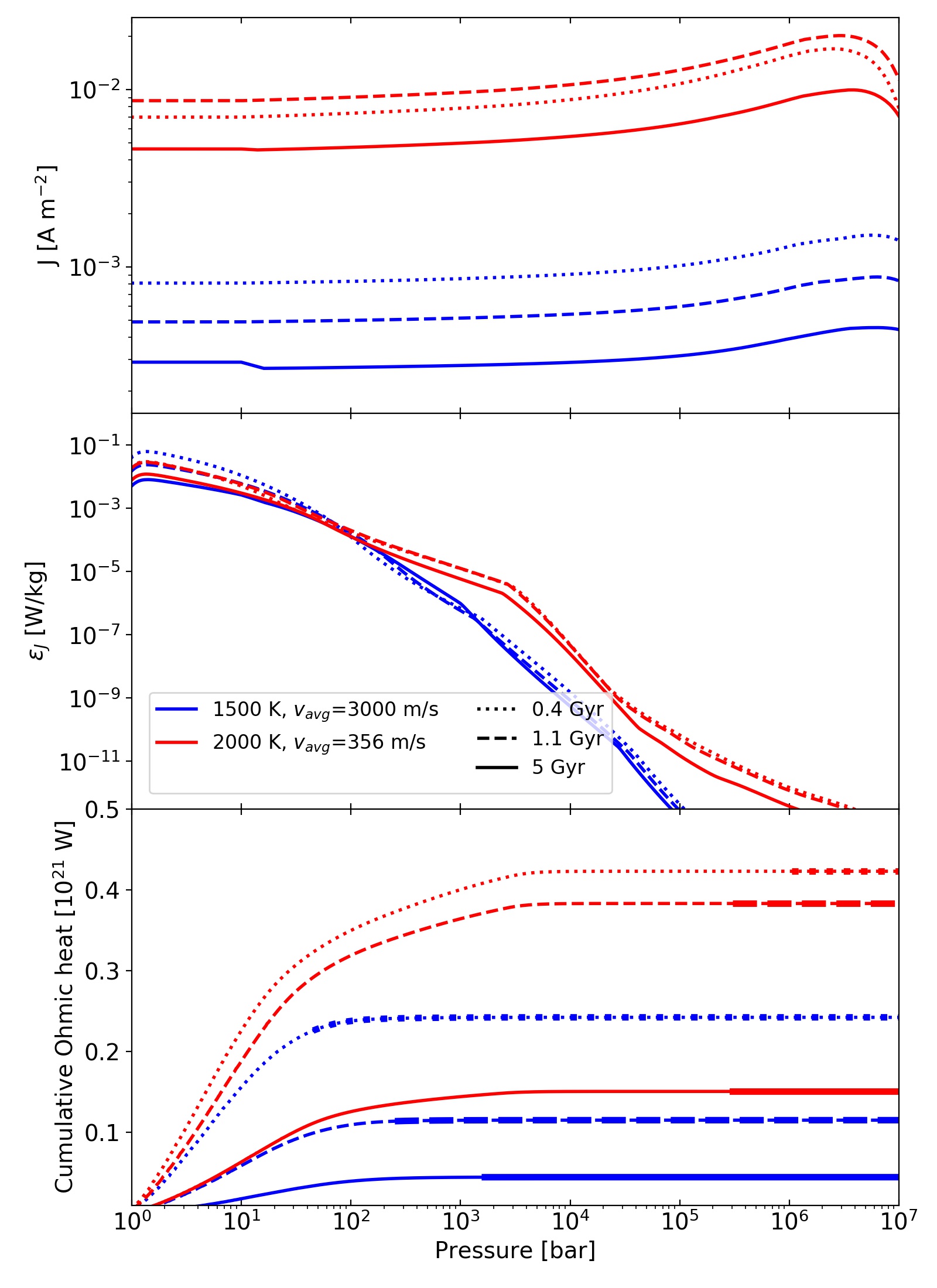}}
\caption{Profiles as a function of pressure of the induced currents, $J$ (top), and Ohmic specific (middle) and cumulative (bottom) heat rate. We show the same heated cases and ages as in Fig.~\ref{fig:cond}. The thick lines in the bottom panel identify the convective regions at each age of each model.} 
\label{fig:Qj_sigma_P}
\end{figure}

\subsubsection{Treatment of the induction in the atmospheric region}\label{sec:induction_atm}

Given the latitudinal differences in the thermodynamical radial profiles, and their deviations from axial symmetry, a fully consistent study of the atmospheric induction in the regions where jets flow should ideally include 3D GCM simulations coupled with information about structure at a given age, from long-term evolution runs. These simulations would allow for the calculation of $\vec{J}$ for given $p(T)$ profiles and boundary conditions, and would take into account the non-linearity of the atmospheric induction equation (see also e.g. the discussion in \citealt{batygin11,batygin13}).

There are several factors that introduce uncertainties and, likely, planet-to-planet variations in the profiles of $J(r)$, in the uppermost layers.
First, the internally generated magnetic field, $\vec{B}_{\rm dyn}$, is usually assumed to be an aligned dipole and orthogonal to $\vec{v}$, but its tilt, and the contributions from higher multipoles, are often non-negligible, as seen both in the Solar system planets \citep{schubert11} and in dynamo simulations (e.g. \citealt{gastine14,wicht19a,wicht19b,elias25}).
Secondly, and related to that, an idealised configuration for which $\vec{v}\times\vec{B}\simeq \vec{v}\times\vec{B}_{\rm dyn}$ in eq. (\ref{eq:current}), implies that either $\vec{B}_{\rm dyn}$ and $\vec{v}$ are perfectly orthogonal, or that the induced field is only a perturbation to the background, $\vec{B}_{\rm ind}\ll \vec{B}_{\rm dyn}$. The latter approximation breaks down at sufficiently high temperatures ($T\gtrsim 1500$ K for reasonable values of $B_{\rm dyn}\sim 10$ G), due to the increasing magnetic Reynolds number $Rm>$ 1 \citep{dietrich22}. Therefore, on one side $\vec{B}$ becomes more complex, and, at the same time, the flow circulation patterns, $\vec{v}$, are distorted by the magnetic field configuration \citep{batygin13,rogers14a,rogers14b}. Third, the value of $\sigma$, which is highly sensitive on the local temperature, shows important latitudinal and longitudinal variations \citep{rogers17} due to the thermal gradients caused by the tidal locking and the impossibility of having a perfectly thermal redistribution of the global circulation. Moreover, the radial profiles of $\sigma$ are greatly affected by the local composition and opacity, as shown for instance by the important differences in the $\sigma(r)$ profiles derived from using different $p(T)$ in modelling a given planet \citep{kumar21}. Finally, the 1D evolutionary calculations typically do not include the uppermost layers ($\lesssim$ mbar), where other effects like dayside photoionisation \citep{koskinen14}, ambipolar diffusion \citep{savel24,soriano25}, current leakage to the magnetosphere can take place and affect the atmospheric electrodynamics, and thermo-resistive instability, which potentially arises from magnetic drag in the dayside and can lead to a thermal runaway as the conductivity increases \citep{menou12b,hardy22,hardy23,hardy25}.

Therefore, due to the expected non-trivial configurations of the magnetic and velocity fields, the tangential variations of $\sigma$, and uncertainties in the detailed profile of $p(T)$, the term $\sigma(\vec{v}\times\vec{B})$ can in general allow for non-trivial combinations of non-zero multipoles $\Phi_{lm}$ of the $\Phi$ expansion, in the solution to eq. (\ref{eq:continuity_j}).

Also for these reasons, and in contrast with previous pioneering efforts \citep{batygin10,batygin11,wu13}, here we do not aim at including a mathematically detailed solution for the atmospheric region in the 1D evolutionary code. Instead, below a certain pressure $p_{\rm atm}$ where the term $\vec{v}\times\vec{B}$ is non-negligible, we estimate the order of magnitude of the sphere-averaged current density profile by the following effective parametrisation:
\begin{equation}\label{eq:j_atm}
    J(p<p_{\rm atm})(t) = \sigma_{\rm atm}(t) v_{\rm avg} B_{\rm atm}(t) \,,
\end{equation}
where $B_{\rm atm}$ is connected to the dynamo field $\vec{B}_{\rm dyn}$ (see Sect.~\ref{sec:mf_estimate} for the prescription), $\sigma_{\rm atm}(t)$ is the value of $\sigma(r)$ averaged in the region $p<p_{\rm atm}$, and we highlighted the dependences on time, $t$. The free parameter $v_{\rm avg}$ is a proxy to the average wind velocities and, comparing the definition of $J(p<p_{\rm atm})$ with eq.~(\ref{eq:current}), is formally defined by the ratio between the real (unknown) value of $|\vec{J}|$ averaged in the outermost region (the volume $V_{\rm atm}$ defined by the outermost layers, with pressure $p<p_{\rm atm}$) and the product $\sigma_{\rm atm}B_{\rm atm}$:
\begin{equation}
    v_{\rm avg} = \frac{\int_{V_{\rm atm}}  |\sigma(\vec{v}\times\vec{B} - \vec{\nabla}\Phi)| dV_{\rm atm}}{V_{\rm atm}\sigma_{\rm atm}B_{\rm atm}}~,
\end{equation}
which can overestimate the real average velocity, $v = (\int_{V_{\rm atm}} |\vec{v}| dV_{\rm atm})/V_{\rm atm}$, if the induced field perpendicular to the velocity is comparable or larger than $B_{\rm atm}$ (non-linear regime). This approach implicitly assumes that, in the wind region $p<p_{\rm atm}$, the $|\vec{v}\times\vec{B}|$ is non-negligible compared to $|\vec{\nabla}\Phi|$. We note that $\sigma_{\rm atm}$ and $B_{\rm atm}$ vary in time, while we assume that $v_{\rm avg}$ is a constant parameter through the evolution, since it is mainly set by the stellar irradiation flux. Therefore, we are implicitly neglecting two minor, opposite effects affecting the long-term evolution: (i) the slow increase of luminosity of the host star, i.e. $F{\rm irr}$, which should slightly increase in time, slowing down more the cooling and favouring more magnetic induction (see also \citealt{lopez16}); (ii) the feedback on velocity (i.e. $v_{\rm avg}$) due to a changing $B_{\rm atm}$, via magnetic drag \citep{menou12}. In Sect. \ref{sec:evolving_luminosity} we relax assumption (i) by studying how the long-term evolution of the irradiation from a Sun-like star affects our results, for a few representative models.

Keeping this in mind, in this work, as a first step, we take $v_{\rm avg}$ as constant in time. Hereafter, we fix $p_{\rm atm}=10$ bar, implicitly working in the regime of shallow circulation, supported by recent studies \citep{knierim22}. Higher values of $p_{\rm atm}$, i.e. including deeper regions where winds are slower, would simply correspond to lower values of $v_{\rm avg}$. 

\subsubsection{Induced currents in deeper regions}\label{sec:J_deep}

In deeper regions, $p\geq p_{\rm atm}$, wind intensity decays quickly, so that we neglect the $\vec{v}\times\vec{B}$ term in eq.~(\ref{eq:continuity_j}), and find a solution for $\Phi$. Moreover, no strong temperature differences are expected, since the $p(T)$ profile converges to the internal adiabat: 1D, spherical symmetry calculations are then appropriate and the conductivity gradient is purely radial. Therefore, eq.~(\ref{eq:continuity_j}) for each multipole, $\Phi_{lm}$, reduces to
\begin{equation}
    \sigma \Phi_{lm}'' + \left(\frac{2\sigma}{r} + \sigma'\right)\Phi_{lm}' - \sigma l(l+1) \frac{\Phi_{lm}}{r^2} = 0~,
    \label{eq:ode_potential}
\end{equation}
where the primes denote radial derivatives, and we made use of the spherical harmonic identities for the angular part of the Laplacian operator to get the $l(l+1)/r^2$ term. In general, eq.~(\ref{eq:ode_potential}) is solved numerically for a given profile of $\sigma$ (see e.g. \cite{batygin10,batygin11,wu13} for specific modes).

Analytical solutions exist if we parametrise the conductivity with a power law, $\sigma \sim r^{-\beta}$, so that
\begin{equation}
    \Phi_{lm}'' + \left(\frac{2 - \beta}{r}\right)\Phi_{lm}' - \frac{l(l+1)}{r^2} \Phi_{lm} = 0~,
    \label{eq:ode_beta}
\end{equation}
which has solutions of the form $\Phi_{lm} \sim r^\alpha$, where 
\begin{equation}
    \alpha(\beta,l)=\frac{1}{2}\left[\beta - 1 \pm \sqrt{(\beta - 1)^2 + 4l(l+1)}\right]~.
    \label{eq:alpha}
\end{equation}
We stick only to the positive branch, since we performed our calculation down to the envelope-core interface, and the negative branch would imply diverging solutions towards the centre of the planet as $r\rightarrow 0$.

Therefore, we obtained the sphere-average radial profile $|\vec{\nabla}\Phi|_s(r) \sim \sum_{l} j_{l} r^{\alpha(\beta,l) - 1}$, where $j_{l}$ represent the relative weights of the contributions from each multipole degree, $l$ (summing all the orders $m\in [-l,l]$ of a given $l$ and averaging over the sphere), and are taken as a free set of effective parameters of the model rather than computed for a precise configuration of $\vec{v}(r,\theta,\phi)$ and $\vec{B}(r,\theta,\phi)$,  as in \citealt{batygin10,batygin11,knierim22}, for example. The current density profile is then given by
\begin{equation}
J(r) = \sigma(r) |\vec{\nabla}\Phi|_s(r) \sim \sum_{l} j_{l} r^{-\beta + \alpha(\beta,l) -1}~.
\label{eq:J}
\end{equation}
In case of having only one multipole $j_l\neq 0$, the heating rate per unit volume is then
\begin{equation}
  Q_{\rm j} \sim j_l^2 r^{2\alpha(\beta,l) - \beta-2} \sim j_l^2 r^{\sqrt{(\beta-1)^2+4l(l+1)}-3}\,.
\end{equation}
For the simplest configuration employed by \cite{batygin10}, $l=2$ is the only mode, $j_2\neq 0$. In the simplified case $\sigma'=\beta=0$, eq.~(\ref{eq:ode_potential}) reduces to the Laplace equation, with solutions $\Phi_{lm} \sim r^l$. Moreover, in the $\beta=0$, $l=2$ case, one recovers the simple case $\alpha=2$, $J\sim r$, and $Q_{\rm j}\sim r^{2}$, assumed by \cite{ginzburg16}. Any other value in the range $\beta > 0$ (which is expected everywhere except close to the irradiation region, see Fig.~\ref{fig:cond}), $j_{l>2}\neq 0$ (complexity in the wind and magnetic field geometries) will give higher values of $\alpha$ in the positive branch, since it is a monotonic function of both $l$ and $\beta$. This means that the profile of $J$ will increase more steeply with $r$ for more complex configurations of the wind and/or the dynamo magnetic field (high $l$), and for steeper conductivity values (high $\beta$). In the limit of large values of $\beta\gg l$ (steep radial gradient of $\sigma$), $\alpha\rightarrow\left[\beta - 1 + \frac{l(l+1)}{\beta-1}\right]$, so that $J\sim r^{-2+l(l+1)/(\beta-1)}$ and $Q_{\rm j}\sim r^{\beta-4 + 2l(l+1)/(\beta-1)}$, i.e. a steeply decreasing Ohmic dissipation profile with decreasing radial distance.

Obviously, a power law is not an accurate description of any realistic $\sigma(r)$ profile (see the complex $\sigma(p)$ profiles of Fig. \ref{fig:cond}). However, we can divide the domain in many thin shells and find the power-law best fits to $\sigma$ in each of them. Considering enough shells (typically, a few tens), such piecewise fits reproduce the original profile $\sigma(r)$ accurately, with local relative errors of a few percent in the worst cases. Therefore, for each radial shell, we consider the best-fitting power law $\beta$, and obtain the corresponding values of $\alpha$, for a given $l$, with eq. (\ref{eq:alpha}). Having the piecewise series of radial slopes $\alpha$, we reconstruct the profile of $J$ connecting each shell solution by continuity, starting from the atmospheric value $J(p<p_{\rm atm})$, eq.~(\ref{eq:j_atm}) at $p=p_{\rm atm}$ and going inward, shell by shell.

This semi-analytical approach allows us to easily take into account all $\Phi_{lm}$, and to test any combination of $j_{l}$, i.e. of the relative contribution of each multipole degree $l$ to $|\vec\nabla\Phi|_s$, given by wind and magnetic field configurations in the atmospheric layers. In order to consider a realistic set of relative weights $j_l$, we need to consider the geometries of the deep-seated magnetic field and the velocity field.

For the former, we need to consider that, outside the dynamo region, $r>R_{\rm dyn}$ (see next subsection), currents that sustain the dynamo field are negligible by definition, and the dynamo magnetic field is potential. Therefore, the contribution of each degree $l$ to the background magnetic field intensity in the atmospheric region ($B_{\rm atm}$, $r\simeq R$) scales as $b_l(R) \sim (R_{\rm dyn}/R)^{(l+2)}$, i.e. higher multipoles matter less and less as one gets far from the dynamo region, which is relevant for inflated planets, where the envelope thickness is substantial. We can take into account different options for the set of weights at the dynamo surface, $b_l(R_{\rm dyn})$, ranging from a simple dipole ($b_l=0$ for every $l\neq 1$), to more realistic combinations of multipoles. For the latter, we can consider that at $r\simeq R_{\rm dyn}$ the magnetic energy spectra has a peak for the dipole, $l=1$, and, for $l\gtrsim 5$, it is approximately flat (see e.g. Lowes spectra measured for Earth, \citealt{lowes74}, and Jupiter, \citealt{connerney22}, and reproduced in dynamo simulations, e.g. \citealt{tsang20,elias25}). This means that, at $r=R_{\rm dyn}$, the contributions to the magnetic field intensity are expected to be fairly constant for $l\gtrsim 5$.

Secondly, the wind geometry arguably shows stronger jets in the equatorial regions, with possible counter-jets at higher latitudes. Usually, an approximate equatorial symmetry is expected. If one takes the simplest possible case like in \cite{batygin10}, $v_\phi\sim \sin\theta$ and an aligned, dipolar field, one has $j_l \ne 0$ for $l=2$, which can be simplistically seen as a combination of the $l_w = 1$ and $l_b=1$ contributions from the wind and magnetic fields, respectively. Generalizing it to a given wind contribution, $l_w$, and a set of magnetic contributions, $b_l(R_{\rm dyn})$, the weights of the currents of eq.~(\ref{eq:J}) can be parametrised by
\begin{equation}
    j_{l+l_w} \sim b_l(R_{\rm dyn})\left(\frac{R_{\rm dyn}}{R}
    \right)^{(l+2)}~.
    \label{eq:jl}
\end{equation}
This expression is an effective, practical proxy to the more rigorous and case-to-case variable decomposition of the atmospheric $\sigma(\vec{v}\times\vec{B})$ term, which depends on the specific, detailed profiles of $\sigma$, velocity and magnetic fields. Below, we consider cases ranging from a pure dipole (only $b_1\neq 0$), to a flat, continuum distribution of magnetic weights ($b_l(R_{\rm dyn})$ constant) up to a maximum multipole $l_{max}$.

\begin{figure}
\centerline{\includegraphics[width=.45\textwidth]{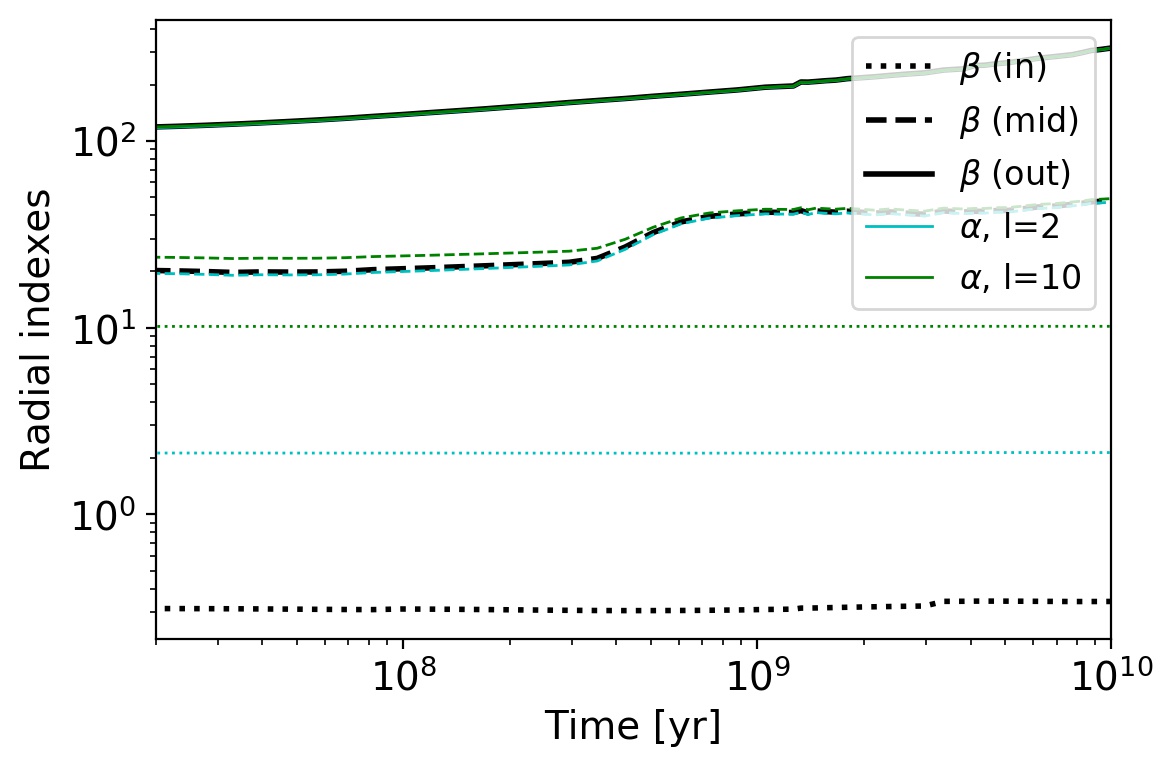}}
\centerline{\includegraphics[width=.45\textwidth]{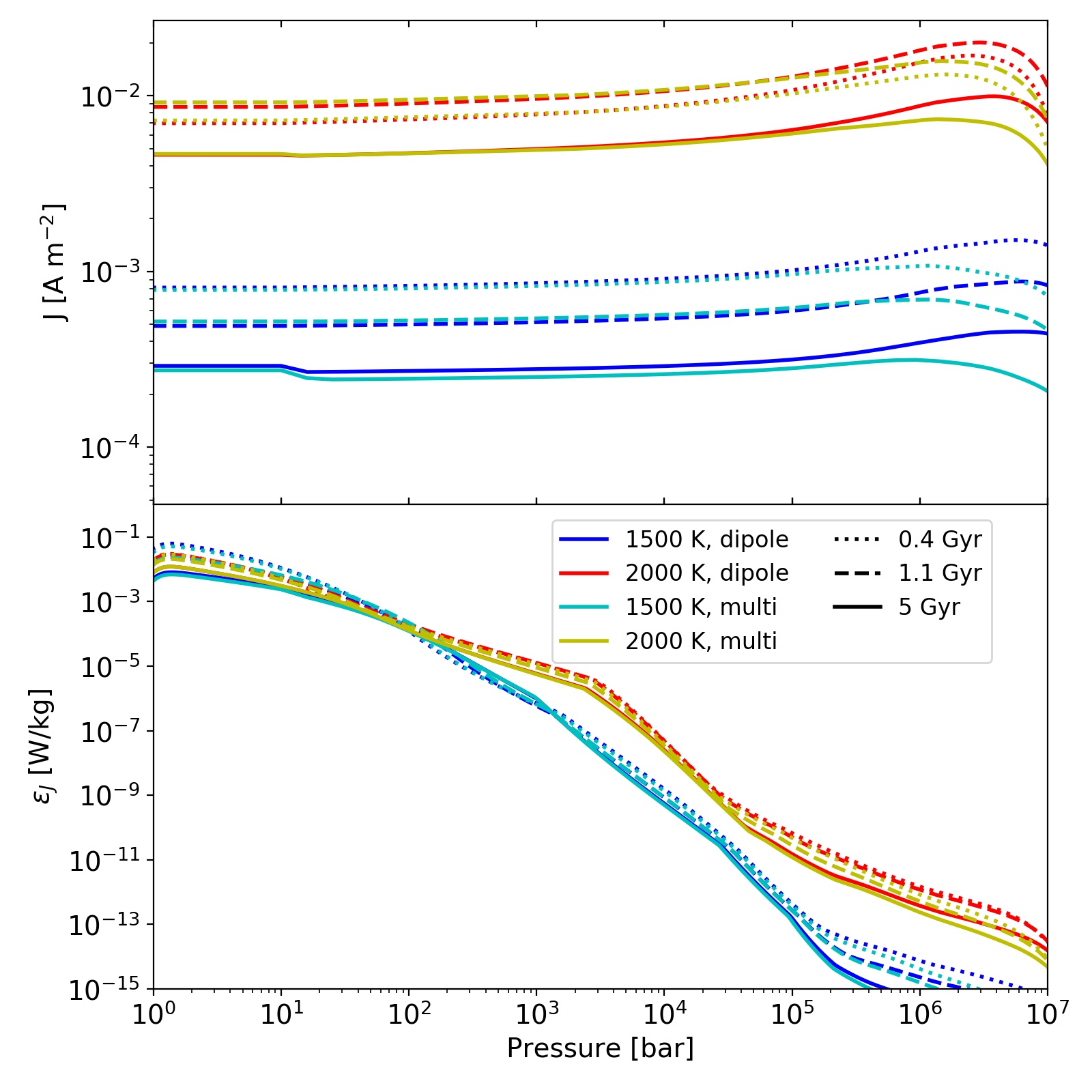}}
\caption{Current reconstruction for the two models with heating considered in Fig.~\ref{fig:cond}. {\em Top:} Evolution of the best-fitting power-law indexes, $\beta$ (black, indicating the steepness of the conductivity profile $\sigma(r)\sim r^{-\beta}$), at shallow ($p\sim 10^2$ bar, solid), intermediate ($p\sim 10^4$ bar, dashed), and deep ($p\sim 10^6-10^7$ bar, dots) layers and the corresponding local indexes, $\alpha$, of the induced current profile, eq.~(\ref{eq:alpha}), for $l=2$ (cyan) and $l=10$ (green). {\em Middle:} Radial profiles of the induced currents for the same model at ages of $t\simeq 0.4$ (dots), 1.1 (dashes), and 5 Gyr (solid lines) for a purely dipolar case, (only $j_2\neq 0$, blue and red) or multipolar, with a flat magnetic spectrum at dynamo surface from $l=2$ to $l=10$ (cyan and golden). {\em Bottom:} Corresponding Ohmic heating rate profiles.} 
\label{fig:j_comparison_l}
\end{figure}

In summary, at each time step, by performing a series of piecewise power-law fits of the conductivity profiles $\sigma(r)$, we can reconstruct the current density profile, testing a broad range of geometries of the magnetic and wind fields. In the top panel of Fig.~\ref{fig:Qj_sigma_P} we show the reconstructed induced currents for the conductivity profiles of Fig.~\ref{fig:cond}, discussed above, for the purely aligned dipolar case (only $j_2\neq0$). In most of the domain, the conductivity increases inward with $\beta\gg l$, so that the average induced current density has a very slight increase inward, following the $J\sim r^{-2}$ limit derived above. As we go inward into the degeneracy-dominated conductivity, $\beta$ is only slightly positive, so that the current density gradually reaches a maximum and then approaches the limit $J\sim r$. Overall, we noticed that $J$ varies only by a factor of a few between the exterior and the beginning of the dynamo region.

In the second panel of Fig.~\ref{fig:Qj_sigma_P}, we show the resulting specific Ohmic heat rate $\epsilon_{\rm j}$, which quickly drops inward due to both the increasing density and the increasing conductivity. The range of values is indeed comparable with e.g. Fig. 4 of \cite{batygin11}. In the bottom panel, we show the corresponding cumulative Ohmic heat, $4\pi\int_{r(p)}^R Q_{\rm j}(p) r^2 dr$. The relevant contributions are confined to the outermost regions, with pressures deeper than $\sim 10^4$ bar providing almost irrelevant contributions.

In Fig.~\ref{fig:j_comparison_l} we consider the same two models with heating considered in Fig.~\ref{fig:cond}. In the top panel, we show the evolution of the radial indexes $\beta$ at three different layers for the $\Teq=1500$ K model, and the corresponding values of $\alpha$ (eq.~\ref{eq:alpha}) for the cases $l=2$ (the simplest, usually considered), and $l=10$ (representative of contributions from high multipoles). In the middle and bottom panels, we show the corresponding solutions for $J$ and $\epsilon_{\rm j}$ for different sets of $j_l$ at three ages, for the $\Teq=1500$ and 2000 K heated models, and the simplest topology $j_2\neq 0$, or a flat distribution at the dynamo, with non-zero, equal contributions $b_l$ for $j\in[1,9]$, and $l_w=1$ (hence, $j_l\neq 0$ for $j\in[2,10]$, eq.~\ref{eq:jl}). The only differences are seen at very deep layers, where, in any case, the dynamo is operating, and therefore, the interaction between the induced and dynamo currents should be formally considered (see discussion below).

The reason for the lack of relevant differences is that, in most of the domain, the solution is determined by the steepness of the conductivity profile, $\beta \gg l$, and only when $\beta \sim l$ the magnetic (and wind) geometry matters. This reinforces and complements what \cite{batygin11} found for exact solutions to different specific wind and magnetic geometries: they found only a minor variability in the induced current profile, mostly driven by the wind region precise solution for a given geometry (which here we instead parametrise with $v_{\rm avg}$). Keeping the same normalisation, the differences in the integrated Ohmic heat (hence, in the Ohmic efficiency and the overall evolution) are negligible. We do not explore this further, and hereafter, all models include only a contribution from $j_2$.

In summary, the profiles of the induced currents are mostly driven by the steepness of the (time-dependent) conductivity profiles, and present much less variation in magnitude than $\sigma$. Therefore, the Ohmic specific rate profile steeply decays inwards, as stressed in previous Ohmic studies \citep{batygin10,batygin11,wu13,knierim22}. Importantly, although the deposited heat in very deep regions is a negligible fraction of the total, with no effects on the radius inflation, using realistic values of $\sigma$ in the entire domain allows us to model more accurately the magnitude of the current density penetrating into the dynamo region.

We note that we have opted for such a semi-analytical approach over a fully numerical solution because it offers a computationally efficient and sufficiently accurate  solution. More importantly, it provides clearer insight 
into the individual contributions of the conductivity gradient and magnetic multipoles to the final solution.

\subsection{Estimates of dynamo magnetic fields}\label{sec:mf_estimate}

Direct measurements of planetary magnetic fields is limited to the Solar system, with a limited model-dependent inference of a few HJs from activity indicators \citep{cauley19}. Therefore, the best guess is based on scaling laws, and there are some studies which indeed relate them to infer HJ magnetic fields \citep{yadav17,kilmetis24,thorngren24}. If one assumes slow rotation (more precisely, Rossby numbers $\gtrsim 0.1$), the dynamo depends on the planetary rotation rate and on diffusivities. Scaling laws have been proposed in this sense \citep{stevens05,sanchezlavega04}, with simple dependences between the magnetic moment, mass, radius and rotation, calibrated on the Solar planets measured values.

For HJs, the rotation rate is usually assumed to be the orbital one (typically a few days) in HJs due to tidal locking (but see e.g. \citealt{wazny25} for a discussion). \cite{elias25b} shows that HJs are most likely fast rotators for a vast range of parameters (Rossby numbers $\lesssim 0.1$). In this regime, both dynamo simulations and observations in stars support a saturation of the dependence between magnetic field and rotation, meaning that scaling laws for fast rotators should not depend on the rotational period.
With this assumption, the most used scaling law is based on \cite{christensen06}, who analysed a broad set of geodynamo models, and proposed correlations between different dynamo numbers. Based on that work, \cite{christensen09} proposed the widely used scaling law, which we also employ here to define the root-mean-square value of the dynamo field at $r=R_{\rm dyn}$:

\begin{equation}
    \frac{B_{\rm dyn}^2}{2\mu_0} \simeq 0.63 f_{\rm ohm}  \frac{1}{V}\int_{r_i}^{r_o} \left(\frac{q_c(r) L(r)}{H_T(r)}\right)^{2/3} \rho(r)^{1/3} 4\pi r^2 dr~,
    \label{eq:bdyn_C09}
\end{equation}
where $V$, $r_i$ and $r_o$ refer to the volume, inner and outer boundary of the convective region, $\mu_0$ is the magnetic permeability, $L={\rm min}(r_o-r_i,H_\rho)$, where $H_\rho$ is the density scale height, $H_T$ is the temperature scale height, $q_c(r)$ is the convective heat flux, and $f_{\rm ohm}$ is the fraction of Ohmic to total dissipation in the dynamo region (which \cite{christensen06} find numerically to be in the range $\sim 0.3$-$0.7$, depending on the simulation).
We evaluate the integral taking $r_i$ as the bottom of the envelope (outside the inert core), and $r_o=R_{\rm dyn}$, i.e. the radius corresponding to a pressure $p_{\rm dyn}:=10^6$ bar, taken as a proxy for the surface of the deep high-conductivity region hosting the dynamo. We note that, strictly speaking, the transition to the metallic hydrogen region corresponds to a $\rho$-$T$ line in the phase diagram, so that $p_{\rm dyn}$ depends slightly on $T$ (or $\rho$). However, since the evaluation of the field is performed via an integral, Eq.~(\ref{eq:bdyn_C09}), a slight change of 
the upper bound $r_o$ implies minor modifications, certainly smaller than the overall intrinsic uncertainties of the scaling law. We checked that, with this recipe, we recover reasonable values of $B_{\rm dyn}\sim 30-50$ G for a Jupiter-like model (negligible irradiation, $M=1~\Mj$), at $4.5$ Gyr-old. In our code, we use $f_{\rm ohm}\sim 0.5$, in the middle of the typical range $f_{\rm ohm}\sim 0.3-0.8$ of e.g. the geodynamo simulations of \cite{christensen06} and the anelastic ones by \cite{gastine21}, and on the upper range of cold Jupiter simulations by \cite{elias25}. In any case, we note that the systematic uncertainties in the absolute values of $B_{\rm dyn}(t)$ (which we can overall quantify of a factor $\sim 2$) affect the inferred values of $v_{\rm avg}$, but do not affect the general trends of the magnetic field evolution. We refer the reader to App.~\ref{app:scaling_laws} for a comparison of different versions of the scaling law \citep{reiners09,reiners10}, and some words of caution about their applicability to the HJ scenario.

Moreover, we estimated the average value of the magnetic field intensity at the surface (close to where winds circulate), which we use in eq.~(\ref{eq:j_atm}) by considering that the weights $b_l(R_{\rm dyn})$ (see Sect.~\ref{sec:J_deep}):
\begin{eqnarray}
    B_{\rm atm} = B_{\rm dyn}\sum_l b_l(R_{\rm dyn})\left(\frac{R_{\rm dyn}}{R}\right)^{l+2} \simeq f_{\rm dip}B_{\rm dyn}\left(\frac{R_{\rm dyn}}{R}\right)^{3}~,\label{eq:bdip}
\end{eqnarray}
where the last equation considers as a only free parameter the dipolarity at the dynamo surface, $f_{\rm dip}$, for simplicity (and given the much larger source of uncertainties in the scaling law). The set of anelastic dynamo simulations of \cite{elias25} found that the dipolarity (calculated over the entire dynamo region, or at its outer surface) is seen to suddenly increase at a few gigayears, switching from a multipolar- to a dipolar-dominated dynamo (see also \citealt{yadav13} for a discussion of these two branches). Here, for simplicity, we employed $f_{\rm dip}=0.1$, constant in time, in the range of found by \cite{elias25}. According to that study, neglecting its time variation might overestimate the surface field at early ages. Using these values of $f_{\rm ohm}$ and $f_{\rm dip}$ allows us to roughly recover Jupiter values of magnetic field at the surface.

\subsection{Numerical details about the heating term}\label{sec:numerical_details}

We gradually activate the Ohmic source term $\epsilon_{\rm j}$, eq.~(\ref{eq:eps_joule}), with a linear increase in time starting from zero at 5~Myr, to have it fully operative at 20~Myr. The reasons to neglect the Ohmic term at earlier ages are: (i) we only consider HJs with $t>100$ Myr, discarding the few younger ones (see Sect. \ref{sec:obs}); (ii) planetary formation, migration, and interaction with the disc are still happening, and the Ohmic model does not account for them; (iii) the internal structure and the radius is still sensitive to the (poorly constrained) initial entropy (i.e. initial radius), something irrelevant after $\sim 100$ Myr; (iv) at $t\lesssim$ Myr, the different versions of the scaling law differ substantially between each other (see App.~\ref{app:scaling_laws}), and eq.~(\ref{eq:bdyn_C09}) easily gives unphysical Ohmic efficiencies due to the large heat flux (which in turn is affected by the choice of the initial entropy).

\begin{figure*}
\includegraphics[width=.33\textwidth]{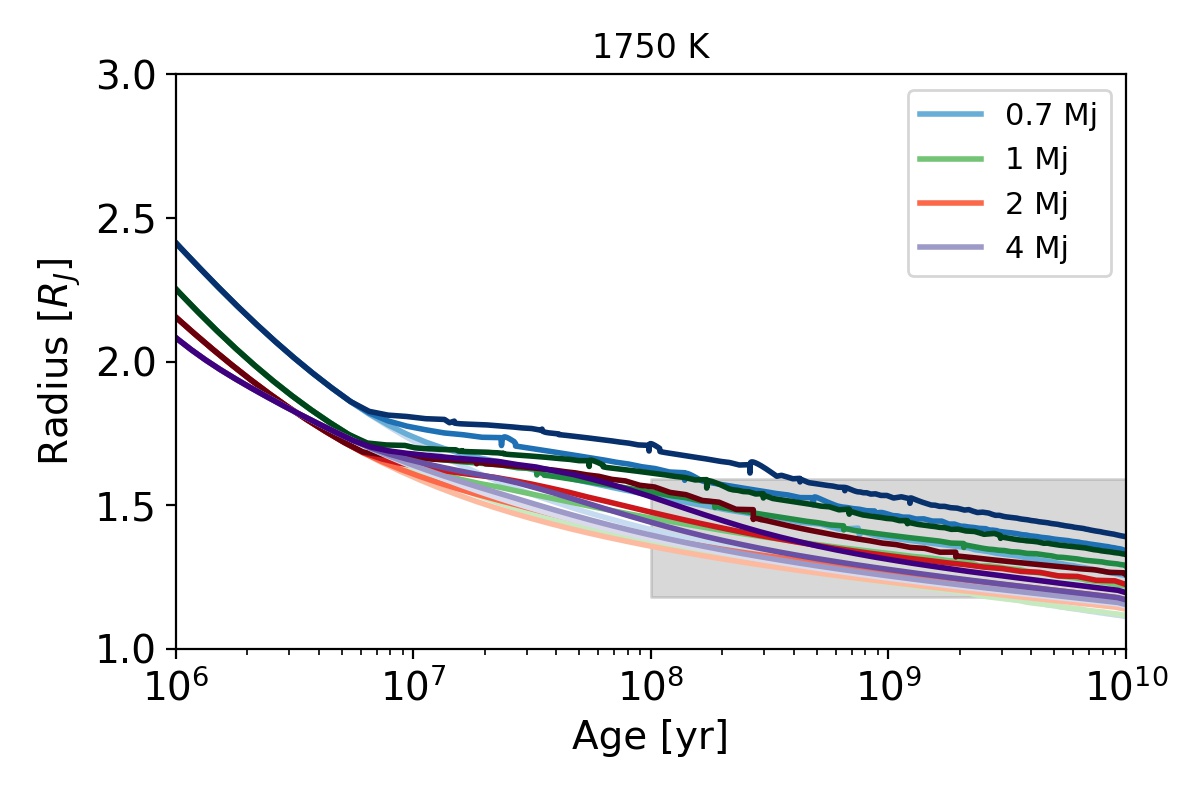}
\includegraphics[width=.33\textwidth]{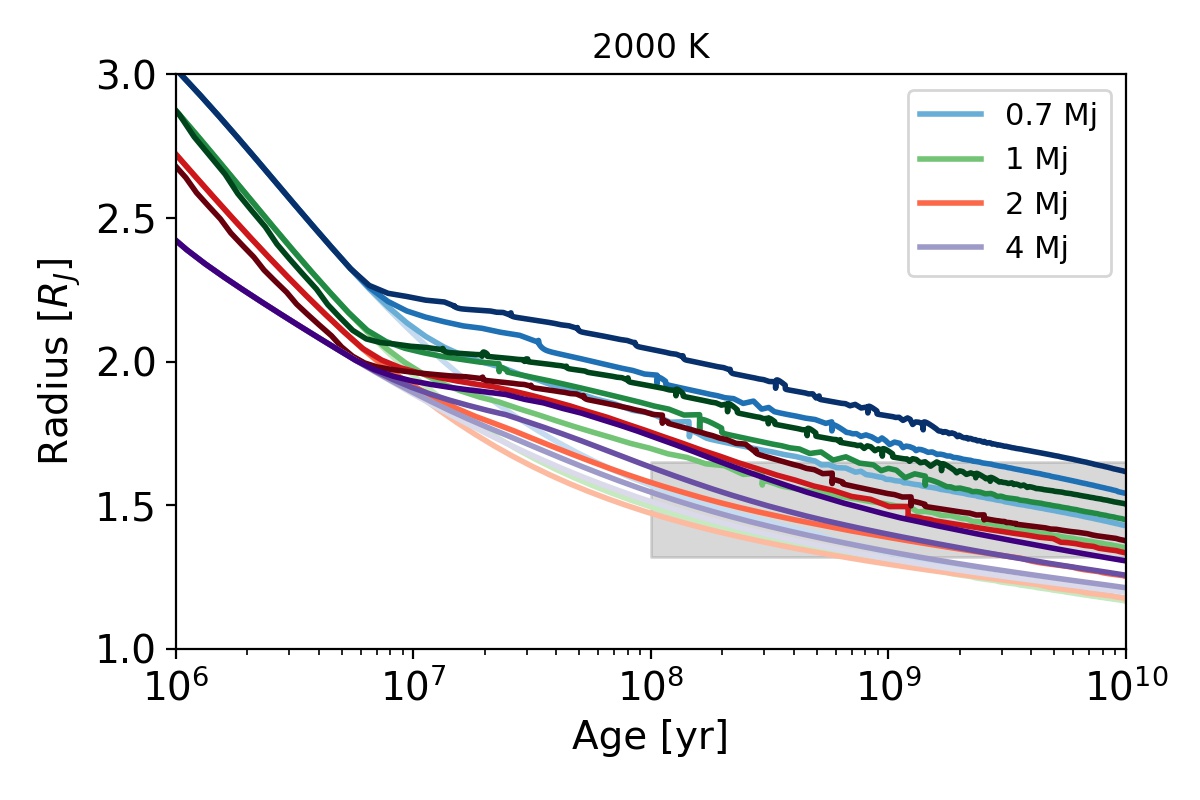}
\includegraphics[width=.33\textwidth]{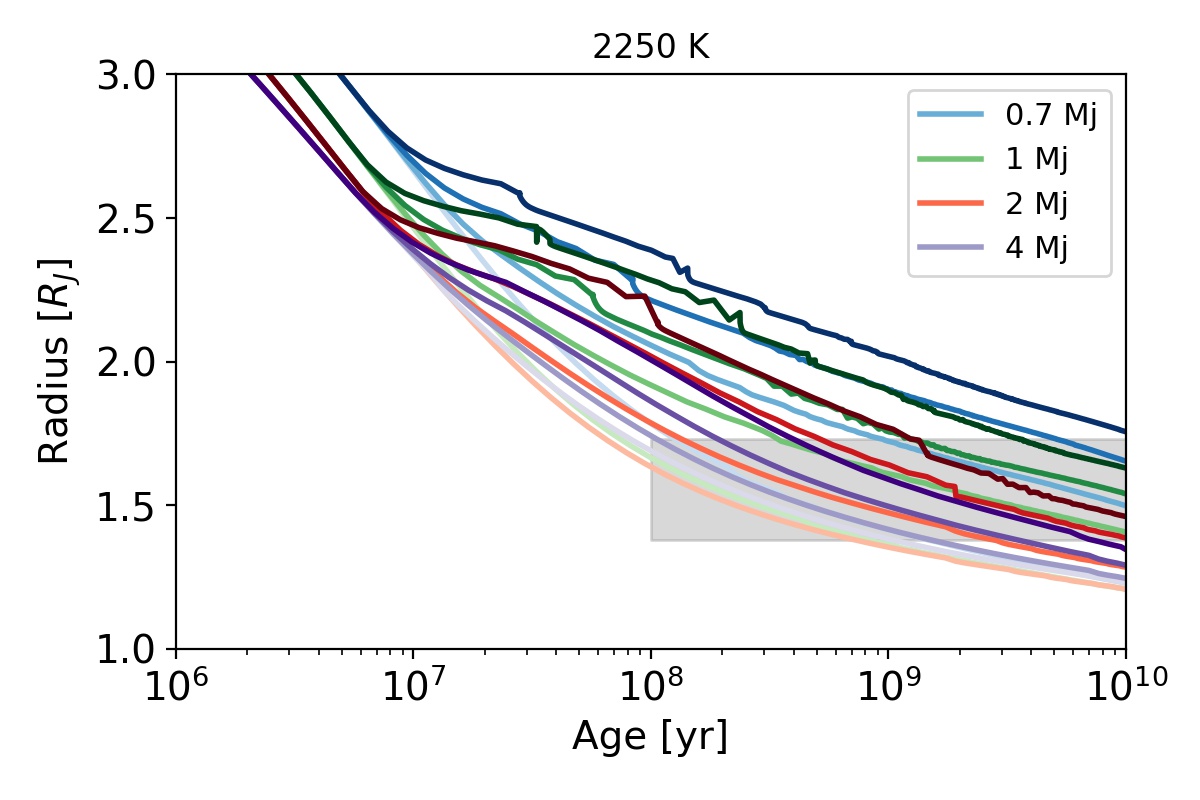}
\caption{Evolution of the radius, $R$, for planets with a given $\Teq=1750,2000,2250$ K (from left to right), different masses ($M=0.7,1,2,4$~$\Mj$ in blue, green, orange, and purple, respectively), and different values of $v_{\rm avg}$, marked by increasingly darker shades: $v_{\rm avg}=\{0, 1000, 2000, 3000\}/[(M/\Mj)(\Teq/1500 {\rm K})^6]$ m/s. In each panel, the grey band indicates the range of observed radii for HJs with $\Teq$ around ($\pm 125$ K), which is the value assumed in the models, $\bar R \pm \sigma_R$, as shown by grey background boxes in Fig.~\ref{fig:obs_hj}.}
\label{fig:radius_main_parameters}
\end{figure*}

Due to the non-trivial feedback between the background field and the internal heat, the code can show instabilities (for both numerical and physical reasons, see below) for high level of irradiation and/or heat deposition, visible as perturbations in the long-term evolutionary curves, and possibly implying early crashes in the simulations. To reduce the numerical artefacts, we use a reduced time and grid resolution, typically 2x or 4x the default one (i.e. {\tt delta\_time\_coeff} and {\tt mesh\_delta\_coeff} 0.5 or 0.25 instead on 1). We also implement an exponential decrease of the heat source for $p\lesssim 1$ bar, which helps the convergence in the structure solution and does not affect the results in terms of inflated radii (the fraction of the heat deposited in these outermost thin layers is minimal).

Additionally, the source, $\epsilon_{\rm j}$, at a given time, $t$, can actually be implemented as the average of the Ohmic heat, $<\frac{Q^n_{\rm j}}{\rho^n}>_{t_{\rm avg}}$, over an interval much smaller than the typical cooling timescales, which are comparable to the age itself  (we checked that such a time average does not affect the overall long-term trend of the radius evolution). This simplified multistep-like method, adapted to our case, smooths out the oscillations in the instantaneous Ohmic heating rate, which are due to intermittent convection and dynamo field (see Sect.~\ref{sec:intermittent}) and stabilise the code. In other words, we numerically mitigate the stiffness of the source term. We notice that in our specific case, the use of the implicit scheme for extra energy sources (with the recently implemented auto differentiation, \citealt{jermyn23}) does not improve the stability due to the intrinsically variable instantaneous heat source. Therefore, we opted for the more manual solution described above.

\section{Results}\label{sec:results}

\subsection{Dependence on $\Teq$, $M$, and $v_{\rm avg}$}

In this section, we focus on the radius evolution and its dependence on $\Teq,M,v_{\rm avg}$. After an extended exploration of parameters, we show the main results for a representative set of models with $M=0.7,1,2,4~\Mj$ and $\Teq=1500, 1750, 2000, 2250$~K spanning the bulk of the observed inflated HJs. For each combination of $M$ and $\Teq$, there is a maximum value, $v_{\rm avg}$, for which the runs are stable. We looked for such values and found an approximately linear dependence with $M$ and a much steeper dependence with $\Teq$, which we have tentatively approximated (for the range of $M$ and $\Teq$ here considered) as

\begin{equation}\label{eq:vavg_max_stability}
    v_{\rm avg}^{\rm stab} \sim 3000\frac{\Mj}{M}\left(\frac{1500\, {\rm K}}{\Teq}\right)^6~\frac{{\rm m}}{{\rm s}}~.
\end{equation}
We note that this is not the inferred maximum value that fits the data and that it does not have any a priori justification. It is merely a functional form that approximately indicates the value of $v_{\rm avg}$ above which the code fails to converge. Runs with values $v_{\rm avg}\lesssim v_{\rm avg}^{\rm stab}$ show pronounced oscillations in, for example, the convective region thickness (see below), but the general trends of radius evolution as a function of the parameters $R(t,M,\Teq,v_{\rm avg})$ are not affected by these details. The instability observed in extreme models is physical in nature and arises from a runaway process triggered by a certain excess of heat, as demonstrated, for example, in \cite{batygin11}. In these cases, the planet undergoes expansion, eventually approaching Roche-lobe overflow (a stage which is not followed up by our simulations).

Each panel of Fig.~\ref{fig:radius_main_parameters} has shows models with a given $\Teq=1750, 2000, 2250$~K (from left to right), and illustrates the radius evolution for various values of $M$ (indicated by different colours). and four values of $v_{\rm avg}$, equally spaced from zero to $v_{\rm avg}^{\rm stab}$, represented by increasingly darker shades. In each panel we show the cases which broadly encompass the observed radii around the value of $\Teq$, indicated by the grey band. Each band delineates the observed mean radius $\bar R$ and its standard deviation $\sigma_R$, for the sample of HJs with a range of $\Teq \pm 125$ K, as in Fig.~\ref{fig:obs_hj}. Here we mark it only for ages $t\geq$ 100 Myr for consistency with our sample selection (there are only a handful of younger HJs anyway). At the beginning of the evolution (after the relaxation phase mentioned at the end of Sect. \ref{sec:irradiation}) tracks start from a value close to the set initial radius, $R=4~\Rj$ (see also App.~\ref{app:benchmark} for general features of the cooling curves). For a given $M$ and $\Teq$, different $v_{\rm avg}$ start showing difference when the Ohmic heating is turned on at 5 Myr.

We note the strong dependence on the combined values of $\Teq$ and $v_{\rm avg}$. The curves with the lightest shades (no Ohmic heating) show only slight inflation, which is larger for increasing $\Teq$, due to the blanketing effect alone (see also benchmark models with no extra heating, App.~\ref{app:benchmark}). For a given pair $(M,T_{\rm eq})$ there is a well-defined range of $v_{\rm avg}$ for which there is a noticeable effect on the radius. In our models, we manage to cover well the hottest HJs ($\Teq \gtrsim 2000$ K) with values of $v_{\rm avg}$ of a few hundred metres per second maximum. If we consider lower values of $\Teq$, we need instead to take the maximum allowed values of $\sim$ km/s, eq.~(\ref{eq:vavg_max_stability}), to cover the observed range of $R$, but models tend to fall short, compared to the more irradiated cases.
We also note that, naturally, in our exploration of parameters, massive planets turn out being less inflated. This is related to the scaling laws and the role played by the dynamo field in the induction, as we discuss below.

\begin{figure*}
\includegraphics[width=.33\textwidth]{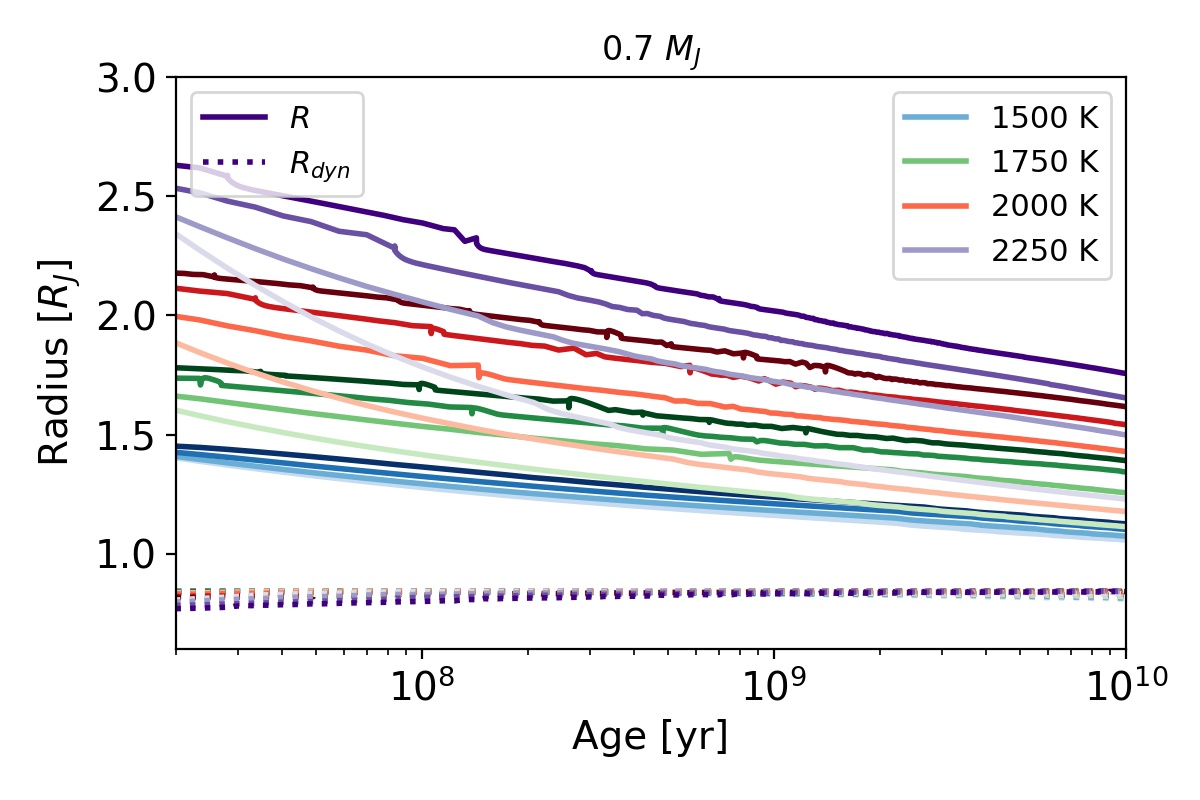}
\includegraphics[width=.33\textwidth]{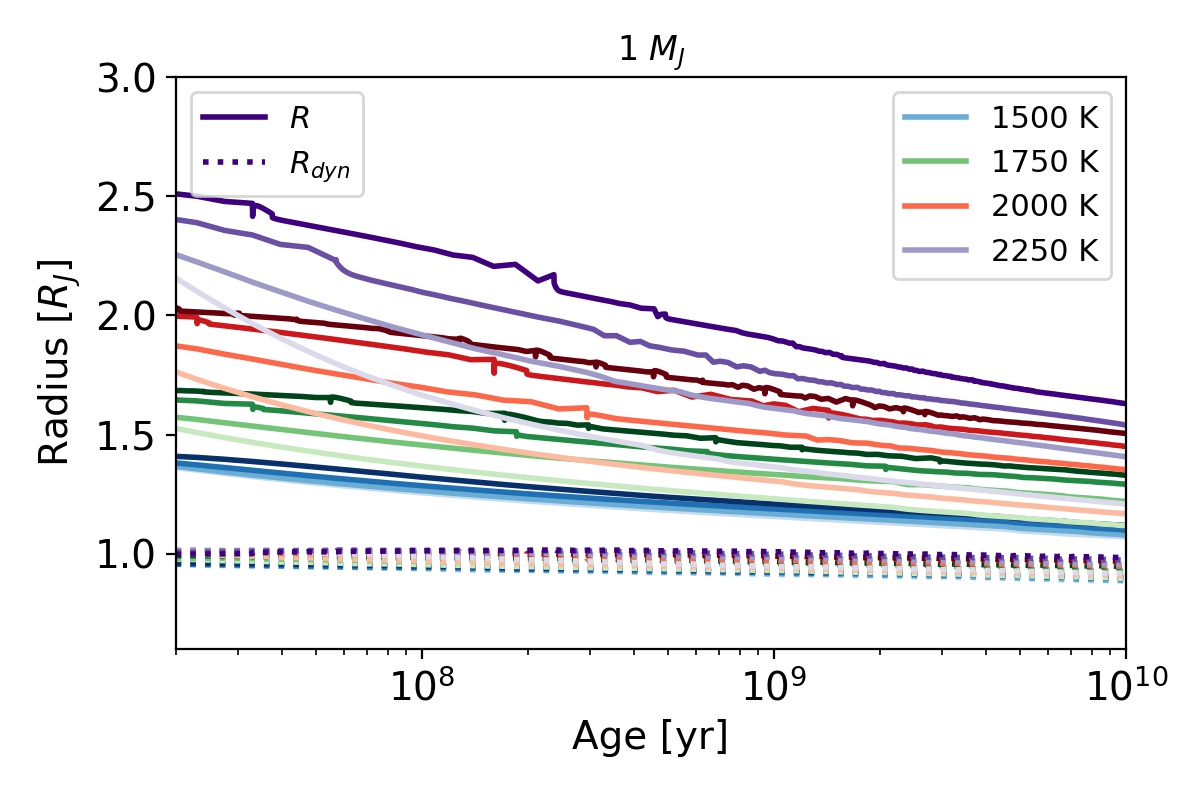}
\includegraphics[width=.33\textwidth]{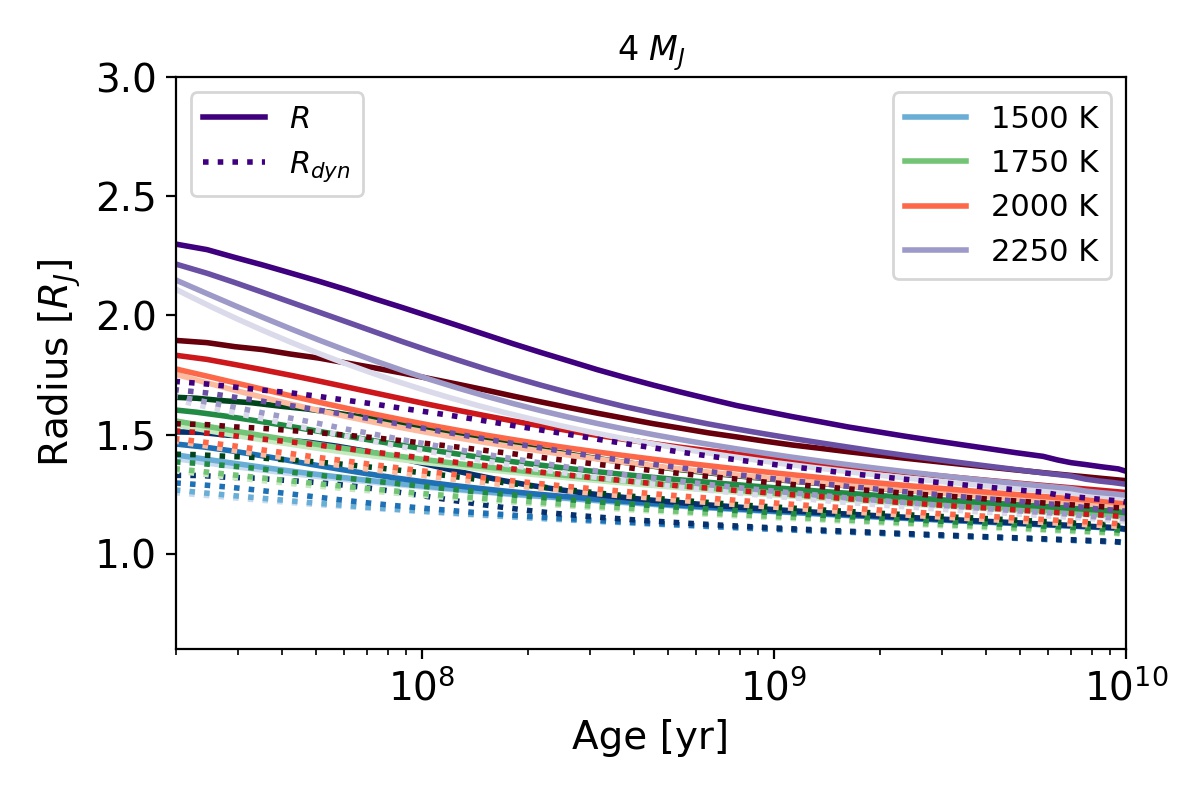}\\
\includegraphics[width=.33\textwidth]{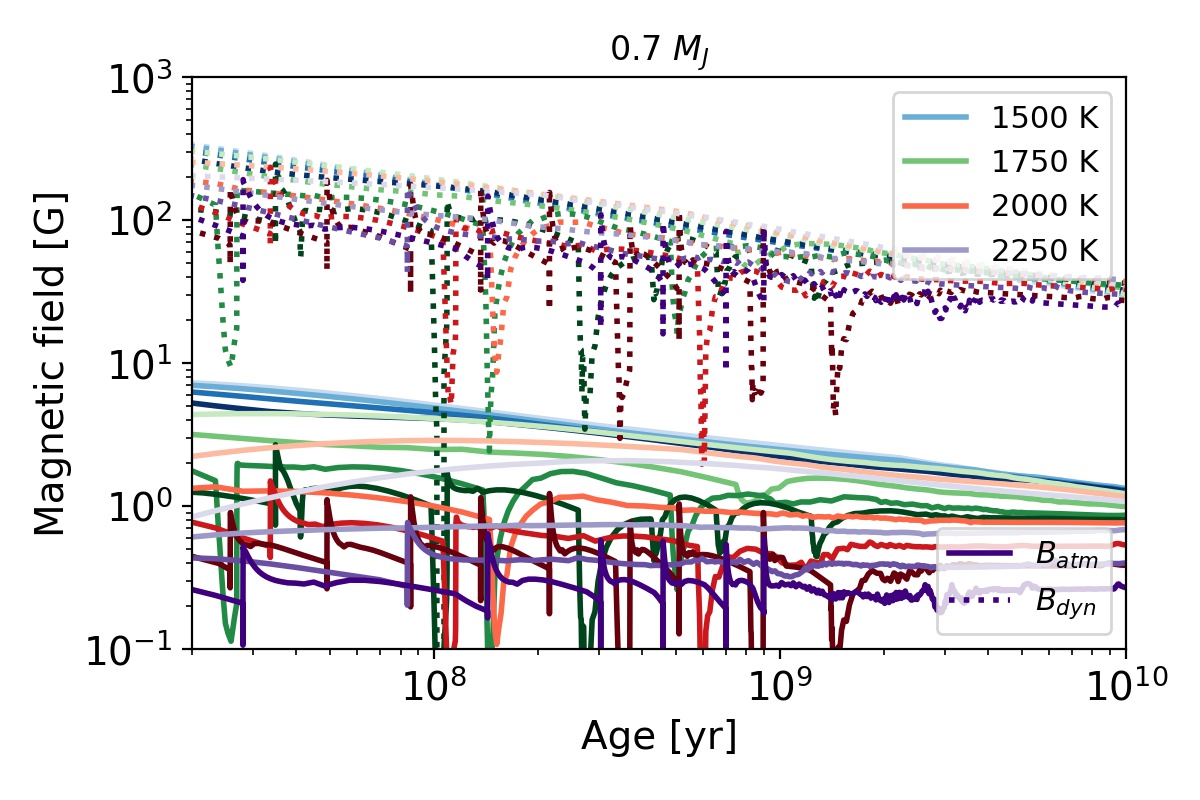}
\includegraphics[width=.33\textwidth]{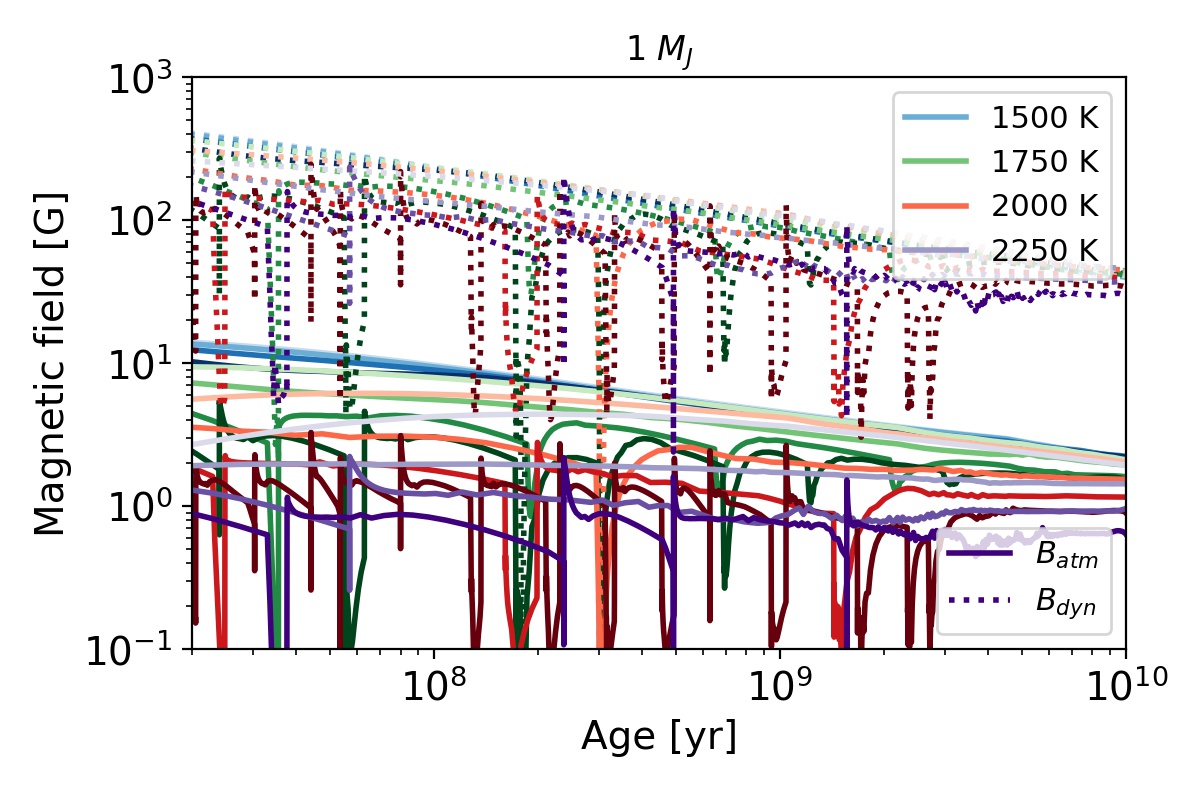}
\includegraphics[width=.33\textwidth]{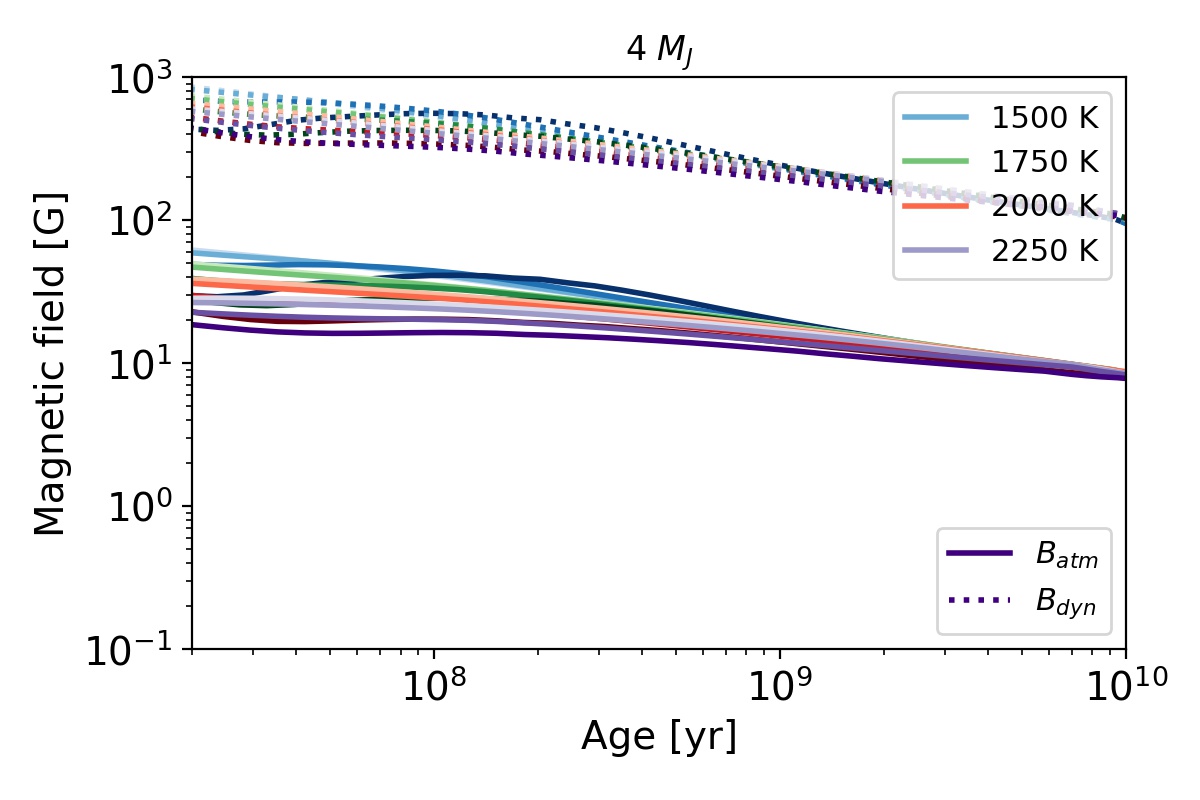}\\
\includegraphics[width=.33\textwidth]{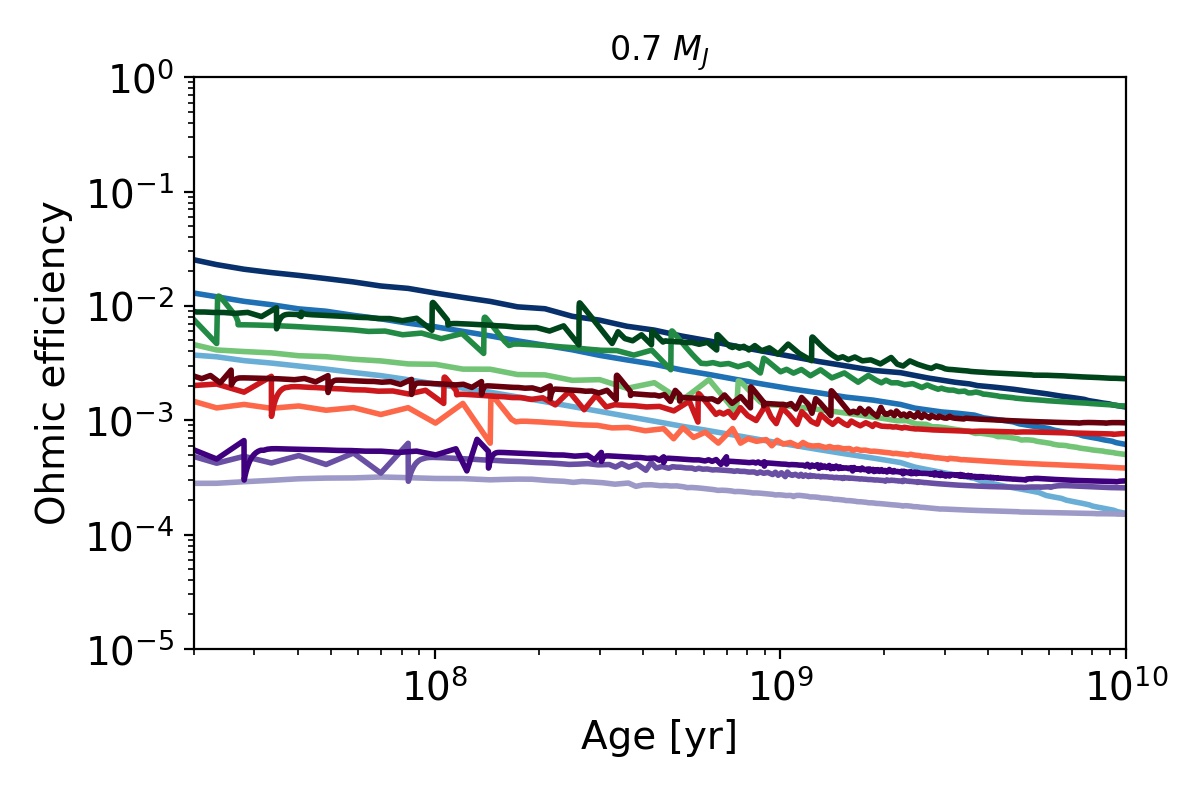}
\includegraphics[width=.33\textwidth]{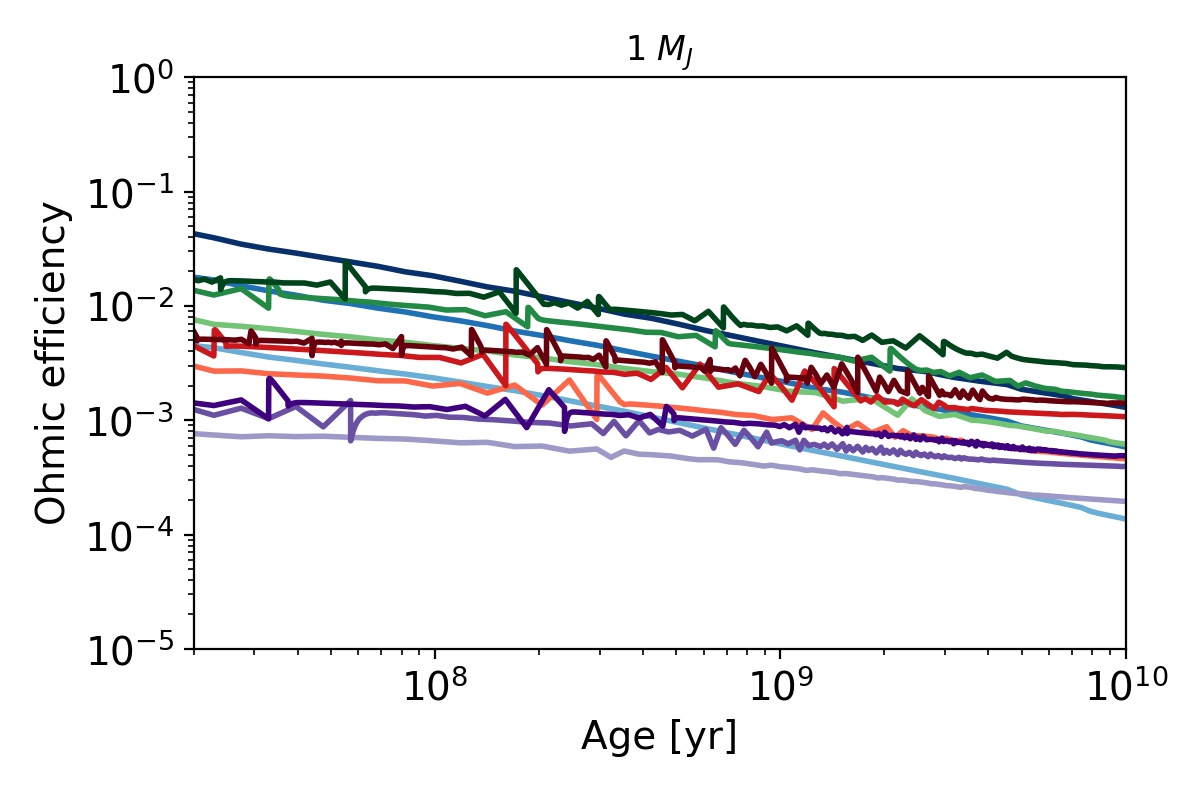}
\includegraphics[width=.33\textwidth]{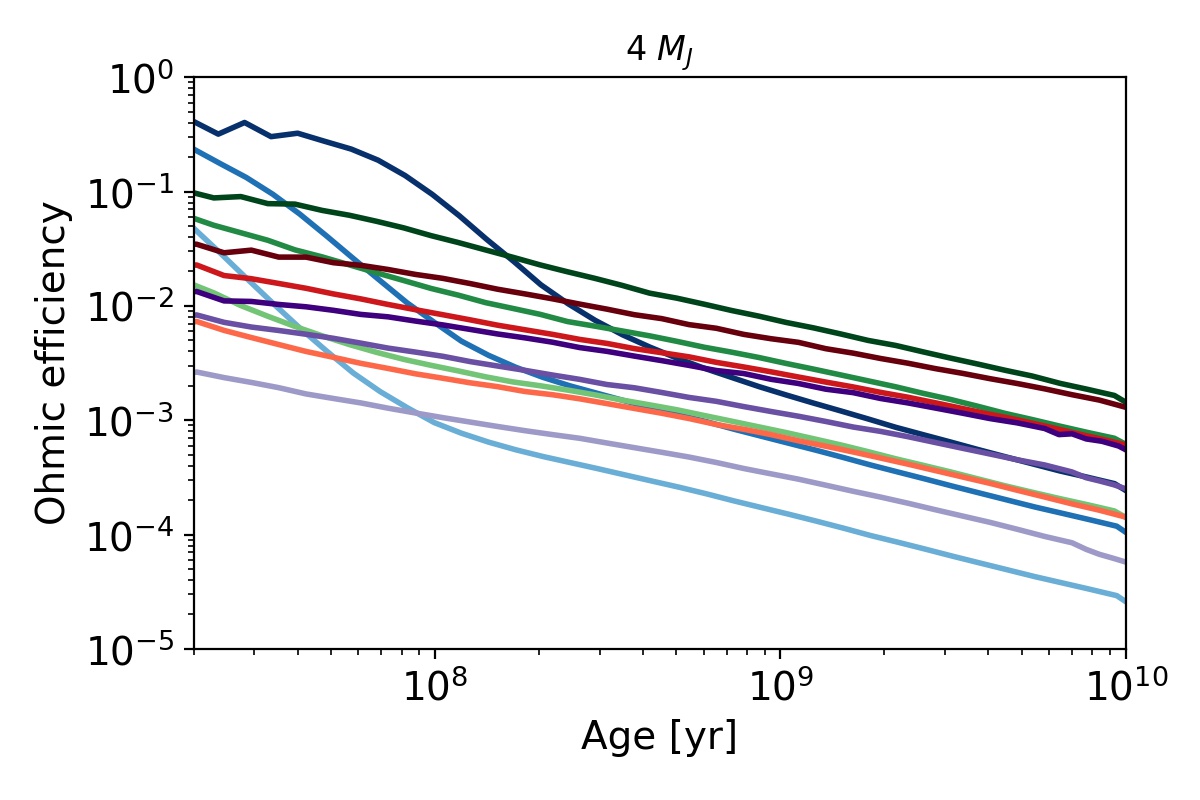}\\
\includegraphics[width=.33\textwidth]{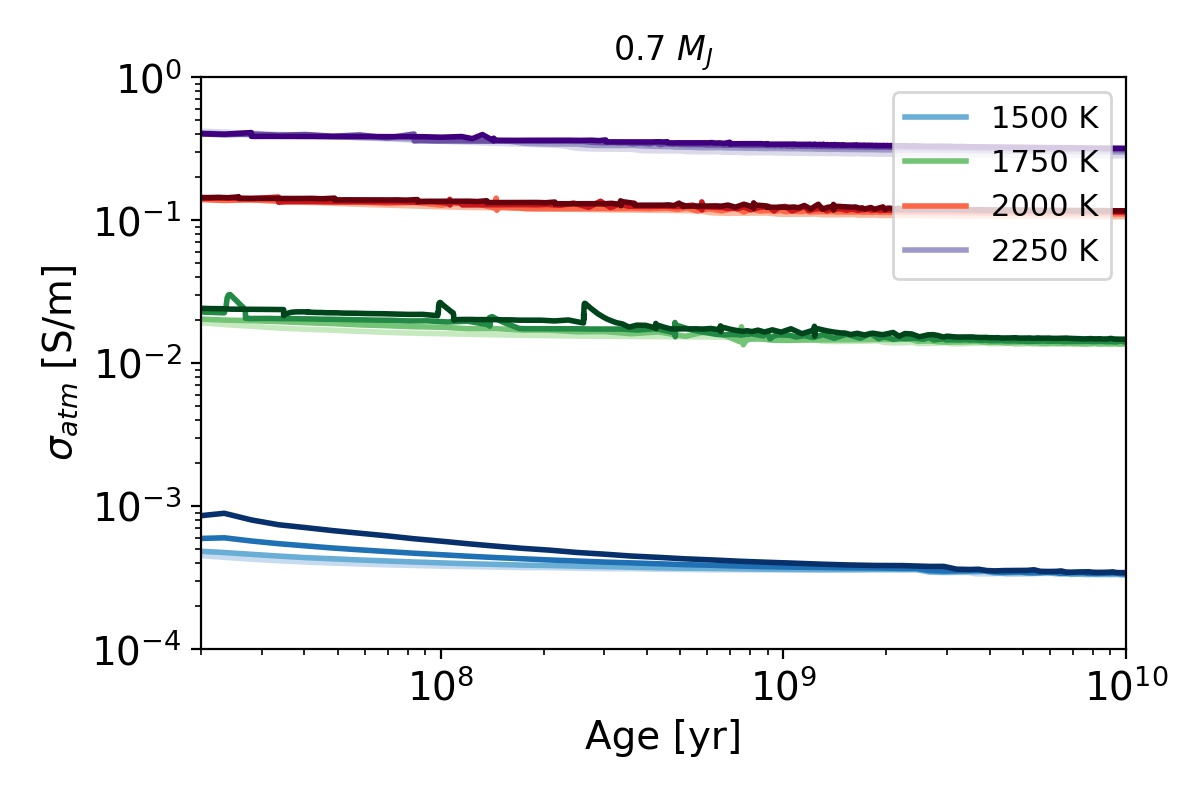}
\includegraphics[width=.33\textwidth]{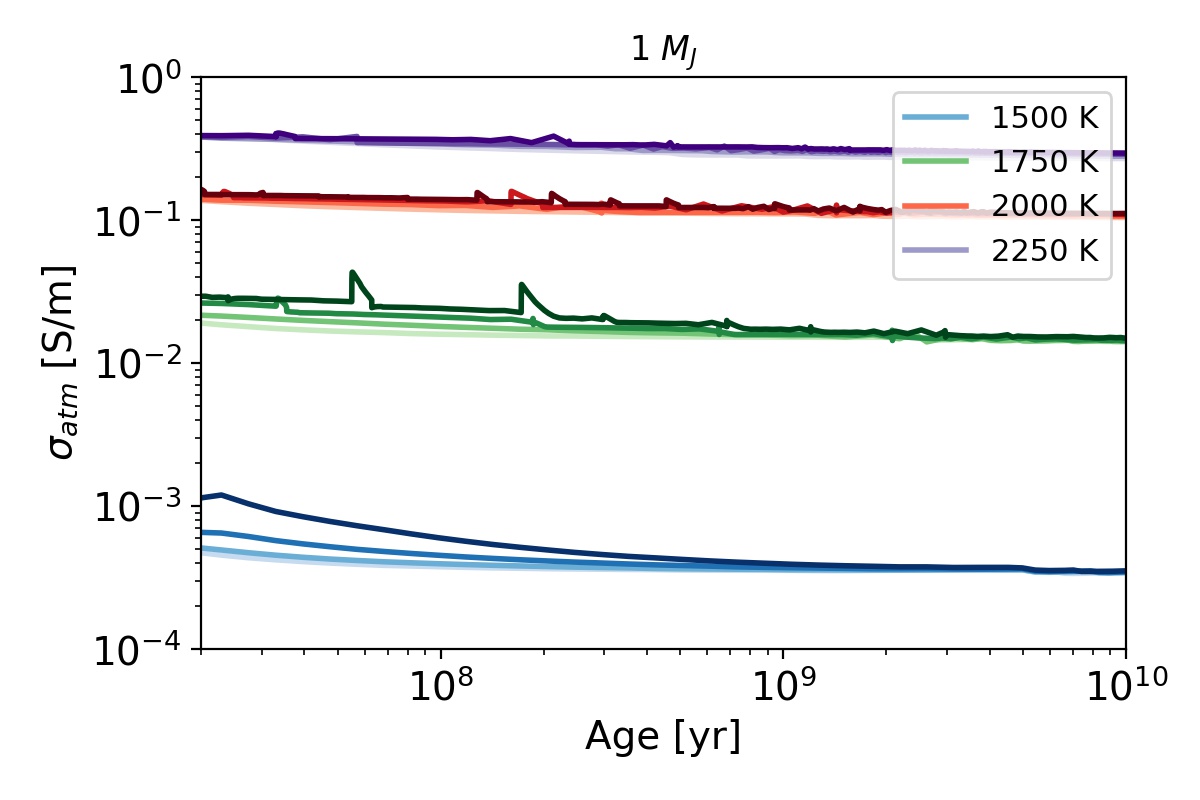}
\includegraphics[width=.33\textwidth]{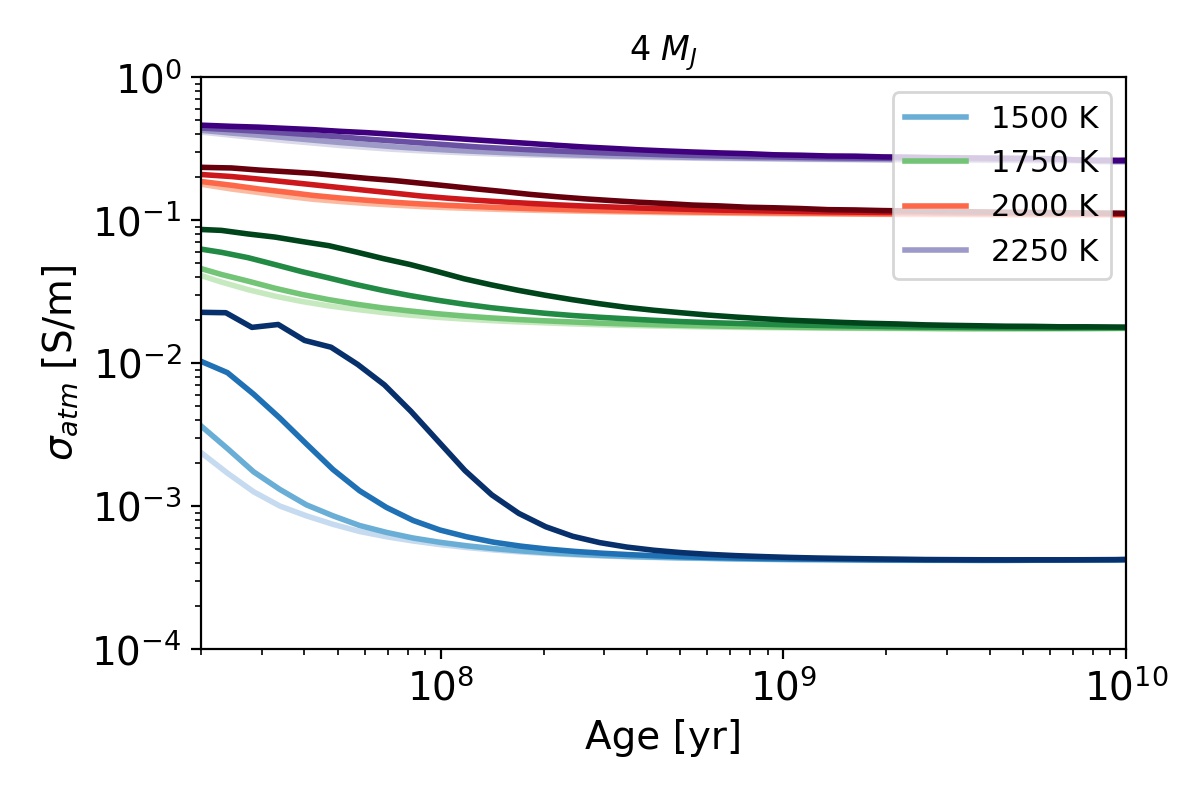}
\caption{Evolution of radii (top row, with dots indicating the dynamo radius and the solid the planetary radius), estimated magnetic field produced in the dynamo region (second row, with dots for $B_{\rm dyn}$ and solid lines for the planetary surface, $B_{\rm atm}$, assuming $f_{\rm dip}=0.1$), Ohmic efficiency (third row), and $\sigma_{\rm atm}$ (bottom row). Each column represent a mass, from left to right: $M=0.7,1,4~\Mj$. In all panels, colours indicate $\Teq=1500,1750,2000,2250$
K (blue, green, orange, respectively), and four values $v_{\rm avg}=\{0, 1, 2, 3\}/[(M/\Mj)(\Teq/1500 {\rm K})^6]$~km/s are indicated by increasingly darker shades. The curves of magnetic fields have been smoothed out by a Savitzky–Golay filter (window length 21, polynomial degree 3) to visually smooth out the frequent oscillations seen in the most heated and irradiated cases in light planets (see text).}
\label{fig:evo_eff_b}
\end{figure*}

To understand more in depth these trends, Fig.~\ref{fig:evo_eff_b} focuses on the most relevant physical quantities for the same set of simulations, considering $M=0.7, 1, 4~\Mj$ (left to right columns),  $\Teq=1500,1750,2000,2250$ K (different colours), and the four same values $v_{\rm avg}$ mentioned above (different shades). In the top panels we show the nominal surface of the dynamo region in dashes, $R_{\rm dyn}=r(p_{\rm dyn})$, together with the planetary radii $R$ (solid lines). For all the cases, the absolute value of the dynamo radius is fairly constant in time and depends on the mass, due to the fact that the planet shrinking ($R$ decreasing) mostly affects the outer, less dense envelope. The difference between $R$ and $R_{\rm dyn}$ is more pronounced for internally hot planets, i.e. at early ages, with high $\Teq$ and/or high Ohmic rate. Heavier planets (e.g. the $4~\Mj$ models, right column) show fewer differences due their higher gravity, with the dynamo radius being quite shallow (as in cold Jupiters) at gigayear ages.

This trend has a consequence on the evolution of the dynamo and surface magnetic field, $B_{\rm dyn}$ and $B_{\rm atm}$, shown with dashed and solid lines, respectively, in the second row of
Fig.~\ref{fig:evo_eff_b}. 
For a given value of $v_{\rm avg}$, $\Teq$, the values of $B_{\rm dyn}$ increase with mass, due to the scaling laws. However, how this translates into a trend between mass and $B_{\rm atm}$ depends on the thickness of the outer envelope, i.e. how expanded are the layers outside the dynamo region, $R-R_{\rm dyn}$, which determines the difference between the dynamo and surface field, eq.~(\ref{eq:bdip}). As a consequence, the surface field $B_{\rm atm}$ is always in the Jupiter range, $1-10$ G. Actually, larger deposited heat have the effect of decrease the surface field, due to the larger $R-R_{\rm dyn}$ and the effect of suppressing convection (causing also the stochastic oscillations in the $B$ curves), described below. Only massive planets, represented by the $4~\Mj$ series (right column) have slightly higher values of $B_{\rm atm}\gtrsim 10$ G at gigayear ages. The trends of $B_{\rm dyn}$ and $B_{\rm atm}$ with mass differ with the studies by \cite{yadav17,thorngren24}, probably because in that case they consider the \cite{reiners09} version of the scaling law, using the Ohmic luminosity as the one setting the convective heat flux (see a discussion in App.~\ref{app:scaling_laws}).

\begin{figure*}
\includegraphics[width=.32\textwidth]{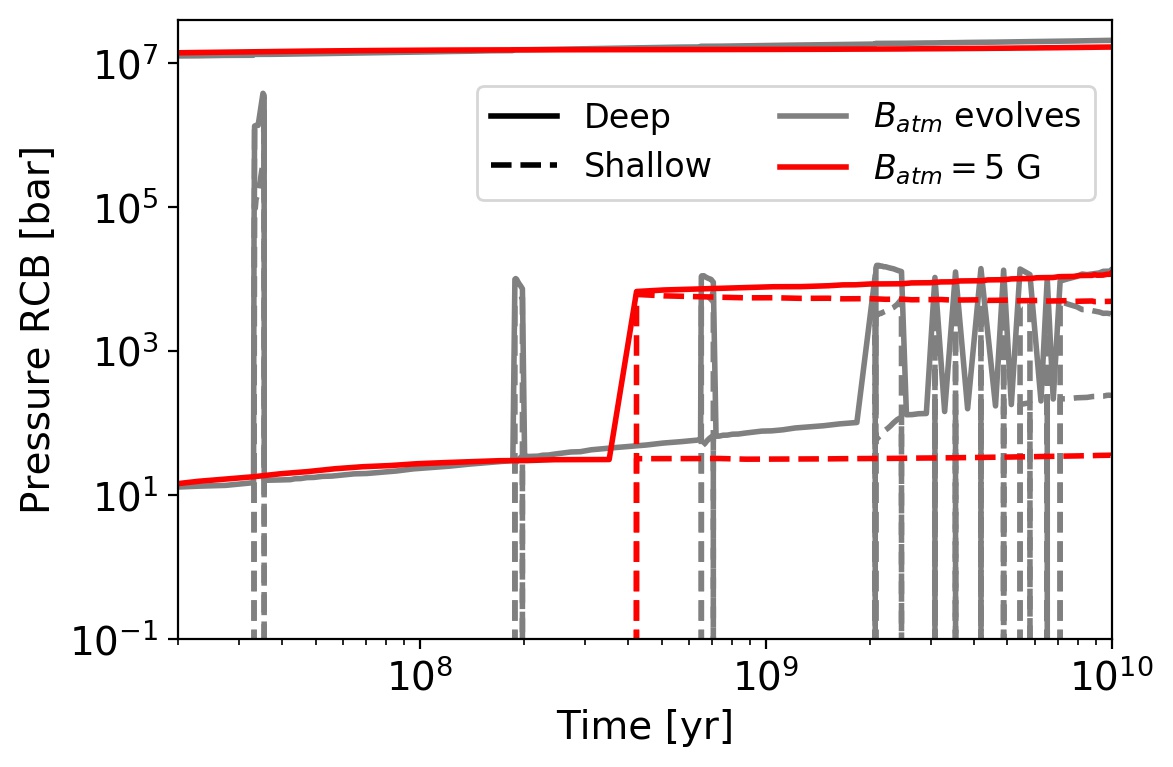}
\includegraphics[width=.32\textwidth]{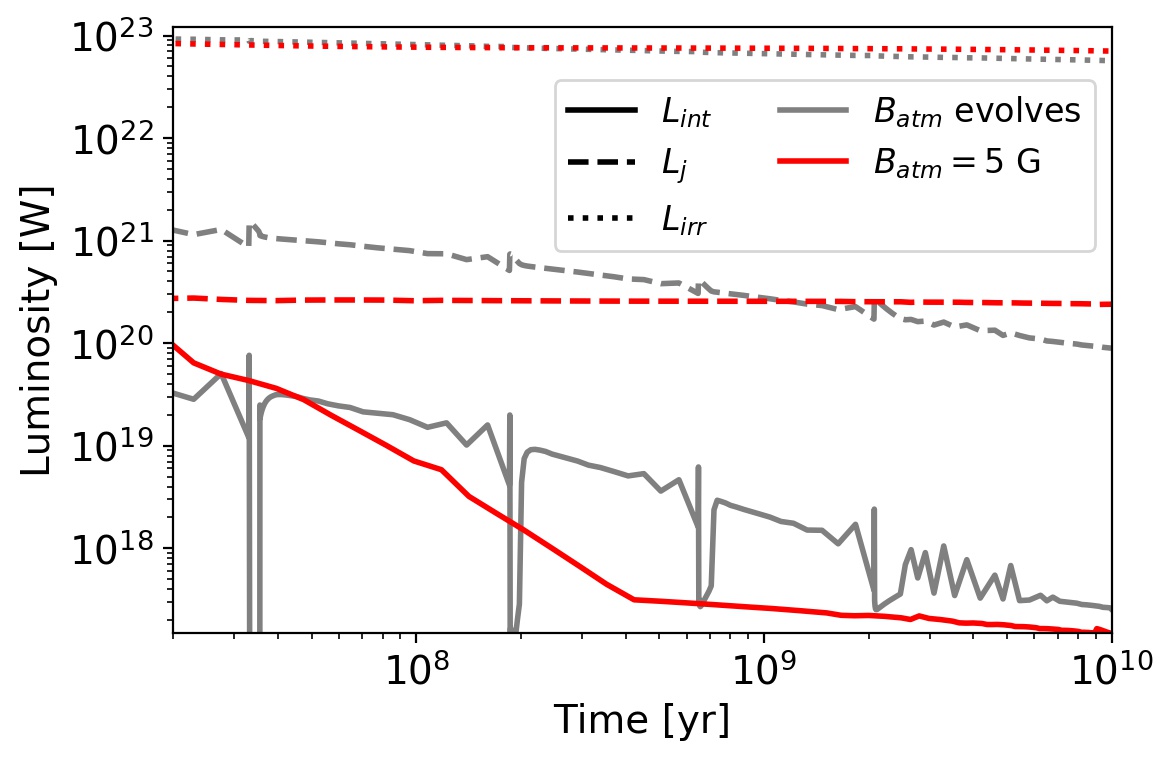}
\includegraphics[width=.32\textwidth]{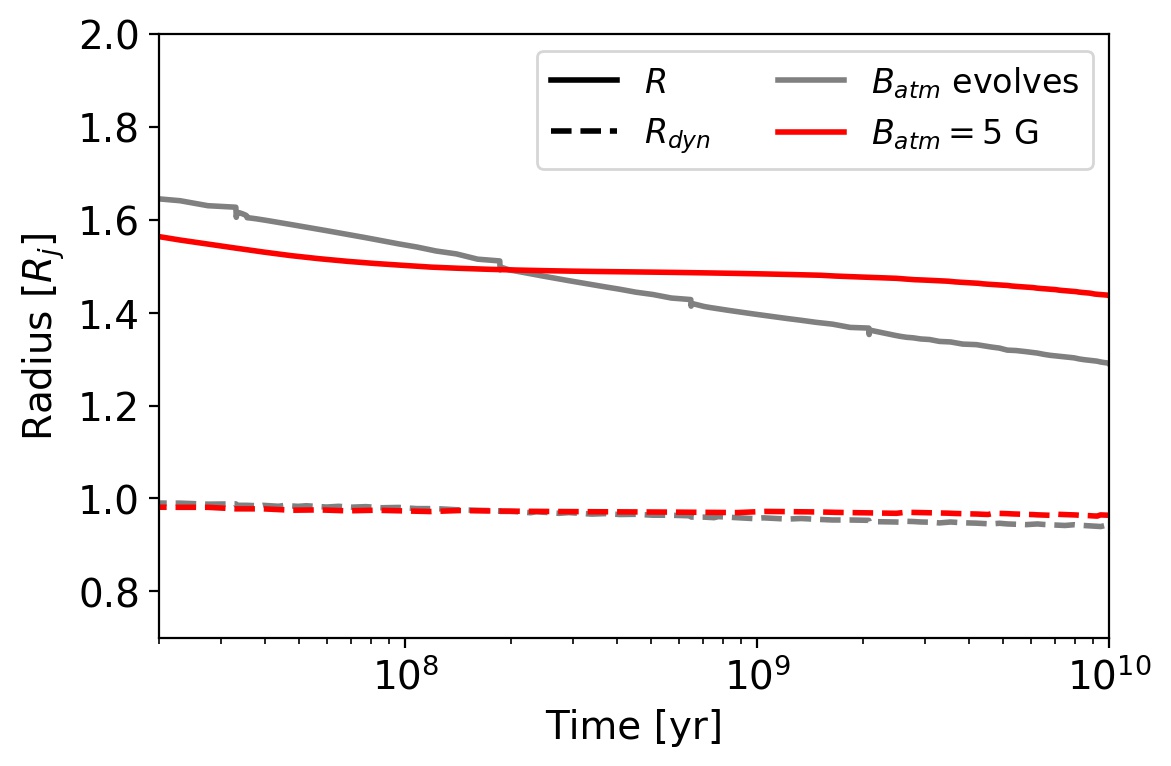}
\caption{Effects of the feedback of $B_{\rm atm}$. We take into account two representative models, both with $M=1\Mj$, $\Teq=1750$ K and $v_{\rm avg}=793$ m/s: one with an evolving $B_{\rm atm}$ (grey) and the other with a fixed $B_{\rm atm}=5$ G. {\em Left:} Evolution of the pressure values delimiting the deep (solid) and shallow (if any, dashes) convective regions. {\em Middle:} Evolution of the internal luminosity, evaluated as $L_{\rm int}=L(p_{\rm dyn})$ (solid), Ohmic luminosity $L_{\rm j}=\int_M \epsilon_{\rm j} ~dm$ (dashes), and irradiation power $L_{\rm irr}=4\pi R^2 F_{\rm irr}$ (dots). {\em Right:} Evolution of the planetary radius (solid) and the dynamo radius (dashes).}
\label{fig:feedback_comparison}
\end{figure*}

A very important consequence of the trend in the magnetic field is the evolution of the Ohmic efficiency, shown in the third row. In all models, it decreases in time, due to the $Q_{\rm j}(t)\propto B_{\rm atm}^2(t)$ dependence, via the normalisation of $J_{\rm atm}$, eq.~(\ref{eq:j_atm}). This feature is in contrast with classical Ohmic models, which assume a constant value of efficiency. As a consequence, our models do not predict the commonly found stalling of the evolution, i.e. the radius remaining more or less constant over the later stages in the evolution (e.g. \citealt{komacek17}), for which the volume-integrated heating balances the cooling. The effect is particularly pronounced for the massive cases, with a drop of $\sim 1.5$ orders of magnitude between $10^8$ and $10^{10}$ yr.

The typical Ohmic efficiency at gigayears as a function of $\Teq$ preserves the same qualitative trend seen in previous statistical studies \citep{thorngren18,sarkis21}, peaking at intermediate temperature (in our case, $T=1750$ K), although a more detailed study is needed for a quantitative comparison.

A consequence of the $B_{\rm atm}(M)$ trend and of the evidence that the radius inflation is observed to decrease with mass (Fig. \ref{fig:obs_hj}), is that much lower values of $v_{\rm avg}$ are needed for heavier HJs. Physically, this is compatible with the fact that in high-mass planets the magnetic drag should be larger than in low-mass ones.

On the other side, the value of $\sigma_{\rm atm}$, which also enters linearly in the definition of $J_{\rm atm}$ barely decreases after $10^8$ yr (bottom row of Fig.~\ref{fig:evo_eff_b}), thus not contributing much the change of Ohmic efficiency.  This is because we consider a wind region of $p\leq p_{\rm atm}=10$ bar, for which $T$ is kept very close to the equilibrium temperature by the irradiated flux. Had we considered a larger value of $p_{\rm atm}$, closer to the radiative-convective boundary (for instance, 100 bar), the change in $\sigma_{\rm atm}$ would have been much more pronounced, since the region over which we average the conductivity would feel the long-term internal cooling.

\subsection{Intermittent dynamo and its feedback on the atmospheric induction}\label{sec:intermittent}

For highly irradiated and highly heated models, the convection region can be reduced in size, or broken into two or more thin convective shell. This can happen when most of the heat is deposited in the outermost layers of the original convective region, so that the normal temperature gradient is reduced, or even inverted. In fact, this has been noticed in past works which deposited heat in outer layers only \citep{komacek17,komacek20,elias25b}, or used Ohmic heating profiles strongly decaying towards the interior \citep{batygin10,batygin11,wu13,knierim22}, similar to what is derived here. In those earlier studies, the consequence of having alternate convective and radiative interior layers was to delay cooling. However, they assumed no feedback on the deposited heat, i.e. they kept a constant-in-time Ohmic efficiency.

In our model, a decrease of the convective region implies a lower average heat flux, which implies a lower dynamo magnetic field, via eq.~(\ref{eq:bdyn_C09}). As a consequence, the deposited heat, $\propto B_{\rm atm}^2$, decreases. Therefore, in time, the convection across a very thick shell (most of the planet) is allowed again, which rises $B_{\rm atm}$. This mechanism implies that the feedback between background field $B_{\rm atm}(L)$ and the Ohmic dissipation rate $Q_{\rm j}(B_{\rm atm})$, generates an unstable mechanism that leads to cyclic rises and falls of the heat flux and therefore of the Ohmic dissipation. This is visible in several models of Fig. \ref{fig:evo_eff_b}.

We illustrate this effect in Fig.~\ref{fig:feedback_comparison}, which considers two representative models with the same $M=1~\Mj$, $\Teq=1750$ K and $v_{\rm avg}=793$ m/s. One is the standard model considered in this paper, with an evolving $B_{\rm atm}$ coupled to the Ohmic heating (coupled model, grey lines); in the other, we manually fix throughout the evolution $B_{\rm atm}=5$ G (a value which is in the range of the evolving $B_{\rm atm}$ case at gigayears; the results would be exactly the same for any given product $B_{\rm atm}v_{\rm avg}$). In the left panel, we show the values of the pressure delimiting the convective regions. The middle panel shows the surface (dots), Ohmic (dashes), and internal luminosity (solid). The latter is evaluated at $p=p_{\rm dyn}$, considering the dynamo region and not the Ohmic contributions in the less-conductive envelope where no dynamo is present, and is a tracer of the convective heat flux $q(r)$ entering in the scaling laws for $B_{\rm dyn}$ (and, consequently, $B_{\rm atm}$). The right panel shows the difference in the planetary and dynamo radii evolution.

We can relate the results illustrated here to the cooling regimes presented by \cite{komacek17} and \cite{ginzburg16}, which correspond to different relative weights of irradiation power, internal luminosity, and total heat rate ($L_{\rm j}$ in our case). Heated HJs generally fall in the case for which $L_{\rm int}\ll L_{\rm j} \ll L_{\rm irr}$, labelled as regime 2 in \cite{komacek17} and stages 3 or 4 in \cite{ginzburg16}. Within this regime, there are qualitative differences, depending on the location and intensity of the heat source. In our models, the heat is always spread continuously over the radiative and convective regions, see Fig. \ref{fig:Qj_sigma_P}. If the heating is confined to radiative layers, with negligible amount in the convective region, there is a perpetual cooling (regime 2d of \citealt{komacek17}), with a progressive deepening of the radiative-convective boundary (RCB), since the internal luminosity decreases in time while the irradiation power stays constant (or increases if the stellar luminosity evolves, see below). However, in many of the models shown here, the relevant scenario is the appearance of a shallow convective region, and a radiative layer just below the region where the bulk of heat is deposited. Such a transition, already identified as appropriate for heated HJs by \cite{komacek17}, is clearly seen in the fixed $B_{\rm atm}$ case (red lines) of Fig. \ref{fig:feedback_comparison}, left panel, described as follows. Before $\sim 0.4$~Gyr, there is only one convective region extending up to tens of bars (regime 2b of \citealt{komacek17}). As the internal luminosity decreases, it becomes comparable or lower than the amount of deposited heat inside the convective region, which is a fraction of the $L_{\rm j}$ indicated in the middle panel: therefore, a radiative region opens around $p\sim 5000$~bar, and slowly expands, splitting the convection in two separated layers below and above it. Physically, this happens because the additional heat, which is concentrated in the outermost regions (Fig.~\ref{fig:Qj_sigma_P}), lowers the local temperature gradient to sub-adiabatic values in a region that is typically located just beneath the main portion of the heated area.

For the coupled model (grey lines in Fig. \ref{fig:feedback_comparison}), however, the scenario is more complicated due to the sudden changes of the heat power. We stress that an exact comparison with \cite{komacek17} is not possible in this case for two reasons. The first reason, valid also for the fixed $B_{\rm atm}$ case, is that they considered a localised heat source at a given pressure entirely in the convective region, while our heat deposition extends from the surface down to the convective region and is more similar to the power laws of \cite{ginzburg16}, who, however, did not consider the relevant regime 2b of \cite{komacek17}. The second reason is that their heat source is constant in time, while ours varies due to the background field coupling (except in the fixed $B_{\rm atm}$ case shown above). Qualitatively, we can affirm that the relevant regimes are the same as the ones described above (2b and 2c) but with multiple transitions between them. 
This cyclic behaviour works as follows: When the heat flux is reduced due to the broken convection, $B_{\rm dyn}$, $B_{\rm atm}$ and the Ohmic power, $L_{\rm j}$, also decreases, although the instantaneous variations are averaged out in time, as described in Sect.~\ref{sec:numerical_details}. Since the heating is lower, the full convection can be restored. This trend is repeated many times, although the time average of the Ohmic source affects the duration of the oscillations, and we do not claim to have properly captured the mechanism. We cannot exclude that the numerical difficulties in converging to a solution contribute to the short-term fluctuations, visible as noise in Ohmic efficiency and radius evolution. However, the qualitative difference between the two models is clear.

Another important difference between the fixed $B_{\rm atm}$ and coupled models is seen when comparing the radius and Ohmic efficiency evolution. At contrast with the coupled models, the Ohmic efficiency is practically constant in the fixed $B_{\rm atm}$ case, the only change can be due to $\sigma_{\rm atm}$, see eq.~(\ref{eq:j_atm}). Consequently, the radius varies much more slowly and maintains larger values, similarly to the models with constant efficiency present in the literature (e.g. \citealt{batygin11,komacek17,thorngren18,knierim22}). This reinforces the peculiarity and complexity of the Ohmic models here presented, considering the evolution of $B_{\rm atm}$ and its coupling to the cooling.

In any case, a word of caution is needed. The two cases mentioned above, almost constant efficiency (fixed $B_{\rm atm}$) versus maximally coupled $B_{\rm atm}$, can be considered the extreme ends: in reality, a sudden change in $B_{\rm atm}$ will also imply less magnetic drag, and an increase in $v_{\rm avg}$ (which we keep constant in this study). Therefore, the product $v_{\rm avg}B_{\rm atm}$, which sets the amount of currents, eq.~(\ref{eq:j_atm}), will vary less than in our study. We leave this further effect for a future work.

\subsection{Effects of evolving the stellar luminosity}\label{sec:evolving_luminosity}

Let us now explore the effect of relaxing the constant irradiation assumption, which might be crucial for the long-term stalling or even re-inflating of HJs, as shown by e.g. \cite{lopez16,komacek20}. Here we consider a specific case of a Sun-like star, parametrizing the evolution of the stellar luminosity as
\begin{equation}
\frac{L_\star(t)}{L_\odot} = 0.75 + \left(\frac{t}{10~{\rm Gyr}}\right)^2~,
\end{equation}
where $L_{\odot}$ is the Solar luminosity at $t=5$ Gyr. This simple formula represents reasonably well the main-sequence evolution as simulated by the recent evolution models of the Dartmouth code (\citealt{dotter08}; see also e.g. \citealt{ribas10} for an explicit plot).

\begin{figure}
\includegraphics[width=.4\textwidth]{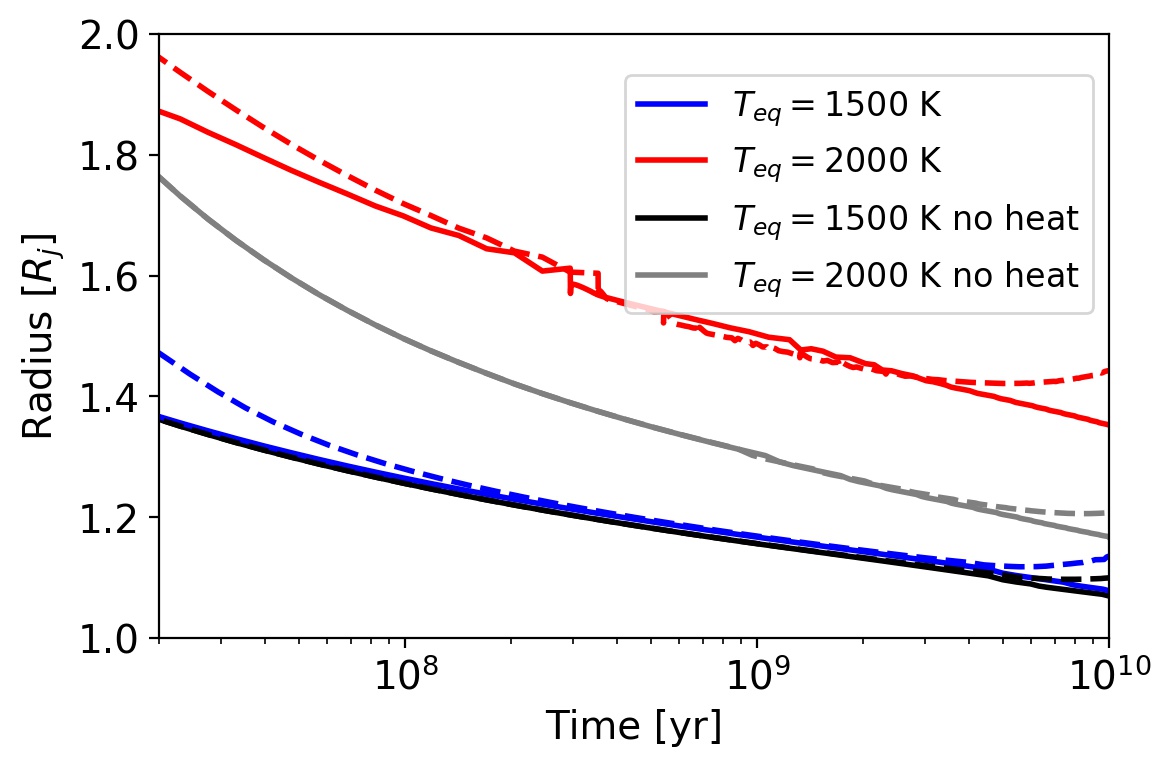}
\includegraphics[width=.4\textwidth]{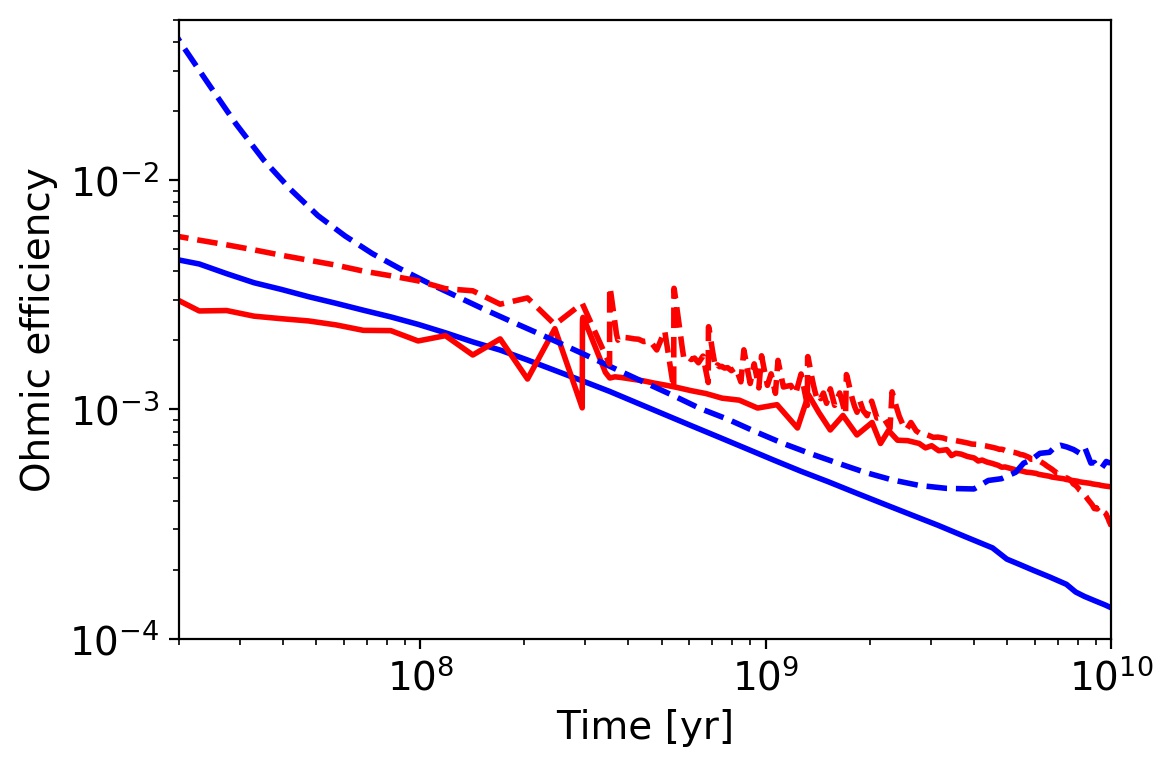}
\includegraphics[width=.4\textwidth]{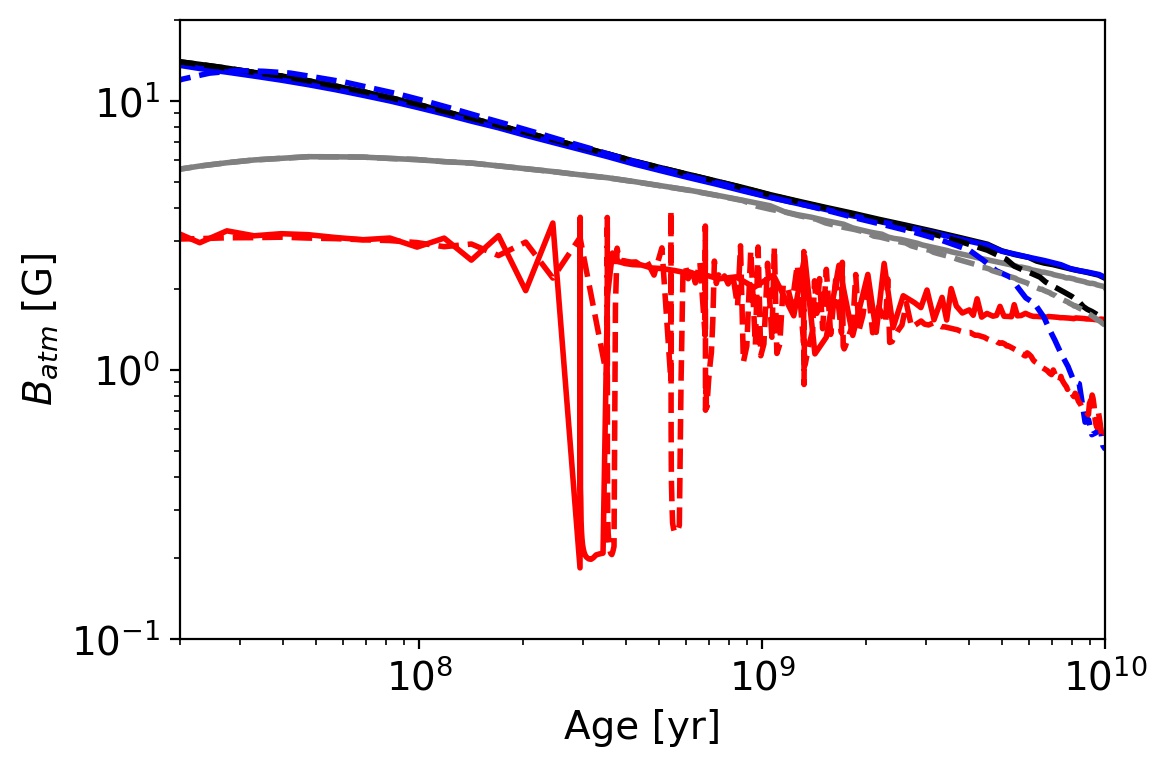}
\caption{Comparison of the planetary radii (top), Ohmic efficiency (middle), and background magnetic field at the surface $B_{\rm atm}$ (bottom) between models with a fixed (solid lines) and varying (dashed) $\Teq$. We consider models with $\Teq=1500$ K (black and blue lines with $v_{\rm avg}=0$ and 1000 m/s, respectively) and 2000 K (grey and red lines with $v_{\rm avg}=0$ and 178 m/s, respectively), where these values of $\Teq$ are at $t=5$ Gyr in the cases with evolving stellar luminosity.}
\label{fig:lstar_evolving}
\end{figure}

We note that the luminosity evolution significantly depends on the host star type. Hence its inclusion effectively breaks the degeneracy between host star type and orbital period, which has been implicit in all the results so far (different combinations of them provide the same irradiation or $\Teq$). Therefore, since the focus of the paper is rather on the effect of coupling the dynamo field to the induced currents, here we only consider a few representative cases, for $1~\Mj$ HJs orbiting a Sun-like star. In Fig. \ref{fig:lstar_evolving}, we show the evolution of radius, Ohmic efficiency and estimated surface dipolar magnetic fields for cases of evolving stellar luminosities (dashed lines) having $\Teq=1500$ K and $2000$ K at $t=5$ Gyr, and we compare them with the cases with the same fixed $\Teq$ (solid lines) shown in the previous sections. We show both the cases without and with heating, for a representative value of $v_{\rm avg}$. The evolution of the radius indicates that, indeed, re-inflation at $t\gtrsim 5$ Gyr is possible. In fact, the overall trends,  such as the decline in Ohmic efficiencies, magnetic field strengths, and the on/off mechanism of the internal convective heat determining the dynamo, are not qualitatively altered by the inclusion of the evolving luminosity during the first few gigayears. However, during the later stages of the main sequence, the increasing luminosity plays a significant role: it can halt planetary contraction even in the absence of internal heating, purely through an enhanced blanketing effect. When heating is present, this impact becomes dramatically more pronounced. Moreover, we noticed how the heated cases with $\Teq(t=5\,{\rm Gyr})=1500$ K or 2000 K behave differently: In the former, the Ohmic efficiency increases after $\sim 4$ Gyr, while in the latter it drops faster than in the corresponding fixed $\Teq$ case. This is due, again, to the powerful effect of suppressing  convection and dynamo in large parts of the interior in case of highly irradiated and highly heated HJs, as evident also from the values of $B_{\rm atm}$ derived from the convective flux (bottom panel of Fig.~\ref{fig:lstar_evolving}).

A thorough investigation of this effect warrants a systematic exploration of the relevant parameters, with the stellar mass (i.e. spectral type) as an additional parameter. Moreover, for the sake of self-consistency, one should also account for the variation of $v_{\rm avg}$, since it is not realistic to allow evolution of the irradiation and of the internal magnetic field while keeping constant the winds. 
Addressing this properly is beyond the scope of this work and should be pursued in future studies. This section is intended
merely to offer a preliminary glimpse into the potentially complex, non-trivial effects of an evolving stellar luminosity.

\subsection{Implications for the internal dynamo region}\label{sec:dynamo}

\begin{figure}
\centerline{\includegraphics[width=\hsize]{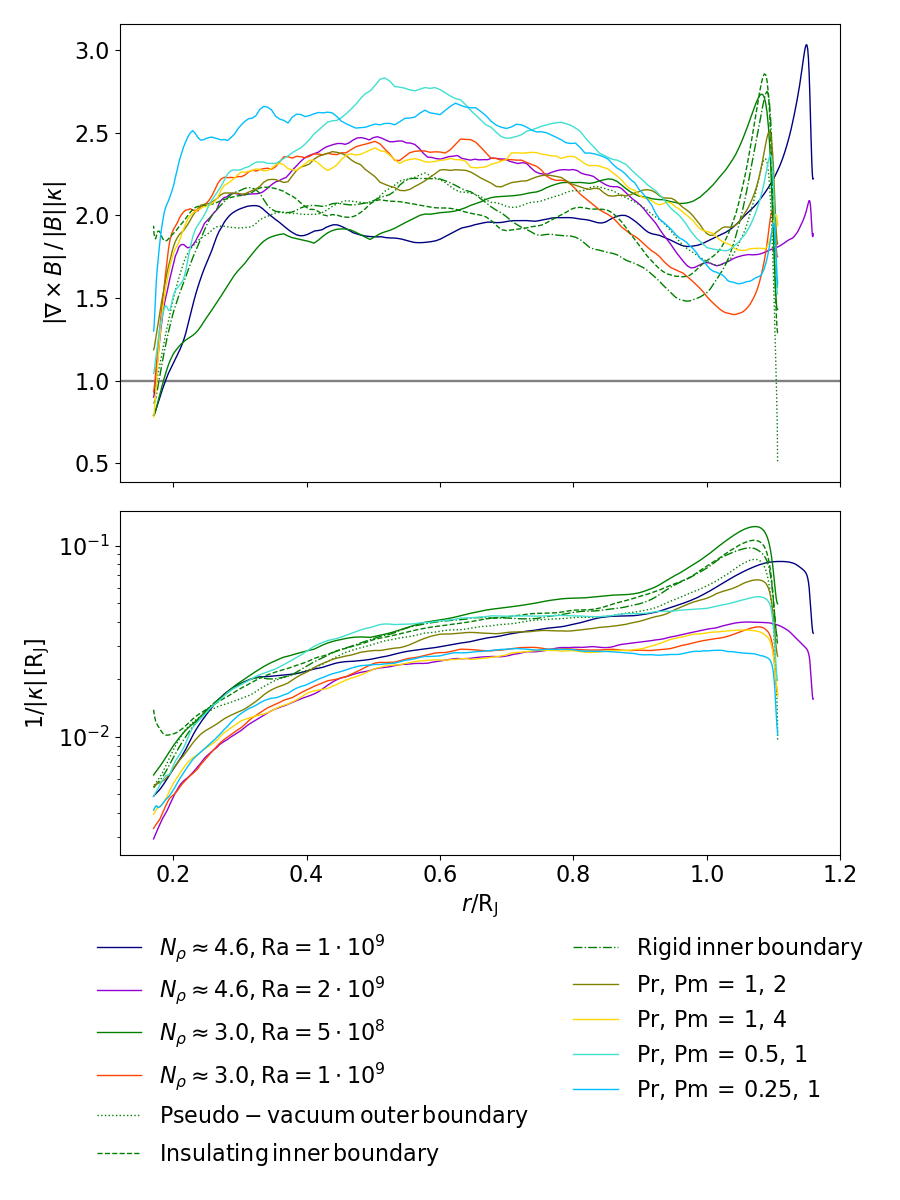}}
\caption{Typical radial profiles of $|\m\nabla \times \m{B}| / (\bar\kappa B)$ (top) and $1/\bar\kappa$ (bottom), averaged over the spherical surface at each radius, from 3D dynamo simulations with the code {\tt MagIC}. In order to explore the sensitivity to the free parameters. All models have the same boundary conditions (i.e.\ stress-free for the fluid and insulating for the magnetic fields outside, stress-free and perfectly conducting inside) except for the cases marked in the legend. The Ekman number is $10^{-5}$, and both Prandtl numbers are set to 1 (except for those also marked in the legend, which have $N_\rho \approx 3.0$ and ${\rm Ra} = 5 \times 10^8$. We note that $1/\bar\kappa$ becomes much smaller close to the external boundary, in order to match with the potential configuration constraint. See App. \ref{app:magic} for more details.}
\label{fig:Magic_Hot_Jupiter_model_curvature_mean}
\end{figure}

For comparison, and to get an insight into the possible interplay between atmospheric and deep-seated dynamos, one can compare the amount of atmospherically induced currents in deep layers with the ones independently generated by the active dynamo. In dynamo simulations, it is not possible to directly predict the absolute value of the magnetic field and currents, since they depend on dynamo parameters like the Rayleigh, Ekman and Prandtl numbers, which are computationally constrained to be many orders of magnitude away from the realistic ones (due to the impossibility of including all relevant scales). However, we can use the information on the typical topology from dynamo simulations, by looking at the ratio between currents and magnetic fields, and re-scale them by using the magnetospheric values measured in the solar planets. In particular, here we make use of the observations of the Jovian field, the one reconstructed with the highest fidelity among gas giants. We have considered the Juno measurements of the Jovian magnetic field above the surface, which is represented in terms of multipolar decomposition of the poloidal and toroidal components, as given by the best-fit model to the data \citep{connerney18,connerney22}. With the set of these best-fitting weights, one can describe the potential magnetic field in the entire volume where there are no currents, down to the radius $R_{\rm dyn}$, below which the dynamo starts to be active and the field is no longer potential. From this, one can then calculate the field line curvature vector, defined by $\m{\kappa} = (\m{b} \cdot \m{\nabla}) \m{b}$, where $\m{b} \equiv \m{B} / B$. We consider the radial profile, taking spatial averages over a sphere at a given radius, $\bar\kappa$. In App. \ref{app:curvature}, we show that, for Jupiter, the inferred values are $\bar\kappa \sim (0.07-0.3~\Rj)^{-1}$, depending on the magnetospheric model used and the depth considered. We adopt this as a lower limit for $\bar\kappa$, since higher values are expected in deeper regions, as a consequence of the turbulent environment in which the magnetic field is continuously regenerated (see bottom panel of Fig.~\ref{fig:Magic_Hot_Jupiter_model_curvature_mean}).

While we can infer the minimum value of $\bar\kappa$ from Juno measurements, we can relate it to the values of currents $J$ by looking at simulations. To do so, we performed MHD dynamo simulations that allowed us to estimate the values of the currents, and to validate the use of the magnetospheric value $\bar\kappa$ as a proxy for the internal $J/B$ ratio. We have used the code {\tt MagIC} \citep{jones11,gastine12} to study the amplification of the magnetic field in HJ-like planets. We used a variety of dynamo parameters and boundary conditions, and a representative structure profile for conductivity, density and other thermodynamic quantities, coming from our irradiated HJ {\tt MESA} models, following the methodology of \cite{elias25} (see App. \ref{app:magic} for details). These simulations provide us with a self-consistent evolution of the magnetic field as amplified in the convective, turbulent interior of a HJ and therefore allow for a robust estimation of the typical length scale $\kappa$. We note that a separate upcoming work will focus particularly on the 3D dynamo solutions for HJs. Here, we only focus on a specific aspect, which helps us to qualitatively compare the intensity of atmospherically induced and dynamo currents.

We show the results in Fig.~\ref{fig:Magic_Hot_Jupiter_model_curvature_mean}, where only some representative models producing a dynamo are shown. Despite some small differences among the various simulations, we obtain radial profiles with the same trend and order of magnitude, both for $|\m\nabla \times \m{B}|/(\bar\kappa B) \sim 1$-$3$ and $1/\bar\kappa \sim (0.01$-$0.1)~\Rj$. For the former (top panel), it confirms that the ratio between the magnetic field intensity and the mean radius of curvature, $B\bar\kappa$, is a very good tracer of the current density, $J$. Regarding $\bar\kappa$ itself (bottom panel), we obtain values similar to the ones inferred from Juno data (App. \ref{app:curvature}). More realistic outer boundary conditions for HJs might potentially alter the curvature $\bar\kappa$ near the external layers, but as seen in Fig.~\ref{fig:Magic_Hot_Jupiter_model_curvature_mean}, the various magnetic boundary conditions tested have no significant impact on the values within the bulk of the simulation.

Therefore, taking typical values $B_{\rm dyn} \sim 10$~G and $\kappa\sim 10^{-6}$ m$^{-1}$, one obtains $J \sim B\kappa/\mu_0 \sim 10^{-3}$ A/m$^2$. This range is the same as (or slightly lower than) the atmospherically induced currents at $r\sim R_{\rm dyn}$ (see top panel of Fig. \ref{fig:Qj_sigma_P}). Therefore, they could be a significant factor to be taken into account to set the boundary conditions in dynamo simulations, which are instead almost always set to be a current-free solution.
Since the geometry of the atmospherically induced currents depends on the wind and background field configurations themselves, this opens to the remarkable possibility of atmospheric processes affecting the deep dynamo solution.

\section{Discussion and conclusions}\label{sec:summary}

In this work we have investigated the evolution of Jupiter-like planets subject to both irradiation and internal Ohmic heating via a suite of numerical simulations with the code {\tt MESA}. We adopted time-dependent radial profiles of the electrical conductivity $\sigma(p,T)$ obtained by interpolations of the latest calculations for thermal alkali ionisation \citep{kumar21} and the degenerate metallic hydrogen contribution evaluated via DFT-MD calculations \citep{bonitz24}. Employing these values further refines the profiles of conductivity presented, for example, by \cite{huang12,wu13}, who accounted for pressure ionisation but neglected strong coupling effects. Compared to those studies, our model yields conductivities in the metallic hydrogen region that are higher by approximately one order of magnitude and in line with \cite{french12}. Although the precise value of $\sigma$ in the deep interior is largely irrelevant, since most of the deposited heat is deposited in much more shallower regions, we have included these values for completeness and to enable a more accurate assessment of the intensity of the induced current intensity.

Given that the problem in the wind region is intrinsically 3D, in our 1D {\tt MESA}-adapted code we parametrised the atmospheric currents as proportional to the time-varying background magnetic field, estimated through usual dynamo scaling laws, the average conductivity in the uppermost layers, and a free parameter, $v_{\rm avg}$, which represents the average of the thermal winds in the outermost layers (eq. \ref{eq:j_atm}).
To prescribe the profile of the current density at larger depths, we followed the classical approach by \cite{batygin10} and \cite{batygin11} to find induction equation solutions, and we have proposed new semi-analytical solutions for the internally induced currents based on piece-wise power-law fits of any conductivity profile and on any multipolar combination of the atmospheric wind flows and magnetic field geometry. The radial profiles of induced currents depend mostly on the radial dependence of the conductivity, which is steep except in the innermost region. The wind and magnetic geometries play a much less significant role, as also mentioned for a few specific simple cases in \citet{batygin11}.

With our model, we explored the combined dependence of the planetary radius evolution on the equilibrium temperature, the planetary mass, and the average velocity, $v_{\rm avg}$ (for a given core mass and composition). As the planet cools down, both the dynamo magnetic field, $B_{\rm dyn}(t)$, and the temperature-controlled atmospheric conductivity decrease, with the net effect of a relevant decrease of the Ohmic efficiency in the long term. Therefore, also the planetary radii decrease in time, and it is hard to match large values of $R\sim 1.5~\Rj$ for moderate irradiation $\Teq\sim 1500$ K. 
This contrasts with previous studies that adopted the Ohmic heating scenario (e.g. \citealt{batygin10,batygin11,knierim22}) or those that left the underlying mechanism unspecified (e.g. \citealt{komacek17,thorngren18}), in which case the heating efficiency was assumed to remain constant over time. In those cases, an equilibrium between the cooling and heating rates is reached, bringing the internal structure to be basically unchanged for many gigayears.

Secondly, the amount of radius inflation is strongly dependent on the parameter $v_{\rm avg}$, and we evaluated the range of values needed to explain the bulk of measured radii for HJs with a given $\Teq$. The inferred decrease of $v_{\rm avg}$ with $\Teq$ is much steeper than the one with $M$. Typically, we need $v_{\rm avg}\sim$ km/s in the lightest, moderately irradiated planets, and one or two orders of magnitude less when we move up to the highest values of $M$ and $\Teq$ considered here.
Keeping in mind that $v_{\rm avg}$ is a proxy of the wind velocity averaged over the uppermost region, $p\lesssim 10$ bar, these average values are compatible with GCMs, which find maximum velocities of a few kilometers per second. It is also compatible with the study by \cite{koll18}.

The decrease of $v_{\rm avg}$ with $M$ is a consequence of the dependence on the mass of the estimated internal field, $B_{\rm dyn}$, which sets the atmospheric background field, $B_{\rm atm}$, which in turn enters in the normalization of the induced currents, $J_{\rm atm}\propto \sigma_{\rm atm} B_{\rm dyn}v_{\rm avg}$. Since there is a lack of highly inflated planets for high masses \citep{sestovic18,thorngren24}, this points to low average velocities for very massive planets. Both this fact and the decrease of $v_{\rm avg}$ with $\Teq$ are consistent with the expected effects of the magnetic drag in slowing down the winds in the high temperature regime. Past works \citep{menou12,knierim22} have also pointed to a steep decrease of wind velocities with $\Teq$. 
Although 1D evolutionary models simulate the bulk structure of the planet and therefore cannot directly predict wind velocities at pressures $\lesssim$ mbar as probed by high-resolution spectroscopy, the trends of the average wind velocities with planetary mass and $\Teq$ can still be tested with a sufficiently large sample.
 
Another interesting point is that the typical induced current densities that penetrate the dynamo region are of the same order as the ones sustaining the deep-seated field, which we have estimated quantitatively by a combination of dynamo simulations and the typical magnetospheric field line curvature extracted from Juno data. This raises the question of how the atmospheric induced currents are going to couple with the ones sustaining the dynamo in the deep regions and calls for dynamo simulations with different boundary conditions (which are usually taken as current-free and designed for cold giants, e.g. \citealt{duarte13,duarte18,wicht19a,wicht19b,elias25}).

In our models there is a strong interplay between the value of the background field, generated via a dynamo in the metallic hydrogen region, and the Ohmic efficiency, due to the proportionality between the induced and background fields. An important effect is that for strongly heated models, the convection region can be strongly suppressed, as noted also by \cite{elias25b}. The drop in the Ohmic efficiency allows the convection region to be recovered, giving cycles of intermittent dynamo and induced field. This feedback effect has been neglected in previous studies, which assume constant values for both the Ohmic efficiency and the background field, with no feedback effects or use of convective heat scaling laws.

The specifics of this effect and an in-depth understanding of the characteristic timescales over which these cycles may operate cannot be properly addressed by this study. In fact, the frequency and timing of these cycles are rather sensitive to both the physical and numerical choices made here, such as the evolution time step, the grid resolution, the form of the scaling laws employed (see App.~\ref{app:scaling_laws}), and the temporal average applied to the  heat source term for numerical reasons (Sect.~\ref{sec:numerical_details}). However, these numerical factors alone are not the ones responsible for the occurrence of such behaviour, which arises regardless of these details (albeit with a variation in form), and has a solid physical explanation: the circular relation between induced Ohmic dissipation, convective heat flux, and dynamo magnetic field. A more quantitative study of this behaviour would require addressing the problem of considering significantly smaller timescales than the ones used here. Although MESA automatically adapts them close to the cycles, they can become so short as to potentially compete with other more dynamical timescales neglected by the long-term cooling code and related, for instance, to convection or dynamo magnetic cycles. In other words, variations over these timescales would require consideration of the effects that are typically neglected in a long-term evolutionary model, which implicitly averages out oscillatory effects acting on short timescales. Moreover, this effect relies on the assumption that the average velocity is constant and is not affected by a change in the magnetic field and on magnetic scaling laws, which also implicitly provide a simple estimate of the average dynamo field, not considering short-term fluctuations.

Related to the last point, this feedback effect would probably be mitigated by the magnetic drag, for which the wind velocity will also quickly respond to compensate for any important decrease or increase of the background temperature, something we do not consider (we assumed the average velocity remains constant over time). Another possible caveat that could have enhanced the effects of the coupling in this study is the linear dependence of the induced currents with the deep-seated field: When the induced magnetic field becomes comparable with the latter (magnetic Reynolds number $Rm\gg 1$), which is expected for a large fraction of HJs \citep{batygin13,dietrich22,soriano25}, the induction mechanism is not linear, and the assumption of linear proportionality, $J\propto B_{\rm dyn}$, is likely simplistic.

We also briefly explored the effects of an evolving stellar luminosity rather than assuming it to be constant. While the main effects discussed here (coupling between dynamo-generated field and induced currents, overall trends, change in Ohmic efficiency) remain largely valid, our results show re-inflation at late stages (several gigayears), in agreement with previous studies \citep{lopez16,komacek20}, and with observational hints \citep{hartman16,thorngren21}. We restricted our analysis to the main-sequence stage of a Sun-like star, leaving a more extensive parameter exploration to future studies. The evolving stellar irradiation further underscores the need to move beyond the simplifying assumption of a  constant $v_{\rm avg}$.

In this sense, future works need to consider with further detail the interplay between the deep-seated dynamo and the atmospheric induction with more detail. Another ingredient to potentially be taken into account in the future is the double-diffusive convection, which might become relevant when the convection is suppressed in highly heated HJs. In any case, our current findings have potential consequences on the elusive radio detectability of HJs \citep{stevens05,griessmeier07,zarka07}, which so far has only given, in the best case, unconfirmed claims (see e.g. \citealt{turner21,turner24}) despite multiple observational campaigns and promising theoretical predictions. Finally, the inferred values of the average velocity as a function of $M$ and $\Teq$ can be put in the context of population synthesis \citep{thorngren18}, considering also other variables (core size and composition, atmospheric modelling), which we took to be constant in this work and might introduce dispersion in the observed radii.

\begin{acknowledgements}
DV, SS, CSG, AEL and TA are supported by the European Research Council (ERC) under the European Union’s Horizon 2020 research and innovation program (ERC Starting Grant ``IMAGINE'' No. 948582, PI: DV) and  by the Spanish program Unidad de Excelencia María de Maeztu, awarded to the Institute of Space Sciences (ICE-CSIC), CEX2020-001058-M. CSG and AEL carried out this work within the framework of the doctoral program in Physics of the Universitat Aut\`{o}noma de Barcelona. We acknowledge the use the MareNostrum and SCAYLE supercomputers of the Spanish Supercomputing Network, via projects RES/BSC Call AECT-2023-2-0013 and RES/BSC Call AECT-2023-2-0034. RP thanks the Institute of Space Sciences for their kind hospitality during the time that part of this project was carried out. We thank Angelica Psaridis and Fabio del Sordo for useful comments.
\end{acknowledgements}

\bibliographystyle{aa}
\bibliography{references}

\begin{thebibliography}{141}
\expandafter\ifx\csname natexlab\endcsname\relax\def\natexlab#1{#1}\fi

\bibitem[{{Arras} \& {Bildsten}(2006)}]{arras06}
{Arras}, P. \& {Bildsten}, L. 2006, ApJ, 650, 394

\bibitem[{{Batygin} {et~al.}(2013){Batygin}, {Stanley}, \& {Stevenson}}]{batygin13}
{Batygin}, K., {Stanley}, S., \& {Stevenson}, D.~J. 2013, ApJ, 776, 53

\bibitem[{{Batygin} \& {Stevenson}(2010)}]{batygin10}
{Batygin}, K. \& {Stevenson}, D.~J. 2010, ApJl, 714, L238

\bibitem[{{Batygin} {et~al.}(2011){Batygin}, {Stevenson}, \& {Bodenheimer}}]{batygin11}
{Batygin}, K., {Stevenson}, D.~J., \& {Bodenheimer}, P.~H. 2011, ApJ, 738, 1

\bibitem[{Bell \& Cowan(2018)}]{bell18}
Bell, T.~J. \& Cowan, N.~B. 2018, The Astrophysical Journal Letters, 857, L20

\bibitem[{{Beltz} {et~al.}(2022){Beltz}, {Rauscher}, {Kempton}, {Malsky}, {Ochs}, {Arora}, \& {Savel}}]{beltz22}
{Beltz}, H., {Rauscher}, E., {Kempton}, E. M.~R., {et~al.} 2022, Astron. J., 164, 140

\bibitem[{{Benavides} {et~al.}(2022){Benavides}, {Burns}, {Gallet}, \& {Flierl}}]{benavides22}
{Benavides}, S.~J., {Burns}, K.~J., {Gallet}, B., \& {Flierl}, G.~R. 2022, ApJ, 938, 92

\bibitem[{{Bodenheimer} {et~al.}(2001){Bodenheimer}, {Lin}, \& {Mardling}}]{bodenheimer01}
{Bodenheimer}, P., {Lin}, D.~N.~C., \& {Mardling}, R.~A. 2001, ApJ, 548, 466

\bibitem[{Bonitz {et~al.}(2024)Bonitz, Vorberger, Bethkenhagen, Böhme, Ceperley, Filinov, Gawne, Graziani, Gregori, Hamann, Hansen, Holzmann, Hu, Kählert, Karasiev, Kleinschmidt, Kordts, Makait, Militzer, Moldabekov, Pierleoni, Preising, Ramakrishna, Redmer, Schwalbe, Svensson, \& Dornheim}]{bonitz24}
Bonitz, M., Vorberger, J., Bethkenhagen, M., {et~al.} 2024, Phys. Plasmas, 31, 110501

\bibitem[{{Brygoo} {et~al.}(2021){Brygoo}, {Loubeyre}, {Millot}, {Rygg}, {Celliers}, {Eggert}, {Jeanloz}, \& {Collins}}]{brygoo21}
{Brygoo}, S., {Loubeyre}, P., {Millot}, M., {et~al.} 2021, \nat, 593, 517

\bibitem[{{Burrows} {et~al.}(2007){Burrows}, {Hubeny}, {Budaj}, \& {Hubbard}}]{burrows07}
{Burrows}, A., {Hubeny}, I., {Budaj}, J., \& {Hubbard}, W.~B. 2007, ApJ, 661, 502

\bibitem[{{Cauley} {et~al.}(2019){Cauley}, {Shkolnik}, {Llama}, \& {Lanza}}]{cauley19}
{Cauley}, P.~W., {Shkolnik}, E.~L., {Llama}, J., \& {Lanza}, A.~F. 2019, Nature Astronomy, 3, 1128

\bibitem[{{Chabrier} \& {Baraffe}(2007)}]{chabrier07}
{Chabrier}, G. \& {Baraffe}, I. 2007, ApJl, 661, L81

\bibitem[{{Cho} {et~al.}(2008){Cho}, {Menou}, {Hansen}, \& {Seager}}]{cho08}
{Cho}, J. Y.~K., {Menou}, K., {Hansen}, B. M.~S., \& {Seager}, S. 2008, ApJ, 675, 817

\bibitem[{{Christensen} \& {Aubert}(2006)}]{christensen06}
{Christensen}, U.~R. \& {Aubert}, J. 2006, Geophysical Journal International, 166, 97

\bibitem[{{Christensen} {et~al.}(2009){Christensen}, {Holzwarth}, \& {Reiners}}]{christensen09}
{Christensen}, U.~R., {Holzwarth}, V., \& {Reiners}, A. 2009, \nat, 457, 167

\bibitem[{Connerney {et~al.}(2018)Connerney, Kotsiaros, Oliversen, Espley, Joergensen, Joergensen, Merayo, Herceg, Bloxham, Moore, Bolton, \& Levin}]{connerney18}
Connerney, J. E.~P., Kotsiaros, S., Oliversen, R.~J., {et~al.} 2018, Geophysical Research Letters, 45, 2590

\bibitem[{Connerney {et~al.}(2022)Connerney, Timmins, Oliversen, Espley, Joergensen, Kotsiaros, Joergensen, Merayo, Herceg, Bloxham, Moore, Mura, Moirano, Bolton, \& Levin}]{connerney22}
Connerney, J. E.~P., Timmins, S., Oliversen, R.~J., {et~al.} 2022, Journal of Geophysical Research: Planets, 127, e2021JE007055, e2021JE007055 2021JE007055

\bibitem[{{Debras} \& {Chabrier}(2019)}]{debras19}
{Debras}, F. \& {Chabrier}, G. 2019, ApJ, 872, 100

\bibitem[{{Dietrich} {et~al.}(2022){Dietrich}, {Kumar}, {Poser}, {French}, {Nettelmann}, {Redmer}, \& {Wicht}}]{dietrich22}
{Dietrich}, W., {Kumar}, S., {Poser}, A.~J., {et~al.} 2022, \mnras, 517, 3113

\bibitem[{{Dobbs-Dixon} \& {Lin}(2008)}]{dobbs08}
{Dobbs-Dixon}, I. \& {Lin}, D.~N.~C. 2008, ApJ, 673, 513

\bibitem[{{Dotter} {et~al.}(2008){Dotter}, {Chaboyer}, {Jevremovi{\'c}}, {Kostov}, {Baron}, \& {Ferguson}}]{dotter08}
{Dotter}, A., {Chaboyer}, B., {Jevremovi{\'c}}, D., {et~al.} 2008, ApJS, 178, 89

\bibitem[{{Draine} {et~al.}(1983){Draine}, {Roberge}, \& {Dalgarno}}]{draine83}
{Draine}, B.~T., {Roberge}, W.~G., \& {Dalgarno}, A. 1983, ApJ, 264, 485

\bibitem[{Duarte {et~al.}(2013)Duarte, Gastine, \& Wicht}]{duarte13}
Duarte, L.~D., Gastine, T., \& Wicht, J. 2013, Physics of the Earth and Planetary Interiors, 222, 22

\bibitem[{Duarte {et~al.}(2018)Duarte, Wicht, \& Gastine}]{duarte18}
Duarte, L.~D., Wicht, J., \& Gastine, T. 2018, Icarus, 299, 206

\bibitem[{{Elias-L{\'o}pez} {et~al.}(2025{\natexlab{a}}){Elias-L{\'o}pez}, {Cantiello}, {Vigan{\`o}}, {Del Sordo}, {Kaur}, \& {Soriano-Guerrero}}]{elias25b}
{Elias-L{\'o}pez}, A., {Cantiello}, M., {Vigan{\`o}}, D., {et~al.} 2025{\natexlab{a}}, arXiv e-prints, arXiv:2507.05202

\bibitem[{{Elias-L{\'o}pez} {et~al.}(2025{\natexlab{b}}){Elias-L{\'o}pez}, {Del Sordo}, {Vigan{\`o}}, {Soriano-Guerrero}, {Akg{\"u}n}, {Reboul-Salze}, \& {Cantiello}}]{elias25}
{Elias-L{\'o}pez}, A., {Del Sordo}, F., {Vigan{\`o}}, D., {et~al.} 2025{\natexlab{b}}, A\&A, 696, A161

\bibitem[{{Enoch} {et~al.}(2012){Enoch}, {Collier Cameron}, \& {Horne}}]{enoch12}
{Enoch}, B., {Collier Cameron}, A., \& {Horne}, K. 2012, A\&A, 540, A99

\bibitem[{{Fortney} {et~al.}(2021){Fortney}, {Dawson}, \& {Komacek}}]{fortney21}
{Fortney}, J.~J., {Dawson}, R.~I., \& {Komacek}, T.~D. 2021, Journal of Geophysical Research (Planets), 126, e06629

\bibitem[{{Fortney} {et~al.}(2008){Fortney}, {Lodders}, {Marley}, \& {Freedman}}]{fortney08}
{Fortney}, J.~J., {Lodders}, K., {Marley}, M.~S., \& {Freedman}, R.~S. 2008, ApJ, 678, 1419

\bibitem[{{Fortney} {et~al.}(2007){Fortney}, {Marley}, \& {Barnes}}]{fortney07}
{Fortney}, J.~J., {Marley}, M.~S., \& {Barnes}, J.~W. 2007, ApJ, 659, 1661

\bibitem[{{French} {et~al.}(2012){French}, {Becker}, {Lorenzen}, {Nettelmann}, {Bethkenhagen}, {Wicht}, \& {Redmer}}]{french12}
{French}, M., {Becker}, A., {Lorenzen}, W., {et~al.} 2012, ApJS, 202, 5

\bibitem[{{Gandhi} \& {Madhusudhan}(2019)}]{gandhi19}
{Gandhi}, S. \& {Madhusudhan}, N. 2019, \mnras, 485, 5817

\bibitem[{{Gandhi} {et~al.}(2020){Gandhi}, {Madhusudhan}, \& {Mandell}}]{gandhi20}
{Gandhi}, S., {Madhusudhan}, N., \& {Mandell}, A. 2020, Astron. J., 159, 232

\bibitem[{Gastine \& Wicht(2012)}]{gastine12}
Gastine, T. \& Wicht, J. 2012, Icarus, 219, 428

\bibitem[{Gastine \& Wicht(2021)}]{gastine21}
Gastine, T. \& Wicht, J. 2021, Icarus, 368, 114514

\bibitem[{{Gastine} {et~al.}(2014){Gastine}, {Wicht}, {Duarte}, {Heimpel}, \& {Becker}}]{gastine14}
{Gastine}, T., {Wicht}, J., {Duarte}, L.~D.~V., {Heimpel}, M., \& {Becker}, A. 2014, \grl, 41, 5410

\bibitem[{{Ginzburg} \& {Sari}(2015)}]{ginzburg15}
{Ginzburg}, S. \& {Sari}, R. 2015, ApJ, 803, 111

\bibitem[{{Ginzburg} \& {Sari}(2016)}]{ginzburg16}
{Ginzburg}, S. \& {Sari}, R. 2016, ApJ, 819, 116

\bibitem[{{Grie{\ss}meier} {et~al.}(2007){Grie{\ss}meier}, {Zarka}, \& {Spreeuw}}]{griessmeier07}
{Grie{\ss}meier}, J.~M., {Zarka}, P., \& {Spreeuw}, H. 2007, A\&A, 475, 359

\bibitem[{{Guillot}(2010)}]{guillot10}
{Guillot}, T. 2010, A\&A, 520, A27

\bibitem[{{Guillot} {et~al.}(1996){Guillot}, {Burrows}, {Hubbard}, {Lunine}, \& {Saumon}}]{guillot96}
{Guillot}, T., {Burrows}, A., {Hubbard}, W.~B., {Lunine}, J.~I., \& {Saumon}, D. 1996, ApJl, 459, L35

\bibitem[{Gómez-Pérez {et~al.}(2010)Gómez-Pérez, Heimpel, \& Wicht}]{gomezperezetal2010}
Gómez-Pérez, N., Heimpel, M., \& Wicht, J. 2010, Physics of the Earth and Planetary Interiors, 181, 42

\bibitem[{{Hardy} {et~al.}(2023){Hardy}, {Charbonneau}, \& {Cumming}}]{hardy23}
{Hardy}, R., {Charbonneau}, P., \& {Cumming}, A. 2023, ApJ, 959, 41

\bibitem[{{Hardy} {et~al.}(2025){Hardy}, {Charbonneau}, \& {Cumming}}]{hardy25}
{Hardy}, R., {Charbonneau}, P., \& {Cumming}, A. 2025, ApJ, 978, 149

\bibitem[{{Hardy} {et~al.}(2022){Hardy}, {Cumming}, \& {Charbonneau}}]{hardy22}
{Hardy}, R., {Cumming}, A., \& {Charbonneau}, P. 2022, ApJ, 940, 123

\bibitem[{{Hartman} {et~al.}(2016){Hartman}, {Bakos}, {Bhatti}, {Penev}, {Bieryla}, {Latham}, {Kov{\'a}cs}, {Torres}, {Csubry}, {de Val-Borro}, {Buchhave}, {Kov{\'a}cs}, {Quinn}, {Howard}, {Isaacson}, {Fulton}, {Everett}, {Esquerdo}, {B{\'e}ky}, {Szklenar}, {Falco}, {Santerne}, {Boisse}, {H{\'e}brard}, {Burrows}, {L{\'a}z{\'a}r}, {Papp}, \& {S{\'a}ri}}]{hartman16}
{Hartman}, J.~D., {Bakos}, G.~{\'A}., {Bhatti}, W., {et~al.} 2016, Astron. J., 152, 182

\bibitem[{{Heng} {et~al.}(2011){Heng}, {Menou}, \& {Phillipps}}]{heng11}
{Heng}, K., {Menou}, K., \& {Phillipps}, P.~J. 2011, \mnras, 413, 2380

\bibitem[{{Heng} \& {Showman}(2015)}]{heng15}
{Heng}, K. \& {Showman}, A.~P. 2015, Annual Review of Earth and Planetary Sciences, 43, 509

\bibitem[{Holst {et~al.}(2011)Holst, French, \& Redmer}]{holst11}
Holst, B., French, M., \& Redmer, R. 2011, Phys. Rev. B, 83, 235120

\bibitem[{{Huang} \& {Cumming}(2012)}]{huang12}
{Huang}, X. \& {Cumming}, A. 2012, ApJ, 757, 47

\bibitem[{{Jermyn} {et~al.}(2023){Jermyn}, {Bauer}, {Schwab}, {Farmer}, {Ball}, {Bellinger}, {Dotter}, {Joyce}, {Marchant}, {Mombarg}, {Wolf}, {Sunny Wong}, {Cinquegrana}, {Farrell}, {Smolec}, {Thoul}, {Cantiello}, {Herwig}, {Toloza}, {Bildsten}, {Townsend}, \& {Timmes}}]{jermyn23}
{Jermyn}, A.~S., {Bauer}, E.~B., {Schwab}, J., {et~al.} 2023, ApJS, 265, 15

\bibitem[{Jones {et~al.}(2011)Jones, Boronski, Brun, Glatzmaier, Gastine, Miesch, \& Wicht}]{jones11}
Jones, C., Boronski, P., Brun, A., {et~al.} 2011, Icarus, 216, 120

\bibitem[{{Kao} {et~al.}(2016){Kao}, {Hallinan}, {Pineda}, {Escala}, {Burgasser}, {Bourke}, \& {Stevenson}}]{kao16}
{Kao}, M.~M., {Hallinan}, G., {Pineda}, J.~S., {et~al.} 2016, ApJ, 818, 24

\bibitem[{{Kao} {et~al.}(2018){Kao}, {Hallinan}, {Pineda}, {Stevenson}, \& {Burgasser}}]{kao18}
{Kao}, M.~M., {Hallinan}, G., {Pineda}, J.~S., {Stevenson}, D., \& {Burgasser}, A. 2018, ApJS, 237, 25

\bibitem[{{Kataria} {et~al.}(2015){Kataria}, {Showman}, {Fortney}, {Stevenson}, {Line}, {Kreidberg}, {Bean}, \& {D{\'e}sert}}]{kataria15}
{Kataria}, T., {Showman}, A.~P., {Fortney}, J.~J., {et~al.} 2015, ApJ, 801, 86

\bibitem[{{Kilmetis} {et~al.}(2024){Kilmetis}, {Vidotto}, {Allan}, \& {Kubyshkina}}]{kilmetis24}
{Kilmetis}, K., {Vidotto}, A.~A., {Allan}, A., \& {Kubyshkina}, D. 2024, \mnras, 535, 3646

\bibitem[{{Knierim} {et~al.}(2022){Knierim}, {Batygin}, \& {Bitsch}}]{knierim22}
{Knierim}, H., {Batygin}, K., \& {Bitsch}, B. 2022, A\&A, 658, L7

\bibitem[{{Koll} \& {Komacek}(2018)}]{koll18}
{Koll}, D. D.~B. \& {Komacek}, T.~D. 2018, ApJ, 853, 133

\bibitem[{{Komacek} {et~al.}(2022){Komacek}, {Tan}, {Gao}, \& {Lee}}]{komacek22}
{Komacek}, T.~D., {Tan}, X., {Gao}, P., \& {Lee}, E. K.~H. 2022, ApJ, 934, 79

\bibitem[{{Komacek} {et~al.}(2020){Komacek}, {Thorngren}, {Lopez}, \& {Ginzburg}}]{komacek20}
{Komacek}, T.~D., {Thorngren}, D.~P., {Lopez}, E.~D., \& {Ginzburg}, S. 2020, ApJ, 893, 36

\bibitem[{{Komacek} \& {Youdin}(2017)}]{komacek17}
{Komacek}, T.~D. \& {Youdin}, A.~N. 2017, ApJ, 844, 94

\bibitem[{{Koskinen} {et~al.}(2014){Koskinen}, {Yelle}, {Lavvas}, \& {Y-K. Cho}}]{koskinen14}
{Koskinen}, T.~T., {Yelle}, R.~V., {Lavvas}, P., \& {Y-K. Cho}, J. 2014, ApJ, 796, 16

\bibitem[{{Kumar} {et~al.}(2021){Kumar}, {Poser}, {Sch{\"o}ttler}, {Kleinschmidt}, {Dietrich}, {Wicht}, {French}, \& {Redmer}}]{kumar21}
{Kumar}, S., {Poser}, A.~J., {Sch{\"o}ttler}, M., {et~al.} 2021, Phys. Rev. E, 103, 063203

\bibitem[{{Kurokawa} \& {Nakamoto}(2014)}]{kurokawa14}
{Kurokawa}, H. \& {Nakamoto}, T. 2014, ApJ, 783, 54

\bibitem[{{Laughlin} {et~al.}(2011){Laughlin}, {Crismani}, \& {Adams}}]{laughlin11}
{Laughlin}, G., {Crismani}, M., \& {Adams}, F.~C. 2011, ApJl, 729, L7

\bibitem[{{Lavie} {et~al.}(2017){Lavie}, {Ehrenreich}, {Bourrier}, {Lecavelier des Etangs}, {Vidal-Madjar}, {Delfosse}, {Gracia Berna}, {Heng}, {Thomas}, {Udry}, \& {Wheatley}}]{lavie17}
{Lavie}, B., {Ehrenreich}, D., {Bourrier}, V., {et~al.} 2017, A\&A, 605, L7

\bibitem[{{Lazovik}(2023)}]{lazovik23}
{Lazovik}, Y.~A. 2023, \mnras, 520, 3749

\bibitem[{{Leconte} \& {Chabrier}(2013)}]{leconte13}
{Leconte}, J. \& {Chabrier}, G. 2013, Nature Geoscience, 6, 347

\bibitem[{{Li} \& {Goodman}(2010)}]{li10}
{Li}, J. \& {Goodman}, J. 2010, ApJ, 725, 1146

\bibitem[{{Liu} {et~al.}(2008){Liu}, {Goldreich}, \& {Stevenson}}]{liu08}
{Liu}, J., {Goldreich}, P.~M., \& {Stevenson}, D.~J. 2008, \icarus, 196, 653

\bibitem[{{Lopez} \& {Fortney}(2016)}]{lopez16}
{Lopez}, E.~D. \& {Fortney}, J.~J. 2016, ApJ, 818, 4

\bibitem[{{Lorenzen} {et~al.}(2009){Lorenzen}, {Holst}, \& {Redmer}}]{lorenzen09}
{Lorenzen}, W., {Holst}, B., \& {Redmer}, R. 2009, \prl, 102, 115701

\bibitem[{{Lowes}(1974)}]{lowes74}
{Lowes}, F.~J. 1974, Geophysical Journal, 36, 717

\bibitem[{{Menou}(2012{\natexlab{a}})}]{menou12}
{Menou}, K. 2012{\natexlab{a}}, ApJ, 745, 138

\bibitem[{{Menou}(2012{\natexlab{b}})}]{menou12b}
{Menou}, K. 2012{\natexlab{b}}, ApJl, 754, L9

\bibitem[{{Morales} {et~al.}(2013){Morales}, {Hamel}, {Caspersen}, \& {Schwegler}}]{morales13}
{Morales}, M.~A., {Hamel}, S., {Caspersen}, K., \& {Schwegler}, E. 2013, \prb, 87, 174105

\bibitem[{{Owen} \& {Wu}(2016)}]{owen16}
{Owen}, J.~E. \& {Wu}, Y. 2016, ApJ, 817, 107

\bibitem[{{Parmentier} {et~al.}(2013){Parmentier}, {Showman}, \& {Lian}}]{parmentier13}
{Parmentier}, V., {Showman}, A.~P., \& {Lian}, Y. 2013, A\&A, 558, A91

\bibitem[{{Paxton} {et~al.}(2011){Paxton}, {Bildsten}, {Dotter}, {Herwig}, {Lesaffre}, \& {Timmes}}]{paxton11}
{Paxton}, B., {Bildsten}, L., {Dotter}, A., {et~al.} 2011, ApJS, 192, 3

\bibitem[{{Paxton} {et~al.}(2013){Paxton}, {Cantiello}, {Arras}, {Bildsten}, {Brown}, {Dotter}, {Mankovich}, {Montgomery}, {Stello}, {Timmes}, \& {Townsend}}]{paxton13}
{Paxton}, B., {Cantiello}, M., {Arras}, P., {et~al.} 2013, ApJS, 208, 4

\bibitem[{{Paxton} {et~al.}(2015){Paxton}, {Marchant}, {Schwab}, {Bauer}, {Bildsten}, {Cantiello}, {Dessart}, {Farmer}, {Hu}, {Langer}, {Townsend}, {Townsley}, \& {Timmes}}]{paxton15}
{Paxton}, B., {Marchant}, P., {Schwab}, J., {et~al.} 2015, ApJS, 220, 15

\bibitem[{{Paxton} {et~al.}(2018){Paxton}, {Schwab}, {Bauer}, {Bildsten}, {Blinnikov}, {Duffell}, {Farmer}, {Goldberg}, {Marchant}, {Sorokina}, {Thoul}, {Townsend}, \& {Timmes}}]{paxton18}
{Paxton}, B., {Schwab}, J., {Bauer}, E.~B., {et~al.} 2018, ApJS, 234, 34

\bibitem[{{Paxton} {et~al.}(2019){Paxton}, {Smolec}, {Schwab}, {Gautschy}, {Bildsten}, {Cantiello}, {Dotter}, {Farmer}, {Goldberg}, {Jermyn}, {Kanbur}, {Marchant}, {Thoul}, {Townsend}, {Wolf}, {Zhang}, \& {Timmes}}]{paxton19}
{Paxton}, B., {Smolec}, R., {Schwab}, J., {et~al.} 2019, ApJS, 243, 10

\bibitem[{{Perez-Becker} \& {Showman}(2013)}]{perez13}
{Perez-Becker}, D. \& {Showman}, A.~P. 2013, ApJ, 776, 134

\bibitem[{{Perna} {et~al.}(2012){Perna}, {Heng}, \& {Pont}}]{perna12}
{Perna}, R., {Heng}, K., \& {Pont}, F. 2012, ApJ, 751, 59

\bibitem[{{Perna} {et~al.}(2010{\natexlab{a}}){Perna}, {Menou}, \& {Rauscher}}]{perna10a}
{Perna}, R., {Menou}, K., \& {Rauscher}, E. 2010{\natexlab{a}}, ApJ, 719, 1421

\bibitem[{{Perna} {et~al.}(2010{\natexlab{b}}){Perna}, {Menou}, \& {Rauscher}}]{perna10b}
{Perna}, R., {Menou}, K., \& {Rauscher}, E. 2010{\natexlab{b}}, ApJ, 724, 313

\bibitem[{{Ramakrishna} {et~al.}(2024){Ramakrishna}, {Lokamani}, \& {Cangi}}]{ramakrishna24}
{Ramakrishna}, K., {Lokamani}, M., \& {Cangi}, A. 2024, Electronic Structure, 6, 045008

\bibitem[{{Rauscher} \& {Menou}(2012)}]{rauscher12}
{Rauscher}, E. \& {Menou}, K. 2012, ApJ, 750, 96

\bibitem[{{Rauscher} \& {Menou}(2013)}]{rauscher13}
{Rauscher}, E. \& {Menou}, K. 2013, ApJ, 764, 103

\bibitem[{{Redmer}(1997)}]{redmer97}
{Redmer}, R. 1997, \physrep, 282, 35

\bibitem[{{Reiners} {et~al.}(2009){Reiners}, {Basri}, \& {Christensen}}]{reiners09}
{Reiners}, A., {Basri}, G., \& {Christensen}, U.~R. 2009, ApJ, 697, 373

\bibitem[{{Reiners} \& {Christensen}(2010)}]{reiners10}
{Reiners}, A. \& {Christensen}, U.~R. 2010, A\&A, 522, A13

\bibitem[{{Ribas}(2010)}]{ribas10}
{Ribas}, I. 2010, in IAU Symposium, Vol. 264, Solar and Stellar Variability: Impact on Earth and Planets, ed. A.~G. {Kosovichev}, A.~H. {Andrei}, \& J.-P. {Rozelot}, 3--18

\bibitem[{{Rogers} \& {Komacek}(2014)}]{rogers14b}
{Rogers}, T.~M. \& {Komacek}, T.~D. 2014, ApJ, 794, 132

\bibitem[{{Rogers} \& {McElwaine}(2017)}]{rogers17}
{Rogers}, T.~M. \& {McElwaine}, J.~N. 2017, ApJl, 841, L26

\bibitem[{{Rogers} \& {Showman}(2014)}]{rogers14a}
{Rogers}, T.~M. \& {Showman}, A.~P. 2014, ApJl, 782, L4

\bibitem[{{Salz} {et~al.}(2015){Salz}, {Schneider}, {Czesla}, \& {Schmitt}}]{salz15}
{Salz}, M., {Schneider}, P.~C., {Czesla}, S., \& {Schmitt}, J.~H.~M.~M. 2015, A\&A, 576, A42

\bibitem[{{S{\'a}nchez-Lavega}(2004)}]{sanchezlavega04}
{S{\'a}nchez-Lavega}, A. 2004, ApJl, 609, L87

\bibitem[{{Sarkis} {et~al.}(2021){Sarkis}, {Mordasini}, {Henning}, {Marleau}, \& {Molli{\`e}re}}]{sarkis21}
{Sarkis}, P., {Mordasini}, C., {Henning}, T., {Marleau}, G.~D., \& {Molli{\`e}re}, P. 2021, A\&A, 645, A79

\bibitem[{{Saumon} {et~al.}(1995){Saumon}, {Chabrier}, \& {van Horn}}]{saumon95}
{Saumon}, D., {Chabrier}, G., \& {van Horn}, H.~M. 1995, ApJS, 99, 713

\bibitem[{{Savel} {et~al.}(2024){Savel}, {Beltz}, {Komacek}, {Tsai}, \& {Kempton}}]{savel24}
{Savel}, A.~B., {Beltz}, H., {Komacek}, T.~D., {Tsai}, S.-M., \& {Kempton}, E. M.~R. 2024, ApJl, 969, L27

\bibitem[{{Schubert} \& {Soderlund}(2011)}]{schubert11}
{Schubert}, G. \& {Soderlund}, K.~M. 2011, Physics of the Earth and Planetary Interiors, 187, 92

\bibitem[{{Sengupta} \& {Sengupta}(2023)}]{sengupta23}
{Sengupta}, S. \& {Sengupta}, S. 2023, \na, 100, 101987

\bibitem[{{Sestovic} {et~al.}(2018){Sestovic}, {Demory}, \& {Queloz}}]{sestovic18}
{Sestovic}, M., {Demory}, B.-O., \& {Queloz}, D. 2018, A\&A, 616, A76

\bibitem[{{Showman} {et~al.}(2009){Showman}, {Fortney}, {Lian}, {Marley}, {Freedman}, {Knutson}, \& {Charbonneau}}]{showman09}
{Showman}, A.~P., {Fortney}, J.~J., {Lian}, Y., {et~al.} 2009, ApJ, 699, 564

\bibitem[{{Showman} \& {Guillot}(2002)}]{showman02}
{Showman}, A.~P. \& {Guillot}, T. 2002, A\&A, 385, 166

\bibitem[{{Showman} {et~al.}(2015){Showman}, {Lewis}, \& {Fortney}}]{showman15}
{Showman}, A.~P., {Lewis}, N.~K., \& {Fortney}, J.~J. 2015, ApJ, 801, 95

\bibitem[{{Smith} {et~al.}(2018){Smith}, {Fratanduono}, {Braun}, {Duffy}, {Wicks}, {Celliers}, {Ali}, {Fernandez-Pa{\~n}ella}, {Kraus}, {Swift}, {Collins}, \& {Eggert}}]{smith18}
{Smith}, R.~F., {Fratanduono}, D.~E., {Braun}, D.~G., {et~al.} 2018, Nature Astronomy, 2, 452

\bibitem[{{Soriano-Guerrero} {et~al.}(2023){Soriano-Guerrero}, {Vigan{\`o}}, {Perna}, {Akg{\"u}n}, \& {Palenzuela}}]{soriano23}
{Soriano-Guerrero}, C., {Vigan{\`o}}, D., {Perna}, R., {Akg{\"u}n}, T., \& {Palenzuela}, C. 2023, \mnras, 525, 626

\bibitem[{{Soriano-Guerrero} {et~al.}(2025){Soriano-Guerrero}, {Vigan{\`o}}, {Perna}, {Elias-L{\'o}pez}, \& {Beltz}}]{soriano25}
{Soriano-Guerrero}, C., {Vigan{\`o}}, D., {Perna}, R., {Elias-L{\'o}pez}, A., \& {Beltz}, H. 2025, \mnras, 540, 1827

\bibitem[{{Spiegel} \& {Burrows}(2013)}]{spiegel13}
{Spiegel}, D.~S. \& {Burrows}, A. 2013, ApJ, 772, 76

\bibitem[{{Stevens}(2005)}]{stevens05}
{Stevens}, I.~R. 2005, \mnras, 356, 1053

\bibitem[{{Stevenson}(1980)}]{stevenson80}
{Stevenson}, D.~J. 1980, Science, 208, 746

\bibitem[{{Stevenson} \& {Salpeter}(1977{\natexlab{a}})}]{stevenson77a}
{Stevenson}, D.~J. \& {Salpeter}, E.~E. 1977{\natexlab{a}}, ApJS, 35, 239

\bibitem[{{Stevenson} \& {Salpeter}(1977{\natexlab{b}})}]{stevenson77b}
{Stevenson}, D.~J. \& {Salpeter}, E.~E. 1977{\natexlab{b}}, ApJS, 35, 221

\bibitem[{{Thorngren} {et~al.}(2019){Thorngren}, {Gao}, \& {Fortney}}]{thorngren19}
{Thorngren}, D., {Gao}, P., \& {Fortney}, J.~J. 2019, ApJl, 884, L6

\bibitem[{{Thorngren}(2024)}]{thorngren24}
{Thorngren}, D.~P. 2024, arXiv e-prints, arXiv:2405.05307

\bibitem[{{Thorngren} \& {Fortney}(2018)}]{thorngren18}
{Thorngren}, D.~P. \& {Fortney}, J.~J. 2018, Astron. J., 155, 214

\bibitem[{{Thorngren} {et~al.}(2021){Thorngren}, {Fortney}, {Lopez}, {Berger}, \& {Huber}}]{thorngren21}
{Thorngren}, D.~P., {Fortney}, J.~J., {Lopez}, E.~D., {Berger}, T.~A., \& {Huber}, D. 2021, ApJl, 909, L16

\bibitem[{{Thorngren} {et~al.}(2016){Thorngren}, {Fortney}, {Murray-Clay}, \& {Lopez}}]{thorngren16}
{Thorngren}, D.~P., {Fortney}, J.~J., {Murray-Clay}, R.~A., \& {Lopez}, E.~D. 2016, ApJ, 831, 64

\bibitem[{{Thorngren} {et~al.}(2023){Thorngren}, {Lee}, \& {Lopez}}]{thorngren23}
{Thorngren}, D.~P., {Lee}, E.~J., \& {Lopez}, E.~D. 2023, ApJl, 945, L36

\bibitem[{{Tremblin} {et~al.}(2017){Tremblin}, {Chabrier}, {Mayne}, {Amundsen}, {Baraffe}, {Debras}, {Drummond}, {Manners}, \& {Fromang}}]{tremblin17}
{Tremblin}, P., {Chabrier}, G., {Mayne}, N.~J., {et~al.} 2017, ApJ, 841, 30

\bibitem[{{Tsang} \& {Jones}(2020)}]{tsang20}
{Tsang}, Y.-K. \& {Jones}, C.~A. 2020, Earth and Planetary Science Letters, 530, 115879

\bibitem[{{Turner} {et~al.}(2024){Turner}, {Grie{\ss}meier}, {Zarka}, {Zhang}, \& {Mauduit}}]{turner24}
{Turner}, J.~D., {Grie{\ss}meier}, J.-M., {Zarka}, P., {Zhang}, X., \& {Mauduit}, E. 2024, A\&A, 688, A66

\bibitem[{{Turner} {et~al.}(2021){Turner}, {Zarka}, {Grie{\ss}meier}, {Lazio}, {Cecconi}, {Emilio Enriquez}, {Girard}, {Jayawardhana}, {Lamy}, {Nichols}, \& {de Pater}}]{turner21}
{Turner}, J.~D., {Zarka}, P., {Grie{\ss}meier}, J.-M., {et~al.} 2021, A\&A, 645, A59

\bibitem[{{Tyler} {et~al.}(2024){Tyler}, {Petigura}, {Oklop{\v{c}}i{\'c}}, \& {David}}]{tyler24}
{Tyler}, D., {Petigura}, E.~A., {Oklop{\v{c}}i{\'c}}, A., \& {David}, T.~J. 2024, ApJ, 960, 123

\bibitem[{{Verma} {et~al.}(2019){Verma}, {Raodeo}, {Basu}, {Silva Aguirre}, {Mazumdar}, {Mosumgaard}, {Lund}, \& {Ranadive}}]{verma19}
{Verma}, K., {Raodeo}, K., {Basu}, S., {et~al.} 2019, \mnras, 483, 4678

\bibitem[{{Wahl} {et~al.}(2017){Wahl}, {Hubbard}, {Militzer}, {Guillot}, {Miguel}, {Movshovitz}, {Kaspi}, {Helled}, {Reese}, {Galanti}, {Levin}, {Connerney}, \& {Bolton}}]{wahl17}
{Wahl}, S.~M., {Hubbard}, W.~B., {Militzer}, B., {et~al.} 2017, \grl, 44, 4649

\bibitem[{{Wang} {et~al.}(2015){Wang}, {Fischer}, {Horch}, \& {Huang}}]{wang15}
{Wang}, J., {Fischer}, D.~A., {Horch}, E.~P., \& {Huang}, X. 2015, ApJ, 799, 229

\bibitem[{{Wazny} \& {Menou}(2025)}]{wazny25}
{Wazny}, M. \& {Menou}, K. 2025, ApJl, 979, L31

\bibitem[{{Wicht} {et~al.}(2019{\natexlab{a}}){Wicht}, {Gastine}, \& {Duarte}}]{wicht19b}
{Wicht}, J., {Gastine}, T., \& {Duarte}, L.~D.~V. 2019{\natexlab{a}}, Journal of Geophysical Research (Planets), 124, 837

\bibitem[{{Wicht} {et~al.}(2019{\natexlab{b}}){Wicht}, {Gastine}, {Duarte}, \& {Dietrich}}]{wicht19a}
{Wicht}, J., {Gastine}, T., {Duarte}, L.~D.~V., \& {Dietrich}, W. 2019{\natexlab{b}}, A\&A, 629, A125

\bibitem[{{Wu} \& {Lithwick}(2013)}]{wu13}
{Wu}, Y. \& {Lithwick}, Y. 2013, ApJ, 763, 13

\bibitem[{Yadav {et~al.}(2022)Yadav, Cao, \& Bloxham}]{yadav22}
Yadav, R.~K., Cao, H., \& Bloxham, J. 2022, The Astrophysical Journal, 940, 185

\bibitem[{{Yadav} {et~al.}(2013){Yadav}, {Gastine}, {Christensen}, \& {Duarte}}]{yadav13}
{Yadav}, R.~K., {Gastine}, T., {Christensen}, U.~R., \& {Duarte}, L. D.~V. 2013, ApJ, 774, 6

\bibitem[{{Yadav} \& {Thorngren}(2017)}]{yadav17}
{Yadav}, R.~K. \& {Thorngren}, D.~P. 2017, ApJl, 849, L12

\bibitem[{{Y{\i}ld{\i}z} {et~al.}(2024){Y{\i}ld{\i}z}, {{\c{C}}elik Orhan}, {{\"O}rtel}, \& {{\c{C}}ak{\i}r}}]{yildiz24}
{Y{\i}ld{\i}z}, M., {{\c{C}}elik Orhan}, Z., {{\"O}rtel}, S., \& {{\c{C}}ak{\i}r}, T. 2024, \mnras, 528, 6881

\bibitem[{{Youdin} \& {Mitchell}(2010)}]{youdin10}
{Youdin}, A.~N. \& {Mitchell}, J.~L. 2010, ApJ, 721, 1113

\bibitem[{{Zarka}(2007)}]{zarka07}
{Zarka}, P. 2007, \planss, 55, 598

\end{thebibliography}

\begin{appendix}
    
\section{On the different scaling laws}\label{app:scaling_laws}

Some words of caution must be spent about the applicability of the dynamo field scaling laws, used here and in several works.
The \cite{christensen09} scaling law employed in this study, eq.~(\ref{eq:bdyn_C09}), has seen to hold in a broad range of masses, from Earth and Jupiter to main-sequence stars, with a $\sim 60\%$ scattering in the proportionality pre-coefficient of the scaling laws \citep{reiners09}. However, we noticed that several brown dwarfs are outliers in both directions: surprisingly low magnetic field for infant, accreting ones \citep{reiners09}, and high (kG) values for rapidly rotating ones, as inferred from coherent, polarised radio emission at GHz \citep{kao16,kao18}. Moreover, \cite{christensen09} explicitly discarded Mercury, Saturn, Uranus and Neptune from the scaling laws, substantially due to the likely absence of a thick convective region. Therefore, despite the remarkable range of heat flux and magnetic field range covered by the correlation, the planetary application is inferred to be non-trivial in the original work, and needs further considerations.

Furthermore, \cite{reiners09,reiners10} cast the original scaling law by \cite{christensen09}, eq.~(\ref{eq:bdyn_C09}), in a more direct relation with the observables $M$, $R$ and with the internal luminosity $L$, calibrating this relation for models of a range of fully convective cool stars:
    \begin{equation}
    B_{\rm dyn}^{R09} \simeq 4.8 \times 10^3 \left( \frac{(M/M_\odot) (L_{\rm int}/L_\odot)^2}{(R/R_\odot)^7} \right)^{1/6} \, {\rm G}~,
    \label{eq:bdyn_R09}
    \end{equation}
where $M_\odot, L_\odot, R_\odot$ are the solar mass, luminosity and radius.

In either version of the scaling law, the HJ case presents ambiguities in defining the region over which evaluating the integral in eq.~(\ref{eq:bdyn_C09}), or the depth at which $L_{\rm int}$ is evaluated in the simplified version (\ref{eq:bdyn_R09}). However, the original \cite{christensen09} version is much less sensitive to such ambiguities, since it relies on a density-weighted integral, while the latter considers a single value of $L_{\rm int}$ and assumes that the relevant parameter is the planetary radius $R$, which can be conceptually not justified in inflated HJs which present a relevant difference between $R$ and $R_{\rm dyn}$. Related to this, a main point to consider is that, in HJs with internal additional heating in models like ours or \cite{batygin10,batygin11,wu13,knierim22}, the internal luminosity increases steeply outwards, in a region which is mostly convective, but is way too poorly conductive to host the dynamo. This is at contrast with the models against which the scaling laws were calibrated, which have a fairly constant value of the internal luminosity, since there are no relevant heating sources concentrated in the outer convective regions. Therefore, the results for HJs depend on the specific assumption about the range over which we perform the integral in eq.~(\ref{eq:bdyn_C09}), or the value of $L_{\rm int}$ (in eq.~\ref{eq:bdyn_R09}). Moreover, in the latter relation, for highly inflated planets, taking $R$ as the planetary or dynamo surface radius can easily bring a factor $\sim 2$ difference compared to the dynamo radius, at contrast with Jupiter.

\begin{figure}
\centerline{\includegraphics[width=\hsize]{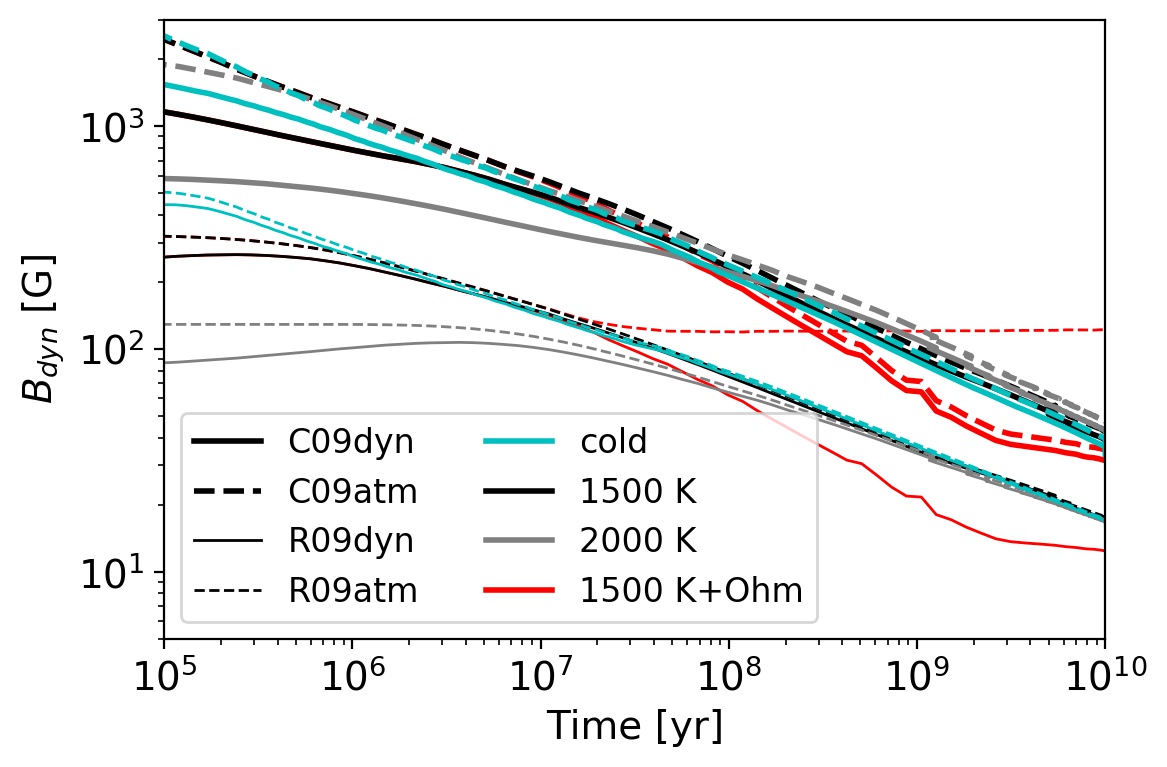}}
\caption{Magnetic field estimates, for a $M=1~\Mj$ planet, with $\Teq=300$ K (a cold Jupiter, cyan), 1500 K (black), and 2000 K (grey), and for $\Teq=1500$ K plus Ohmic heating with $v_{\rm avg}=3$ km/s, but fixing $B_{\rm atm}=10$ G (red, see text for the motivation). The different line styles correspond to the estimates by \cite{christensen09} (thick lines), eq.~(\ref{eq:bdyn_C09}) with $f_{\rm ohm}=0.5$, using the integral in the nominal dynamo region only ($p>p_{\rm dyn}=10^6$ bar, solid lines) or in the entire convective region (dashes), and by \cite{reiners09}, eq.~(\ref{eq:bdyn_R09}) (thin lines), using the internal luminosity at $p=10^6$ bar (solid) or at $p=10^3$ bar (dashed).}
\label{fig:B_estimates}
\end{figure}

In Fig.~\ref{fig:B_estimates} we show the different magnetic field estimates $B_{\rm dyn}$, for a $M=1~\Mj$ planet. We consider four cases: $\Teq=300,1500,2000$ K without additional heating ($\epsilon_{\rm j}=0$), and for $\Teq=1500$ K plus Ohmic heating. The latter is run with $v_{\rm avg}=3$ km/s, but fixing $B_{\rm atm}=10$ G in the $J$ normalisation, eq.~(\ref{eq:j_atm}), instead of evolving it with the same estimate of $B_{\rm dyn}$ via eq.~(\ref{eq:bdip}), in order to decouple the differences in the estimates from the feedback of $B_{\rm dyn}$ on the induced $J$ and on the cooling, discussed in the main text.

Several things must be noticed. First of all, large differences are seen at very early ages, which is related to the discussion in Sect.~\ref{sec:numerical_details}. More importantly, at times relevant for this study, $t\gtrsim 10^8$ yr, the plot shows that the formula by \cite{christensen09} is always pretty insensitive to the minimum pressure from which the integral in eq.~(\ref{eq:bdyn_C09}) is evaluated. This is due to the convective heat flux $q_c(r)$, which is zero in the radiative region, and, together with the density, gives the bulk of the contribution to the integral from the dynamo region. Secondly, the simplified formula (\ref{eq:bdyn_R09}) evaluated at $p_{\rm dyn}$ is seen to systematically provide a factor $\sim 3$ less than the original one, in a broad range of planetary models (including different planetary masses, $\Teq$ and $v_{\rm avg}$, here not shown). The two curves would be compatible between each other for a smaller choice of $f_{\rm dip}\sim 0.06$ (but we noticed that in the original work, \cite{christensen09} employed $f_{\rm ohm}=1$, and numerical simulations typically provide $f_{\rm ohm}\sim 0.3-0.7$, \citealt{christensen06}).

Besides this, the most important result is that, in case of a heated HJ (red curves), the choice of eq.~(\ref{eq:bdyn_R09}) evaluated at $p=10^3$ bar, provides a radically different behaviour, being almost constant in time after the Ohmic term is turned on. This is due to the fact that the additionally heat in the outer convective region, with time, becomes the most relevant contribution and provides a constant-in-time internal luminosity to balance the radiated energy. We note that \cite{yadav17} use the deposited heat as an input value for the scaling law, and they do not specify neither the mechanism nor its specific heating profile. If their approach is applied to the Ohmic dissipation, it presents conceptual issues: (i) the extra luminosity originates from the dissipation of the atmospherically induced field in the shallow layers, while the heat flux in the \cite{christensen09} scaling law has to be evaluated in the deep-seated, highly conductive dynamo region alone; (ii) having Ohmic luminosity both as a dissipation of magnetic field and a source of it via convective flux (even if in different regions) is unclear in terms of energy conservation.
In our models, which do not fix $B_{\rm atm}$, the use of eq.~(\ref{eq:bdyn_R09}) evaluated at $p=10^3$ bar would provide a more complicated curve, due to the coupling, eqs.~(\ref{eq:j_atm}) and (\ref{eq:bdip}), but the arguments above still hold.

Finally, keep in mind an additional caveat which applies to any version of the scaling law. As discussed in the main text and shown by previous works \citep{komacek17,komacek20,elias25b} a strong deposition of heat in Ohmic models (with the heat concentrated in the outer region) can lead to a fragmentation of the large-scale convection into thinner layer(s). This kind of dynamo deviates from the almost fully convective objects against which the scaling laws were calibrated \citep{christensen09}, in a similar way to the cases of the aforementioned majority of Solar planets with thin dynamos, excluded a priori from the original correlation.

Considering all this, and the lack of alternative laws calibrated specifically for HJs (which lack observational data on magnetism), we employ the original scaling law by \cite{christensen09}, integrating over the nominal dynamo region, $p>p_{\rm dyn}=10^6$  bar.

\section{Estimation of currents in the dynamo region}

Given the absence of direct magnetic field measurements beyond the Solar System, we use Jupiter as a case study to investigate typical magnetic field length scales. Our analysis combines in-situ observations with MHD dynamo simulations. While HJs may exhibit different magnetic topologies, \cite{elias25b} demonstrated that they are most likely in the fast-rotation regime, similar to Jupiter, implying no major deviations in the multipolar character of the dynamo. In any case, the purpose of these calculations is to estimate the order of magnitude of the internal currents, enabling comparison with those induced by atmospheric processes.

\subsection{Jovian magnetic field curvature}\label{app:curvature}

The Jovian magnetic field has been accurately measured by the Juno spacecraft and published in 2018 \citep{connerney18} from the first 9 orbits and refined in 2021 after mission completion \citep{connerney22}. In the area between the planet's dynamo region and the magnetosphere/magneto-disk, where there are no free currents, the magnetic field can be expressed as the gradient of a scalar potential,
    \begin{equation}
	\m{B} = -\m{\nabla} V = - \m{\nabla} (V_{\text{int}}+V_{\text{ext}}) \, ,
    \end{equation}
where we have separated the internal (dynamo-generated) and external (magnetosphere/magneto-disk) components. Public data is given in terms of the set of constants $g_n^m$, $h_n^m$, $G_n^m$ and $H_n^m$ (the Schmidt coefficients) which define the spherical harmonic expansions up to some specific multipole degree using the Schmidt quasi-normalised associated Legendre polynomials $P_n^m(\cos\theta)$,
    \begin{equation}
    V_{\text{int}} = a \sum_{n=1}^{n_{\rm max}} \left( \frac{a}{r} \right)^{n+1} \sum_{m=0}^{n} P_n^m(\cos\theta) \left[ g_n^m \cos(m\phi)+h_n^m \sin(m\phi) \right] \, ,
    \end{equation}
and
    \begin{equation}
    V_{\text{ext}} = a \sum_{n=1}^{n_{\rm max}} \left( \frac{r}{a} \right)^{n} \sum_{m=0}^{n} P_n^m(\cos\theta) \left[ G_n^m \cos(m\phi)+H_n^m \sin(m\phi) \right] \, .
    \end{equation}
Analytical expressions can be used to obtain all components of $\m{B}$ in spherical coordinates at any given radius. The 2018 model, JRM09, has a maximum degree $n_{\rm max}$ of 9 for the internal component only. On the other hand, the 2021 model, JRM33, has reasonably well resolved coefficients all the way up to $n_{\rm max} = 13$ (with some useful information up to $n_{\rm max} = 18$) for the internal component, and $n_{\rm max} = 2$ for the external component.

\begin{figure}
\centerline{\includegraphics[width=\hsize]{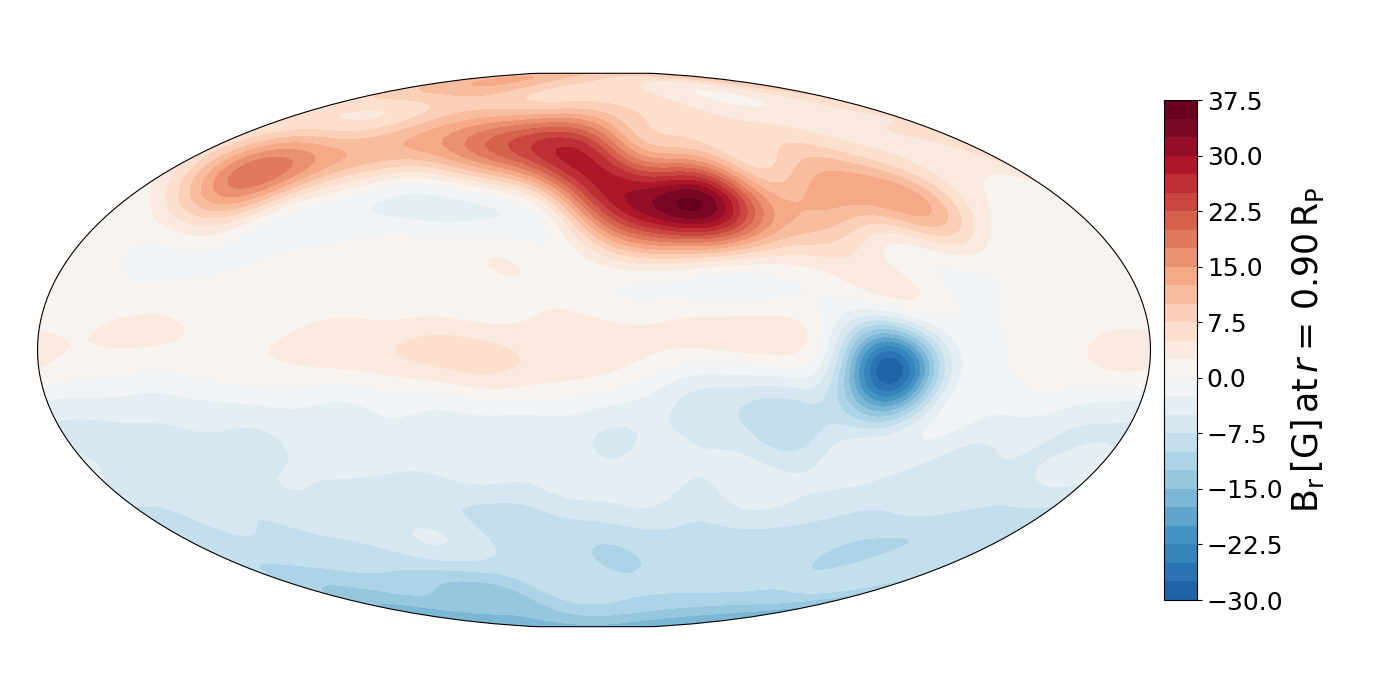}}
\centerline{\includegraphics[width=\hsize]{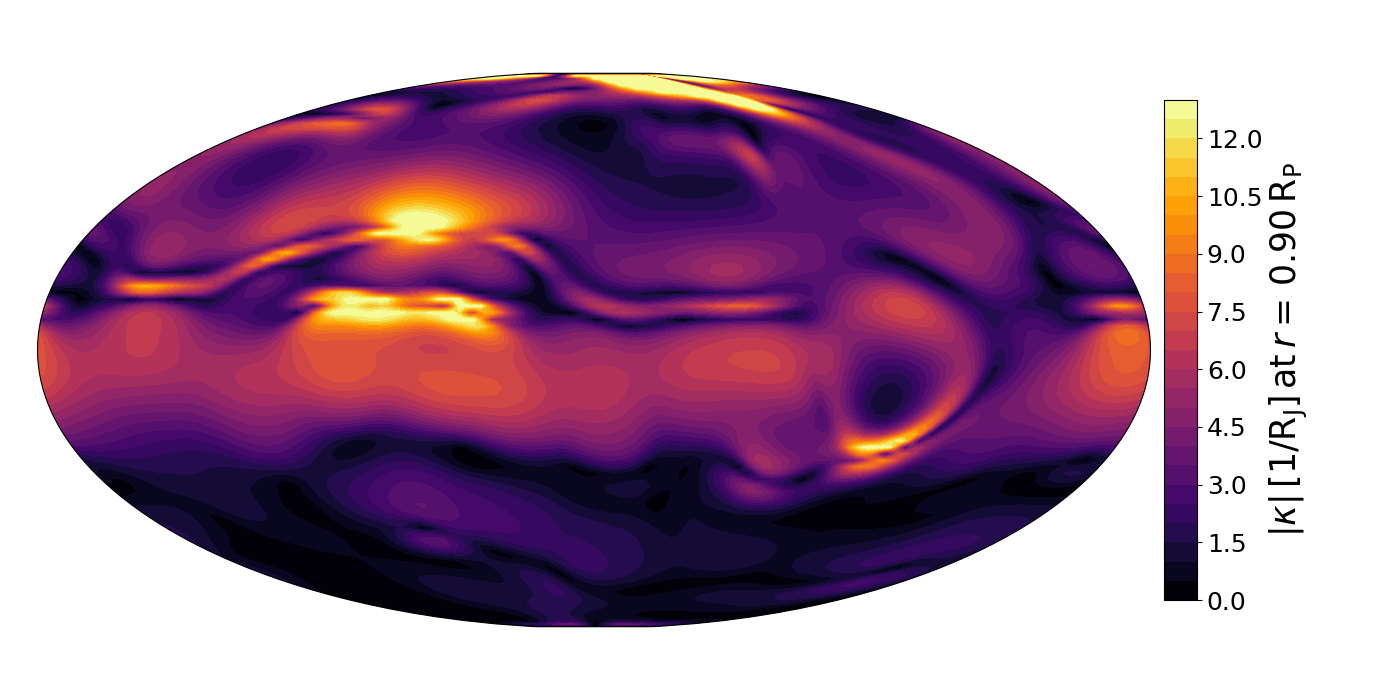}}
\caption{Radial component (top) and curvature modulus (bottom) of Jupiter's magnetic field at 0.9 planetary radius in the Mollweide projection, reconstructed with the model JRM33 ($n_{\rm max}=13$).}
\label{Jupiter_radial_field_curvature}
\end{figure}

The curvature of the magnetic field is defined as $\m{\kappa} = (\m{b} \cdot \m{\nabla}) \m{b}$, with the corresponding unitary vector field $\m{b} = \m{B} / |\m{B}|$. The mean of the modulus of the curvature at a given radius is,
    \begin{equation}
	\langle |\boldsymbol{\kappa}(r=r_{\kappa})| \rangle = \frac{\int |\boldsymbol{\kappa}(\theta, \phi; r=r_{\kappa})| \sin\theta d\theta d\phi}{\int \sin\theta d\theta d\phi} \, .
    \label{mean_curvature}
    \end{equation}

\begin{figure}
\centerline{\includegraphics[width=\hsize]{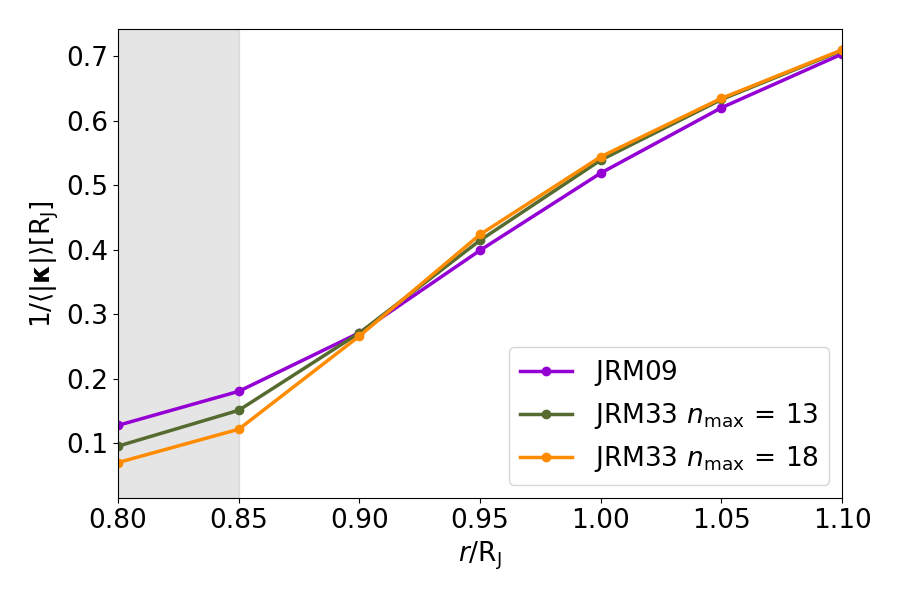}}
\caption{Inverse of the mean curvature modulus for different Jupiter magnetic field models as a function of radius, obtained using a discretised version of equation (\ref{mean_curvature}). The JRM33 models do not significantly change if the external multipoles are included.}
\label{Jupiter_curvature_mean}
\end{figure}

\noindent Fig.~\ref{Jupiter_curvature_mean} shows the inverse of the mean curvature at different radii for both models, including two different cut-offs for JRM33. The gray zone approximately marks the possible start of the conductive zone where the gradual appearance of metallic hydrogen would invalidate the potential solution. The inverse curvature has a mean value of the order of $1/\kappa\sim 0.1~\Rj$ at $r\sim (0.8-0.9)~\Rj$ (where the dynamo surface should be), which can be taken as a lower limit of the magnetic field curvature inside the Jovian dynamo.

\subsection{Details of MHD dynamo simulations}\label{app:magic}

The code {\tt MagIC}\footnote{\url{https://github.com/magic-sph/magic}} \citep{jones11,gastine12} has been used to solve the MHD equations in a spherical shell under the anelastic approximation with the {\tt MESA} radial structure as background profiles. As explained in detail in \cite{elias25}, we implement the radial dependence of density, temperature, gravity, thermal expansion coefficient and the Gr\"uneisen parameter with a very high degree polynomial, fitting them better than $\lesssim 1\%$. We assume constant thermal and viscous diffusivities and we adopt the conductivity profile (and thus a corresponding magnetic diffusivity) first defined in \cite{gomezperezetal2010} which consists of an approximately constant conductivity in the hydrogen metallic region joined (at $r=r_m$ with $\sigma(r)=\sigma_m$) via a polynomial to an exponentially decaying outer molecular region,
    \begin{equation}
    \tilde{\sigma}(r) = \begin{cases} 1+ (\sigma_m-1)\left(\dfrac{r-r_i}{r_m-r_i}\right)^a \quad \hbox{for}\quad r<r_m~, \\
    \sigma_m \exp \left[a \left(\dfrac{r-r_m}{r_m-r_i}\right)\dfrac{\sigma_m-1}{\sigma_m}\right]
    \quad\hbox{for}\quad r\geq r_m~. \end{cases} 
    \end{equation}
Many works have employed this profile for gas giant convection and dynamo modelling \citep{duarte13,duarte18,wicht19a,wicht19b,gastine21,yadav22} with values of $\sigma_m$ and $a$ ranging approximately from 0.9 to 0.01 and from 1 to 25, respectively. In our models, we have arbitrarily chosen $\sigma_m=0.6$ and $a=11$, and we fix $r_m=R_{\rm dyn}$, i.e. the radius of $10^6$ bar. Due to numerical limitations, dynamo models cannot cover the entire density range of the dynamo region and we have to consider a more limited density contrast (ratio between the innermost and outermost densities) $N_\rho = \ln \rho_i/\rho_o \lesssim 4.6$. We thus externally cut the profiles of {\tt MESA} to achieve a density contrast $N_\rho$ of either 3.0 or 4.6.

For example, for the models shown in Fig.~\ref{fig:Magic_Hot_Jupiter_model_curvature_mean}, the background {\tt MESA} profiles correspond to an irradiated $1 \Mj$ planet with internal Ohmic heating at 10~Gyr. At that time, the simulated planetary radius is $1.218~\Rj$, corresponding to an external pressure of and 64 and 5.4~kbar, respectively. We used resolutions of $(N_r,N_{\theta},N_{\phi}) = (193,256,512)$ for the cases with $N_\rho=3.0$, and $(N_r,N_{\theta},N_{\phi}) = (241,320,640)$ for the cases with $N_\rho=4.6$. In order to explore how sensitive the magnitudes in Fig.~\ref{fig:Magic_Hot_Jupiter_model_curvature_mean} are, we vary the Prandtl, magnetic Prandtl, and Rayleigh numbers, the value of $N_\rho$, and the boundary conditions. Generally speaking, the models shown above reach a steady state dynamo with equatorial jets. More details and the main results of these simulations will be presented in an upcoming paper dedicated to the dynamo in HJs, closely following the methodology employed in \cite{elias25b,elias25}. For the specific purpose of this work, we are only interested in the profiles shown in Fig.~\ref{fig:Magic_Hot_Jupiter_model_curvature_mean}.

\section{Results from standard cooling models}\label{app:benchmark}

\begin{figure*}
\centerline{\includegraphics[width=.36\textwidth]{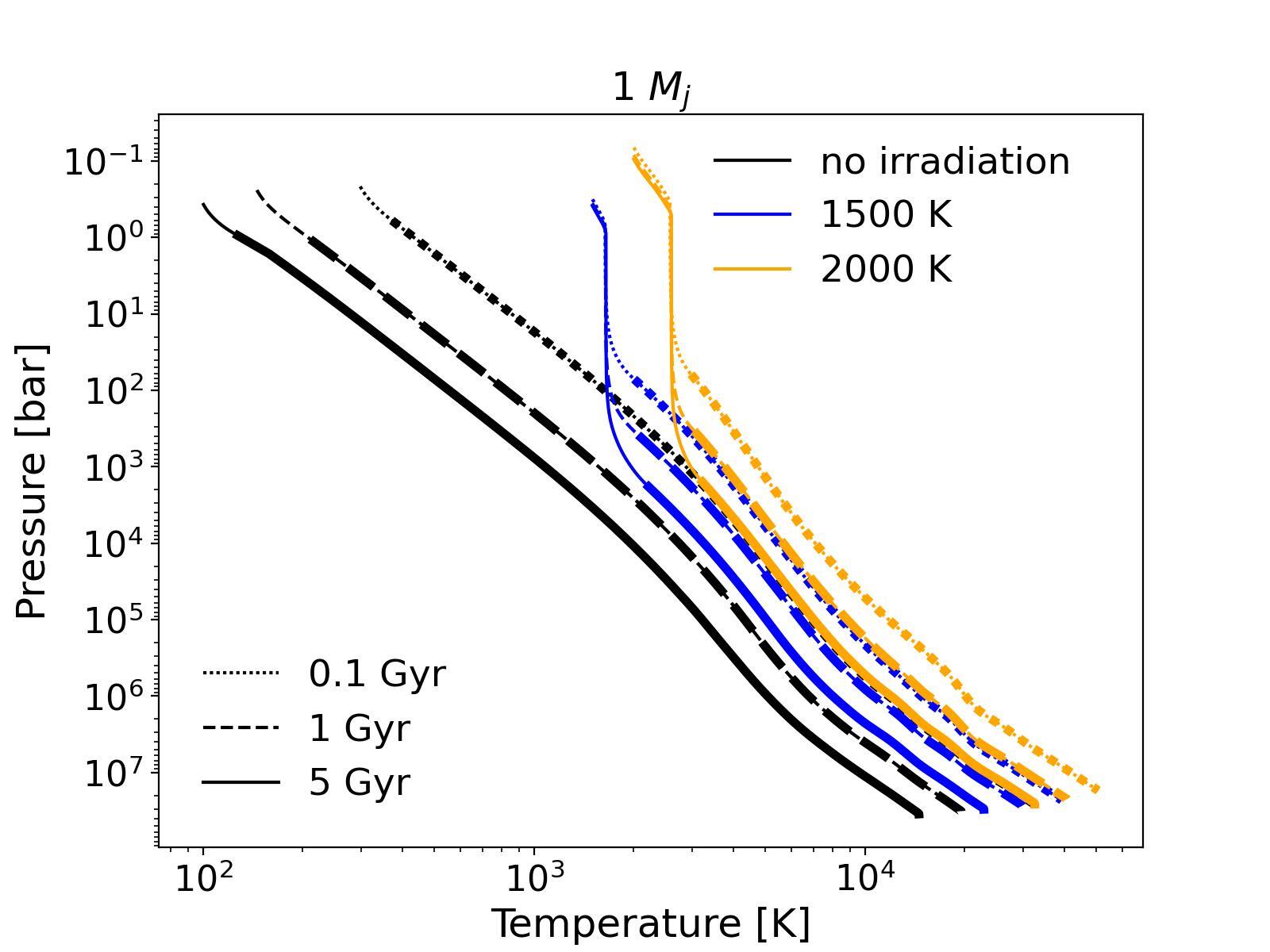}
\includegraphics[width=.32\textwidth]{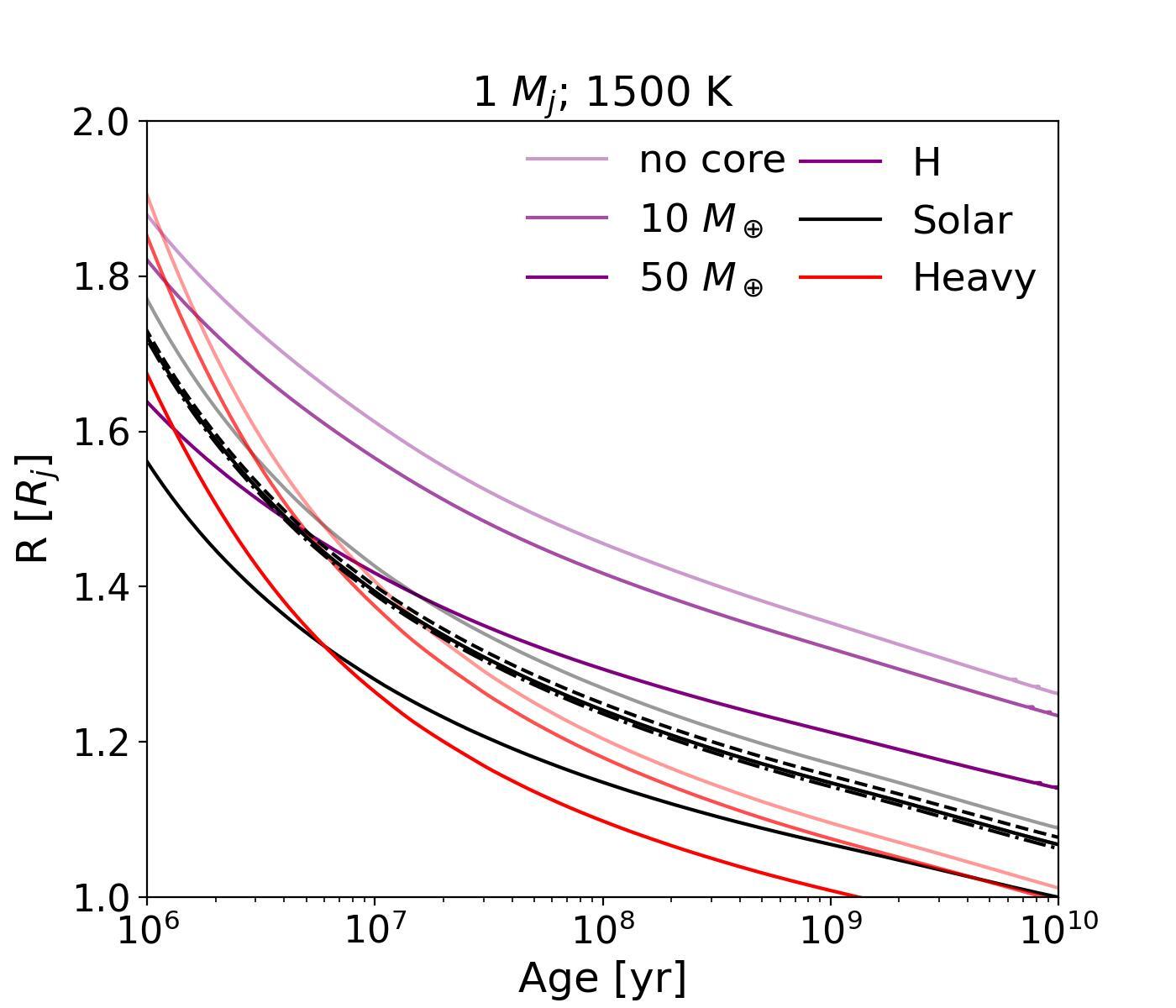}
\includegraphics[width=.32\textwidth]{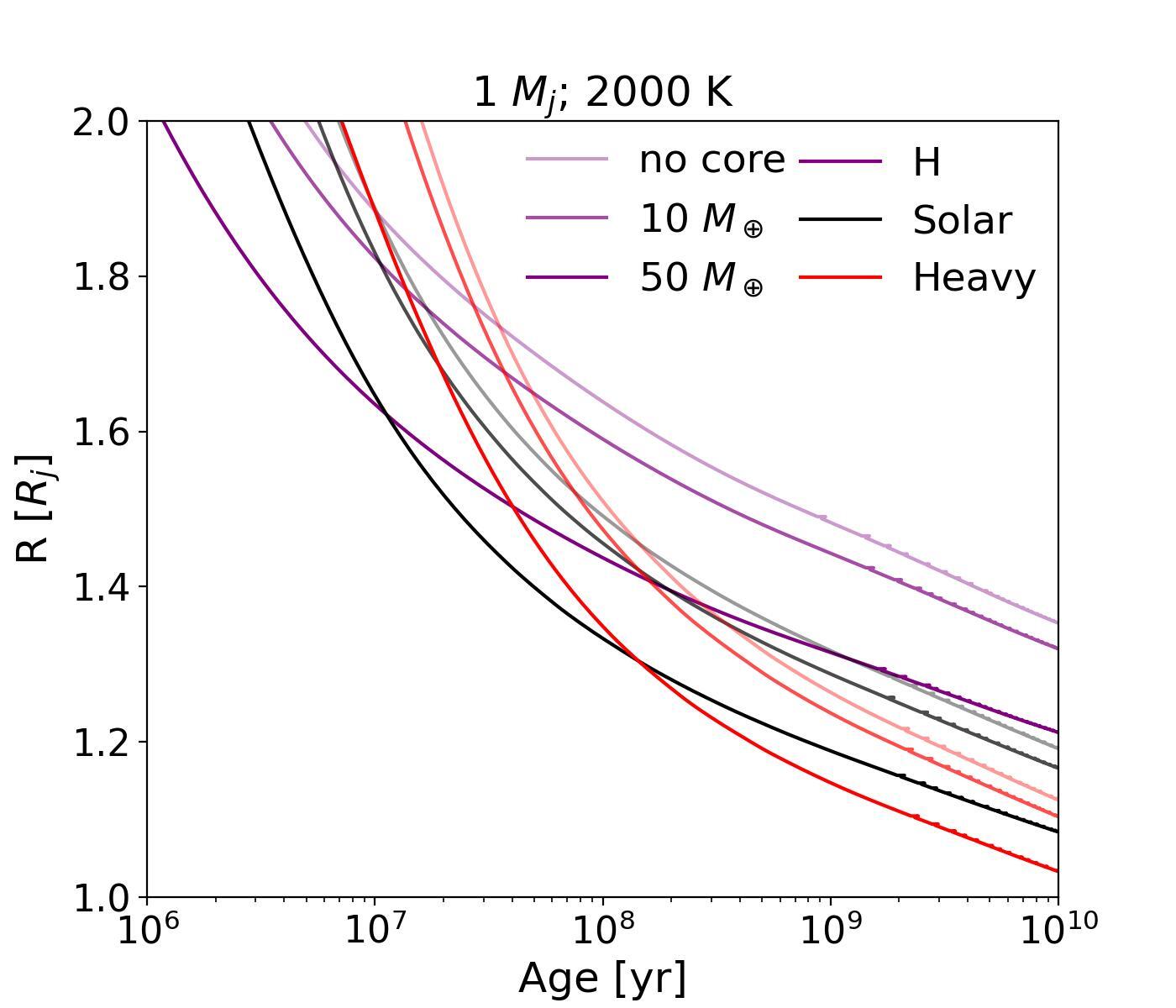}}
\centerline{\includegraphics[width=.36\textwidth]{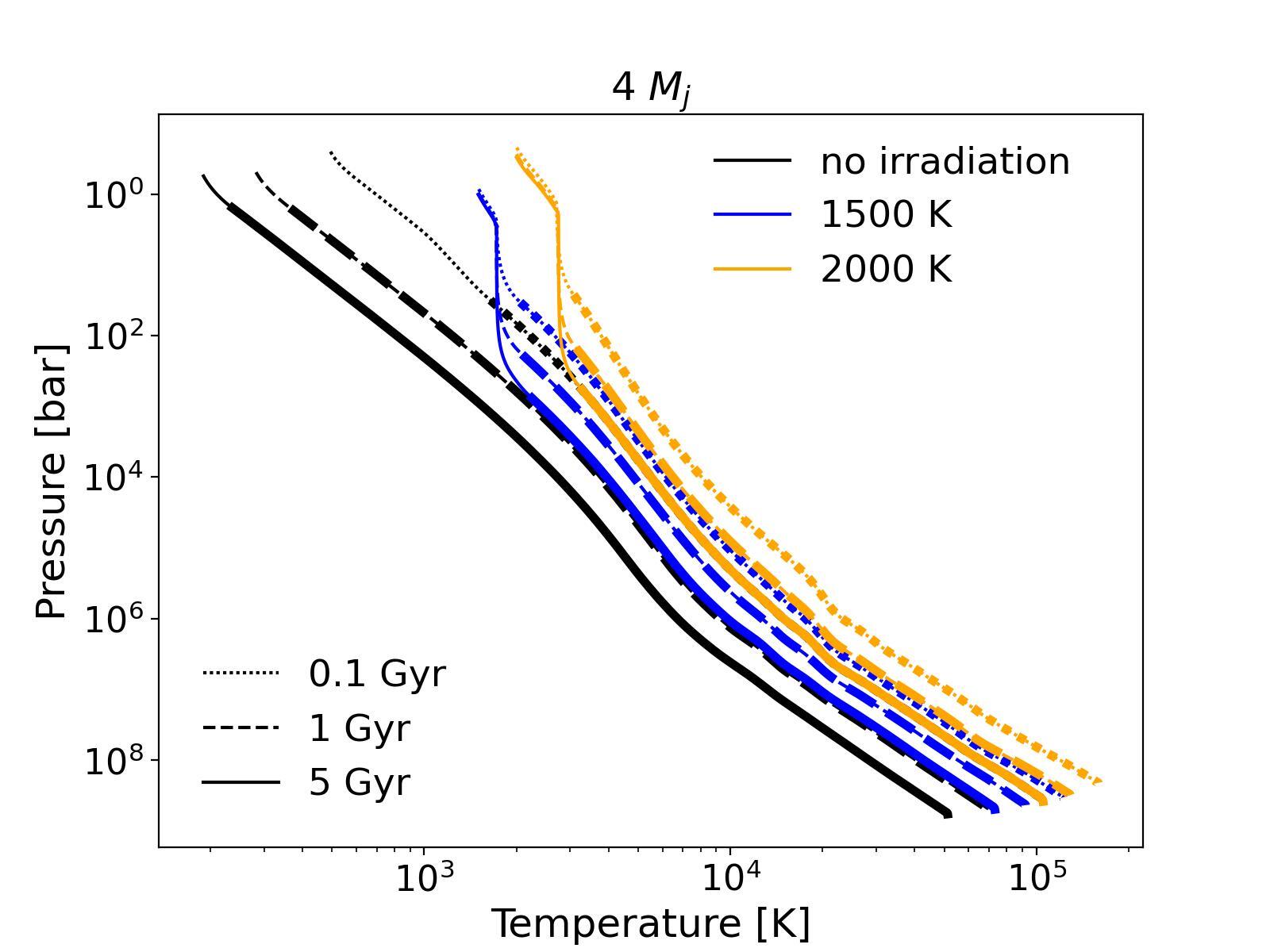}
\includegraphics[width=.32\textwidth]{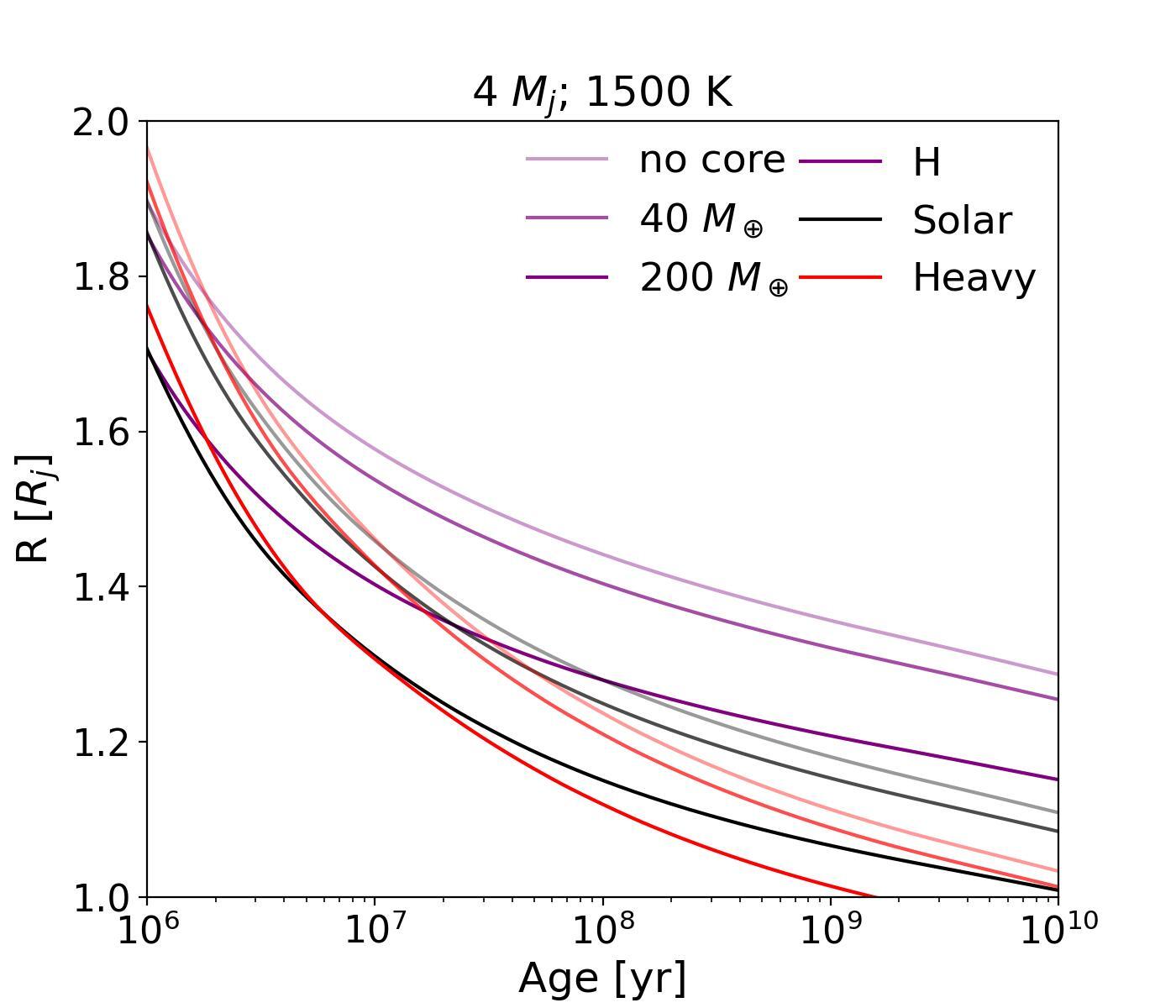}
\includegraphics[width=.32\textwidth]{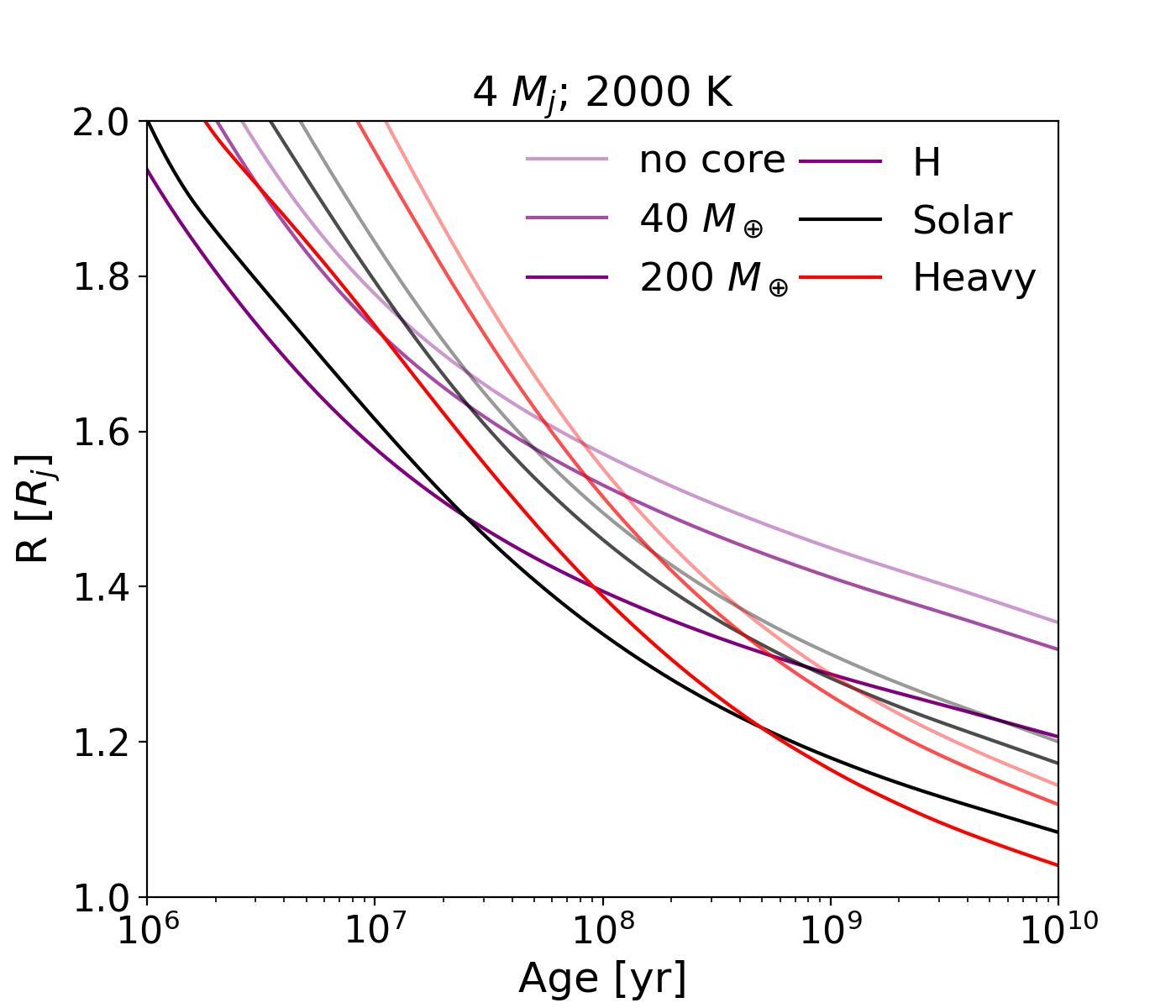}}
\caption{Evolution of irradiated HJs with no additional heating, $\epsilon_{\rm j}=0$, for models with $M=1~\Mj$, $\rho_c=10$ \gcc~(top row) or $M=4~\Mj$, $\rho_c=45$ \gcc~(bottom row). {\em Left panels:} $p(T)$ profiles at ages $t\sim$ 0.1, 1 and 5 Gyr (dots, dashes, solid lines, respectively), comparing the cases without (black) and with irradiation, $\Teq = 1500$~K (blue) and 2000~K (orange), all with solar composition. The thicker lines correspond to the convective zone, which extends from the RCB ($p\sim 1-10^3$ bar) to the outer core ($p > 10^7$ bar). {\em Central panels:} Radius evolution of planets with $\Teq=1500$ K and different core masses ($M_c=0, 10~(M/\Mj), 50~(M/\Mj) ~\Me$, increasing for darker shades, all with solar composition) and composition of the envelope: pure hydrogen (purple), solar (black) and higher helium and heavier elements content, $(Y,Z)=(0.35,0.035)$ (purple, black and red, respectively, all with $M_c=10~\Me$). In the top central panel we also show the effects of variation of $\rho_c=5$\gcc~(dashes) and $\rho_c=20$\gcc~(dot-dashed lines) for the representative $M_c=10~\Me$, solar composition case: since the effect is rather small compared to varying other parameters, we show it only in that particular panel. {\em Right panels:} Same as the central panels, but for $\Teq=2000$ K.}
\label{fig:standard_cooling}
\end{figure*}

As a reference and comparison with well-known results of classical cooling, we briefly show illustrative examples for the planetary structure with no extra heating, $\epsilon_{\rm j} = 0$ in eq.~(\ref{eq_luminosity}). In Fig.~\ref{fig:standard_cooling} we summarise the main results for a few representative cases with $M=1~\Mj$ (top panels) and $M=4~\Mj$ (bottom panels). In the left panels we compare the $p(T)$ profiles of the non-irradiated case (black), and HJs with $T_{\rm eq}=1500$~K (blue) and 2000 K (orange), at three different ages: 0.1, 1 and 5 Gyr (different line styles). In the same plot, we indicate the position of the convective (thick lines) and radiative layers (thin lines). In all cases, as the internal heat decreases over time, the $p(T)$ lines naturally move downwards. For a cold Jupiter, the radiative-convective boundary (RCB) always remains very shallow ($\sim 1$ bar), and additional radiative regions at intermediate pressure can appear as the internal entropy drops \citep{ginzburg15,komacek17}. For HJs, instead, irradiation causes much thicker outermost layers to become radiative. The specific profiles depend on the atmospheric modelling: in the simple {\tt MESA} recipes, the profiles are isothermal, with a drop in temperature in the region $m<\Sigma_\star$; non-grey atmospheric models provide non-isothermal, longitude-dependent profiles, including thermal inversion, for example, \cite{gandhi19}. Such details are of primary importance for atmospheric modelling and data interpretation, but are not relevant for this work, since HJ radius inflation has to be driven from much deeper layers.

Regardless of these details, the irradiation pushes the RCB to much deeper levels, typically $p\sim 10^3$ bar at gigayear ages. Moreover, when comparing the $p(T)$ curves of different $\Teq$, the blanketing effect of the highly irradiated layers is evident: the interior is kept warmer for HJs with higher $\Teq$. These preliminary considerations on irradiation and RCB are in agreement with what has been widely discussed in previous works, e.g. \cite{arras06, fortney07,ginzburg15, komacek17,thorngren19}.

In the central panels of Fig.~\ref{fig:standard_cooling}, we show how the irradiation (the same cases as the left panel), core mass and density impact the radius evolution. We vary the core mass: $M_c=(0,10,50)~\Me$. For the composition, we vary the baseline values (solar), considering two additional cases: pure hydrogen, $Y=Z=0$, and heavy composition, with $(Y,Z)=(0.35,0.035)$, covering the upper range of the dispersion observed in stellar helium abundances and metallicities \citep{verma19}. Higher values of metallicities are not discarded \citep{thorngren16}, and have the effect of further contracting the planet. Since we are focusing on explaining the large inflated radii, we do not explore them further. We also show the negligible effect of varying the core density alone by a factor of 2 (dashed and dot-dashed lines in the top central panel): In the rest of the paper, we use fixed standard values of $\rho_c(M)$, as presented above.

On one side, as expected, we find positive trends between radius inflation and irradiation. The radius can increase up to $\sim 20\%$ at gigayear ages for $\Teq=2000$ K, for $M=1~\Mj$, $M_c=10~\Me$, $\rho_c=10$~\gcc, and solar composition. On the other hand, higher values of $M_c$ or $\rho_c$ make the planet more compact, while higher metallicity makes the planets cool faster.
When we look at heavier planets, of which we show the $M=4~\Mj$ cases in the bottom rows of Fig.~\ref{fig:standard_cooling}, the dispersion in the cooling curve using the same ranges of parameters is similar to the $M=1~\Mj$ HJs. This is due to the total higher inertia (and the lower relative weight of the core for the range considered here).

In summary, the total spread of $R(t)$ for HJs at gigayear ages is in the range of $\sim 1-1.4~\Rj$, with typical values around $1.2~\Rj$. This is due to the combined effects of, in order of importance, $\Teq$, $M$, $M_c$, and composition, and is compatible with earlier results by e.g. \cite{guillot96,fortney07,thorngren18}. These results represent the baseline cooling model, on top of which we add the Ohmic heating model, as described next.

\begin{figure}
\centerline{\includegraphics[width=.4\textwidth]{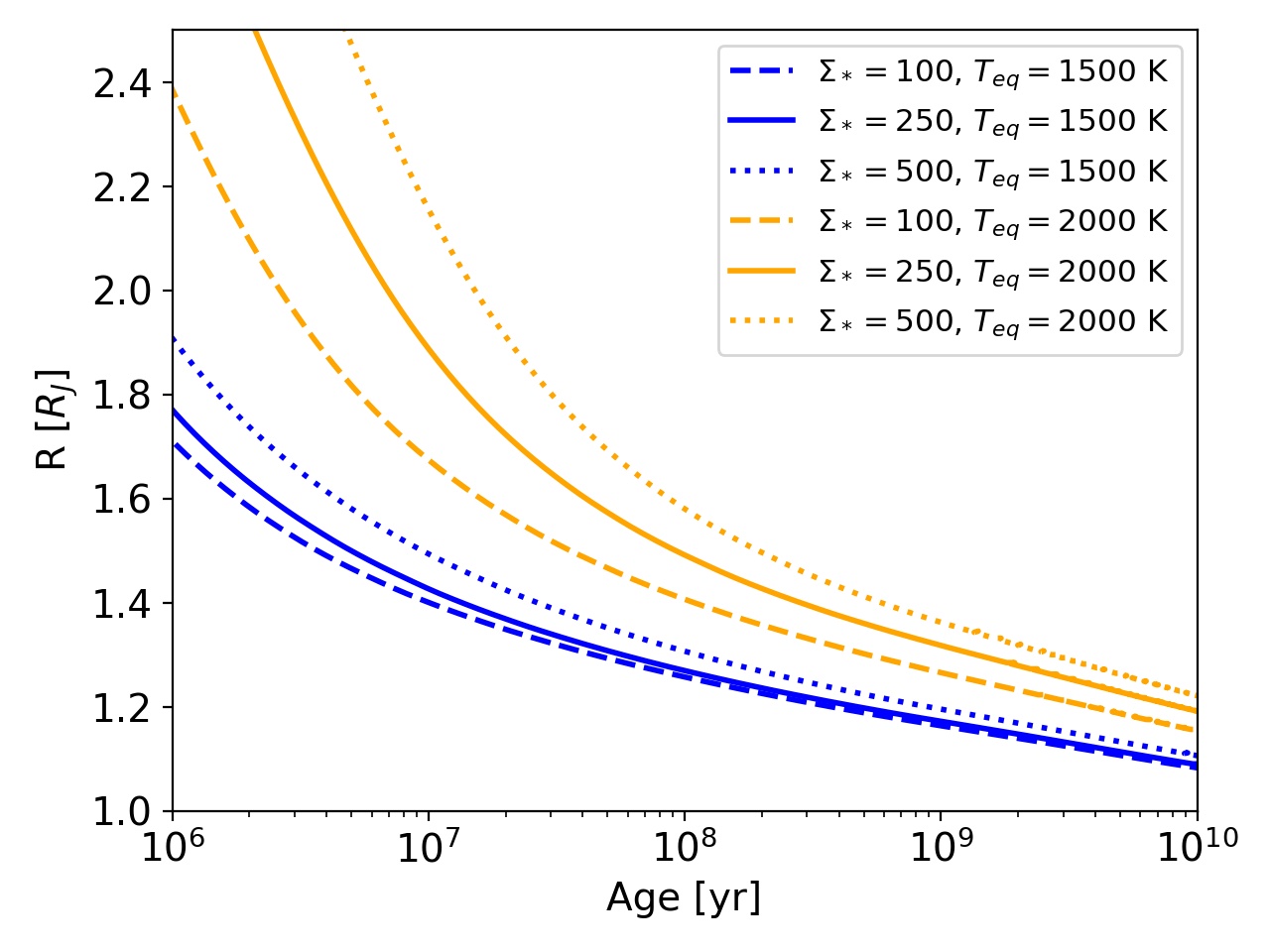}}
\caption{Evolution of the radius for models with $M=1~\Mj$, $M_c=10~\Me$, $\rho_c=10$\gcc, solar composition, no Joule heating, comparing cases with $\Teq=1500$ K (blue) and 2000 K (orange), and $\Sigma_\star=100,250,500$~\Su~(dashes, solid lines, dots, respectively).}
\label{fig:rhocore}
\end{figure}

In Fig.~\ref{fig:rhocore} we show the sensitivity of the models on a range of values $\Sigma_\star\in[100,500]$~\Su, in case of $\Teq=1500$ and 2000 K. This value approximately corresponding to homogeneous (per unit mass) absorption of the irradiation down to pressures from $\sim$0.5 bar to $\sim$1.2 bar for gigayear ages. Remember that $\Sigma_\star$ effectively models in a simple way how deeply the radiation is absorbed, which in reality depends on the specific stellar spectrum and the wavelength-dependent opacities (i.e. composition and atmospheric structure) of each planet. We see an important effect for early ages, but at gigayear ages, the impact is of a few percent at most. We note that \cite{owen16}, in the context of atmospheric mass loss by highly irradiated super Earths, proved a similar limited sensitivity of results on a broader range $\Sigma_\star\in[50,1250]$~\Su.

\end{appendix}

\end{document}